\documentclass[twocolumn]{aastex63}
\usepackage[normalem]{ulem}
\usepackage[version=4]{mhchem}
\usepackage{multirow}
\usepackage{appendix}
\usepackage{color}
\usepackage{xcolor}
\usepackage{txfonts}
\usepackage{placeins}
\usepackage{verbatim}
\usepackage{array}
\graphicspath{{./}{Figures/}}

\received{...}
\revised{...}
\accepted{...}
\shorttitle{iSHELL $M$-band disk survey}

\begin{document}

\title{Scanning disk rings and winds in CO at 0.01--10~au: a high-resolution $M$-band spectroscopy survey with IRTF-iSHELL}

\author{Andrea Banzatti}

\author{Kirsten M. Abernathy}
\affil{Department of Physics, Texas State University, 749 N Comanche Street, San Marcos, TX 78666, USA}

\author{Sean Brittain}
\affiliation{Department of Physics \& Astronomy, 118 Kinard Laboratory, Clemson University, Clemson, SC 29634, USA}

\author{Arthur D. Bosman}
\affil{Department of Astronomy, University of Michigan, 1085 S. University Ave, Ann Arbor, MI 48109}

\author{Klaus M. Pontoppidan}
\affiliation{Space Telescope Science Institute 3700 San Martin Drive Baltimore, MD 21218, USA}

\author{Adwin Boogert}
\affiliation{Institute for Astronomy, University of Hawaii, 2680 Woodlawn Drive, Honolulu, HI, USA 96822}

\author{Stanley Jensen}
\affiliation{Department of Physics \& Astronomy, 118 Kinard Laboratory, Clemson University, Clemson, SC 29634, USA}

\author{John Carr}
\affiliation{Department of Astronomy, University of Maryland, College Park, MD 20742, USA}

\author{Joan Najita}
\affiliation{NSF’s NOIRLab, 950 N. Cherry Avenue, Tucson, AZ 85719, USA}

\author{Sierra Grant}
\affiliation{Max-Planck-Institut für extraterrestrische Physik,
Gießenbachstraße 1, 85748 Garching bei München}

\author{Rocio M. Sigler}
\author{Michael A. Sanchez}
\affil{Department of Physics, Texas State University, 749 N Comanche Street, San Marcos, TX 78666, USA}

\author{Joshua Kern}
\affiliation{Department of Physics \& Astronomy, 118 Kinard Laboratory, Clemson University, Clemson, SC 29634, USA}

\author{John T. Rayner}
\affiliation{Institute for Astronomy, University of Hawaii, 2680 Woodlawn Drive, Honolulu, HI, USA 96822}

\begin{abstract}

We present an overview and first results from a $M$-band spectroscopic survey of planet-forming disks performed with iSHELL on IRTF, using two slits that provide resolving power R~$\approx$~60,000--92,000 (5--3.3~km/s). iSHELL provides a nearly complete coverage at 4.52--5.24~$\mu$m in one shot, covering $>50$ lines from the R and P branches of \ce{^{12}CO} and \ce{^{13}CO} for each of multiple vibrational levels, and providing unprecedented information on the excitation of multiple emission and absorption components. Some of the most notable new findings of this survey are: 1) the detection of two CO Keplerian rings at $<2$~au (in HD~259431), 2) the detection of \ce{H2O} ro-vibrational lines at 5~$\mu$m (in AS~205~N), and 3) the common kinematic variability of CO lines over timescales of 1--14~years. By homogeneously analyzing this survey together with a previous VLT-CRIRES survey of cooler stars, we discuss a unified view of CO spectra where emission and absorption components scan the disk surface across radii from a dust-free region within dust sublimation out to $\approx10$~au. We classify two fundamental types of CO line shapes interpreted as emission from Keplerian rings (double-peak lines) and a disk surface plus a low-velocity part of a wind (triangular lines), where CO excitation reflects different emitting regions (and their gas-to-dust ratio) rather than just the irradiation spectrum. A disk+wind interpretation for the triangular lines naturally explains several properties observed in CO spectra, including the line blue-shifts, line shapes that turn into narrow absorption at high inclinations, and the frequency of disk winds as a function of stellar type.

\end{abstract}


\section{Introduction} \label{sec: intro}
While ALMA observations trace cold dust and gas in protoplanetary disk at $>$~10~au and reveal large-scale radial structures in great detail \citep[][for a review]{andrews20}, the higher temperature, density, and irradiation of inner disks at $<$~10~au results in the excitation of rotational, ro-vibrational, or electronic bands of several molecules (\ce{H2}, \ce{CO}, \ce{H2O}, OH, HCN, \ce{C2H2}, \ce{CO2}) that can be studied using high-resolution spectroscopy at infrared and shorter wavelengths \citep[e.g.][]{najita07,cn08,france12,pont14}. Of these molecules, CO is the most easily and commonly observed in disks.

\begin{table*}[ht]
    \centering
    \caption{Summary of 20 years of high-resolution $M$- and $L$-band spectroscopy of protoplanetary disks, reporting only the largest samples obtained with the three leading instruments. }
    \label{tab: surveys}
    \begin{tabular}{| p{1.7cm} | p{1.25cm} | p{1.6cm}| p{11cm} |}
    \hline
    \textbf{Instrument} & \textbf{Resolving power} & \textbf{Spectral coverage ($\mu$m)} & \textbf{Samples and references} \\
    \hline
    \hline
    \multirow{2}{4em}{Keck-NIRSPEC} & 25,000 & 4.65 -- 4.78 4.96 -- 5.1 & 16 T~Tauri \citep{najita03}, $\approx150$ T~Tauri and Herbig Ae/Be \citep{blake04,salyk09,salyk11,salyk13}, 24 T~Tauri \citep{doppmann17}, 14 Herbig Ae/Be \citep{brittain07} \\
     & 25,000 & $2.9-3.7$  (with gaps) & $\approx50$ T~Tauri and Herbig Ae/Be (PI: Blake), 7 Herbig Ae/Be \citep{brittain16} \\
    
    \hline
    \multirow{2}{4em}{VLT-CRIRES} & 94,000 & 4.65 -- 4.75 & 69 mostly T~Tauri \citep[][; *]{pont11,brown13}, 13 Herbig Ae/Be \citep{vdplas15}, 6 Herbig Ae/Be \citep{heinbert16} \\
     & 94,000 & 2.91 -- 2.97 & 30 T~Tauri \citep{banz17}, 11 Herbig Ae/Be \citep{fedele11} \\

    \hline
    \multirow{2}{4em}{IRTF-iSHELL} & 60,000 -- 92,000 & 4.52 -- 5.24 & 31 mostly Herbig Ae/Be \citep[][and this work]{banz18,brittain18,abernathy21} \\
     & 60,000 & 2.7 -- 3.0 & 12 Herbig Ae/Be and T~Tauri (PIs: Brittain, Banzatti) \\
    \hline
    \hline
    \end{tabular}
    \tablecomments{The third column reports the coverage of spectral settings most used; in a few cases, a larger coverage was obtained by observing multiple settings. Additional references: * \citet{bast11,herczeg11,smith15}. Smaller samples ($<10$~objects) that are not included in the table have been obtained with CSHELL or IRCS with $R \approx$~20,000, and Phoenix with $R \approx$~50,000 \citep[][]{carr01,najita03,goto06,brittain16}}
\end{table*}

The energy of CO fundamental ($\Delta v = 1$) transitions correspond to infrared wavelengths around 4.6~$\mu$m (in the $M$ band), in a spectral region that is relatively transparent in Earth's atmosphere and that can be observed from the ground. Thanks to that, the advent of the first sensitive spectrographs providing high resolving power ($R > 20,000$) three decades ago, especially NIRSPEC on the Keck telescopes \citep{nirspec}, marked the beginning of a series of successful observations of the fundamental CO ro-vibrational bands in protoplanetary disks.
The strong potential of these observations to study molecular gas in inner disks was soon identified \citep{najita03,najita07}: the relatively high Einstein A-coefficients and low excitation temperature of fundamental CO transitions, together with the resistance of the CO molecule to thermal and photo-dissociation \citep[e.g.][]{mitchell84,bruderer13,bosman19}, make these bands the most ubiquitous tracer of warm molecular gas in disks over a broad range of temperatures ($\approx$~100 -- 3000~K) and column densities ($\approx$~$10^{15}$ -- $10^{20}$~cm$^{-2}$). In fact, fundamental CO emission has been observed in disks across at least three orders of magnitude in stellar accretion rates from the young embedded phases of class I disks \citep{herczeg11} down to the dispersal phases of disks with inner dust cavities and class III disks \citep{salyk09,doppmann17}, and in disks around stars spanning a factor $>10$ in mass \citep[0.3 -- 4 $M_{\odot}$, e.g.][]{brown13,vdplas15,banz18}, for a total sample today of more than 200 disks, mostly in nearby star-forming regions within 200~pc.
Table \ref{tab: surveys} reports a summary of the instruments and surveys that have been most productive in high-resolution $M$-band spectroscopy of disks in the past 20 years.
For completeness, we include in the table also a summary of high-resolution $L$-band spectroscopy; these spectra are obtained with the same instruments and present a similar richness and complexity in emission from other molecules, including \ce{H2O}, \ce{OH}, \ce{HCN}, and \ce{C2H2} \citep[e.g.][]{fedele11,mandell12,brittain16,banz17,adams19}.

Following the pioneering work done in the first decade mostly with Keck-NIRSPEC, a new transformative step forward has been enabled with the advent of CRIRES on the VLT \citep{crires}, providing the very high resolving power of $R \approx 94,000$ with its narrowest slit (0".2). 
$M$-band spectroscopy with CRIRES soon demonstrated the strong potential of CO fundamental emission to spectrally-resolve to high detail the gas kinematics in emission and absorption components, revealing features that have full-width-at-half-maximum (FWHM) as narrow as the resolution limits ($\approx$~3~km/s) and an extremely rich variety of line shapes with kinematic structures and asymmetries that can be best observed at high resolutions of $R > 50,000$ \citep{pont08,brown13}.
The higher resolving power also increased the contrast and detections of narrow lines from CO isotopologues (\ce{^{13}C^{16}O}, \ce{^{12}C^{17}O}, and \ce{^{12}C^{18}O}) enabling studies of their abundance ratios \citep{smith15}.

$M$-band spectroscopy of fundamental CO emission has proven to be key also in revealing irradiation and excitation conditions in different inner disk regions and through different mechanisms. Collisional excitation in LTE will populate transitions following a Boltzmann distribution reflecting the local kinetic gas temperature \citep[e.g.][]{brittain07}, generally producing lower vibrational ratios $v2/v1 < 0.3$ unless high temperatures or large column densities are observed \citep[$> 1000$~K and $> 10^{18}$~cm$^{-2}$, e.g.][]{bosman19}. 
Radiative pumping from stellar and local dust IR photons populates CO excited states from the ``bottom-up'', i.e. populating the $v=1-0$ transitions first, while UV pumping from the stellar or accretion spectrum populates an excited electronic state that will then cascade ``top-down" into a number of highly excited vibrational levels whose relative population is expected to reflect the temperature of the impinging UV field \citep{scoville80,brittain03,thi13}. Both excitation mechanisms will result in non-LTE vibrational populations, while the rotational levels within each band are less easily affected and can be closer to thermalization \citep{thi13}. Evidence for all these excitation mechanisms has been found in $M$-band spectra from protoplanetary disks before \citep{brittain07,vdplas15,brown13,thi13}.
In fact, models show that the excitation and kinematics of fundamental CO emission should be very sensitive to several properties, including the local gas-to-dust ratio and dust opacity, the radial size of an inner dust cavity, the UV flux, and the vertical disk scale height, suggesting a key role for CO spectra in revealing different inner disk structures \citep[e.g.][]{woitke16,heinbert16,bosman19,antonellini20}.
Recently, observations revealed a tight correlation between the CO vibrational $v2/v1$ ratio (spanning almost a factor 100, between 0.02 and 1) and the IR excess emission from hot dust in inner disks \citep{banz18}, further supporting a key role for dust mixing and suggesting that the observed CO spectra are excited in disk regions with a wide range in gas-to-dust ratios \citep[from ISM levels of $\sim 100$ up to $>10,000$;][]{bosman19}.

Thanks to the flexible excitation properties that trace a wide range of conditions, and the use of high-resolution spectrographs on large telescopes that resolve in detail the velocity structure of the emission, the analysis of CO spectra has by now set a number of milestones in our understanding of inner disk molecular gas. 
The velocity-resolved kinematic profile of CO spectral lines has enabled to trace the emitting regions in the disk, suggesting emission as radially close to the central star as inside the dust sublimation radius and as far as tens of au in some disks, therefore tracing gas over the entire region of observed exoplanets \citep[e.g.][]{najita03,salyk11,bp15}.
Moreover, the relatively high Einstein $A$-coefficients of fundamental CO $v=1-0$ transitions have proven effective in tracing the depletion of gas in inner disks, providing a complementary tool to ALMA to study disk evolution and planet formation in the innermost disk region
\citep{brittain03,salyk09, bp15, banz17, banz18, bosman19, antonellini20}, in a few cases by directly spatially resolving gas in inner dust cavities through spectro-astrometry \citep{pont08}.
Modeling of the gas surface density drop in inner dust cavities from the CO emission observed in some disks has been found consistent with dynamical interactions with massive planets \citep{brittain07,carmona14,carmona17,bosman19}.
Variable kinematic asymmetry monitored in one disk provided evidence for an orbiting hot-spot of emission at the inner cavity wall, possibly connected to a gap-opening protoplanet \citep{brittain13,brittain14,brittain19}.
The combination of IR CO spectra and dust emission tracers showed that CO traces dust depletion in inner disks and that the excitation and kinematics of CO spectra trace different inner disk cavity structures \citep{banz18, bosman19}.
$M$-band CO spectra have also been found to trace molecular gas outflowing at moderate velocity from disks, with multiple and variable absorption components blue-shifted by 10--100~km/s observed in class I and/or episodically outbursting systems \citep[][]{brittain07b,goto11,herczeg11,banz15}. An asymmetric spectro-astrometric signal also provided evidence for fundamental CO emission to trace a slow inner disk wind in at least a few highly-accreting class II disks \citep{pont11}.

\begin{figure*}
\centering
\includegraphics[width=1\textwidth]{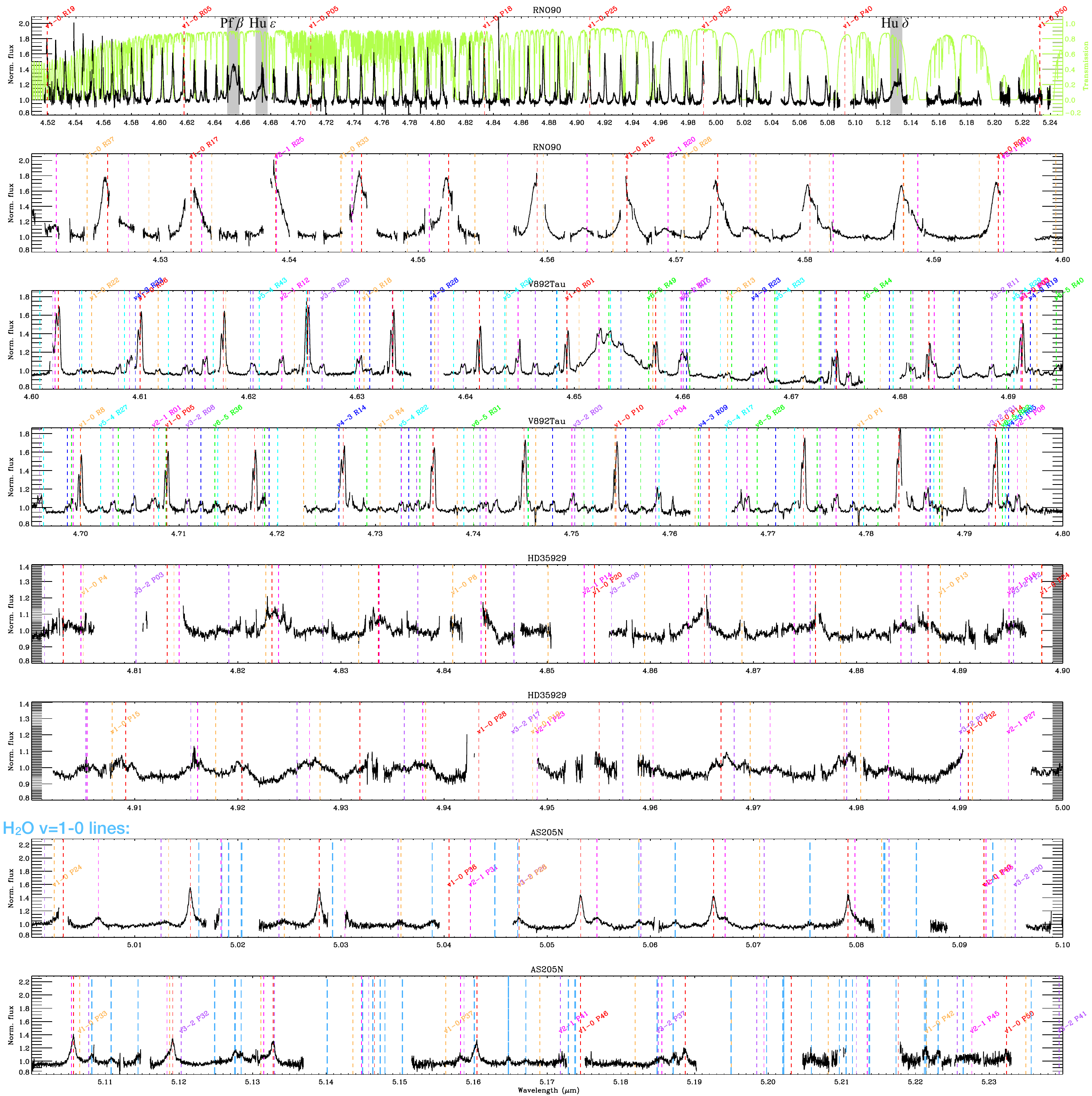} 
\caption{Overview of $M$-band spectral emission observed in this survey. The top panel shows one full spectrum to illustrate the completeness in coverage obtained with iSHELL in one shot; shaded regions mark HI emission lines and the atmospheric transmission is reported for reference in green. In the other panels, the full spectrum is broken up into portions from different targets (as indicated above each panel) to illustrate both richness and diversity of the spectra: from some of the narrowest emission lines with high vibrational excitation (up to v=6-5, in V892~Tau), to some of the broadest lines (in HD~35929). In each panel some CO lines are labeled for guidance, in orange for \ce{^{13}CO} and all other colors for \ce{^{12}CO}. AS~205~N in the last two panels illustrates the first detection of ro-vibrational water lines near $5 \mu$m (marked in light blue). }
\label{fig: spec_overview}
\end{figure*}

The first two decades of high-resolution $M$-band spectroscopy have seen pioneering observations demonstrating the power in revealing the kinematic structure, excitation conditions, and physical evolution of inner disks at $<10$~au. One of the most exciting opportunities for the next decade is to address the heterogeneity, incompleteness, and different samples obtained to date and work towards a global view of molecular gas in inner disks under multiple dimensions across stellar and accretion properties (TTauri versus Herbig AeBe stars), disk emitting regions (disk rim versus disk surface, dust-rich versus dust-poor regions, gas in disks versus winds), and disk structures (full disks versus disks with dust and gas cavities and/or rings). 
As it happened before with NIRSPEC first and then CRIRES, a new phase opens up as a consequence of technical advancements. The new survey presented in this work was motivated by the quality and unprecendented spectral coverage provided by the iSHELL spectrograph \citep{ishell} at the NASA Infrared Telescope Facility (IRTF), covering in one single shot the entire $M$-band between 4.52 and 5.24~$\mu$m with very narrow gaps between the orders (see Figure \ref{fig: spec_overview}). 

While this survey is still ongoing, in this first paper we set the goal to provide an overview of known and newly discovered global trends present in the CO spectra as observed across stellar properties from T~Tauri to Herbig~Ae/Be systems. In particular, we focus on how a study of CO line shapes observed at the highest-resolution ($R\approx60,000-95,000$) combined with their excitation across the $M$-band can advance our understanding of the structure and evolution of inner disks in the next years. Building on previous work, we discuss and re-address fundamental questions on the location, excitation, and disk/wind structures traced by the CO ro-vibrational lines in the framework of a global picture of inner disks, and outline some promising directions for future work in the era of JWST and future high-resolution spectrographs on extremely large telescopes.

Given the yet unparalleled spectral completeness of iSHELL $M$-band spectra, before moving into the rest of the paper we provide here a brief review of gas tracers included at these wavelengths for guidance.

\subsection{Gas emission in $M$-band spectra} \label{sec: data}
As shown in previous work, high-resolution $M$-band spectra observed towards protoplanetary disks are very rich and diverse, with multiple emission and absorption components and measured line widths from a few 100~km/s down to the spectral resolution limits \citep[e.g.][]{najita03,salyk11,brown13}. In Figure \ref{fig: spec_overview} we show a full iSHELL spectrum broken up into portions, to highlight the richness and complexity of spectral emission at these wavelengths. For guidance, we also briefly review here the gas emission/absorption components that these spectra include (from most to least commonly detected in disks):

\paragraph{Carbon Monoxide}
\ce{^{12}CO} is commonly observed in emission with lines in the $v=1-0$ and $v=2-1$ ro-vibrational branches, and in a some disks from higher vibrational levels up to at least $v=6$ \citep{brittain07,vdplas15}; these lines may have one or two velocity components \citep{bp15}. \ce{^{13}CO} is often detected in emission in the $v=1$ level only. Both isotopologues are also sometimes observed in absorption in the ro-vibrational $v=1-0$ branch only. Less abundant isotopologues (e.g. \ce{C^{18}O}) are much more rarely detected \citep[e.g.][]{smith15}. Previous work found that CO lines generally trace gas in inner disks and possibly winds (in emission) and the outer disk or envelopes and outflows (in absorption), see e.g. \citet{brown13}.

\paragraph{Atomic Hydrogen}
Up to three HI lines are detected in the $M$ band: Pfund~$\beta$ ($n=7-5$) at 4.65378~$\mu$m, Humphreys~$\epsilon$ ($n = 11-6$) at 4.67251~$\mu$m, and Humphreys~$\delta$ ($n = 10-6$) at 5.12865~$\mu$m; these lines are typically broad (FWHM~$>200$~km/s, and several times broader than CO emission in the same disk) and trace stellar accretion \citep{salyk13}. In this work, we only provide a gallery of HI line profiles as compared to CO in the Appendix for reference. 

\paragraph{Molecular Hydrogen}
The \ce{H2} S(9) line at 4.6947~$\mu$m has been detected and spectrally resolved in a small sample of disks, mostly in class-I objects; this line may trace disk gas or outflows interacting with surrounding gas \citep{bitner08,herczeg11}, but a spurious slightly red-shifted feature can be introduced by division with the standard spectrum during telluric correction, and should be carefully checked against that effect.

\paragraph{Water}
$M$-band spectra also cover \ce{H2O} emission lines from the ro-vibrational R branch of the bending mode, but these lines, to our knowledge, have never been reported in previous work. These lines are the short-wavelength tail of ro-vibrational water emission bands observed at low resolution ($R\approx120$) with Spitzer-IRS at 6.6~$\mu$m by \citet{sargent14} in disks, and earlier with SWS by \citet{gonzalf98}. In this iSHELL survey, approximately 30 $v=1-0$ lines ($E_{up} = 5000$--9000~K, $A_{up} =2$--20~s$^{-1}$) are detected and spectrally resolved for the first time at wavelengths of 4.6--5.24~$\mu$m.

\begin{figure}
\centering
\includegraphics[width=0.35\textwidth]{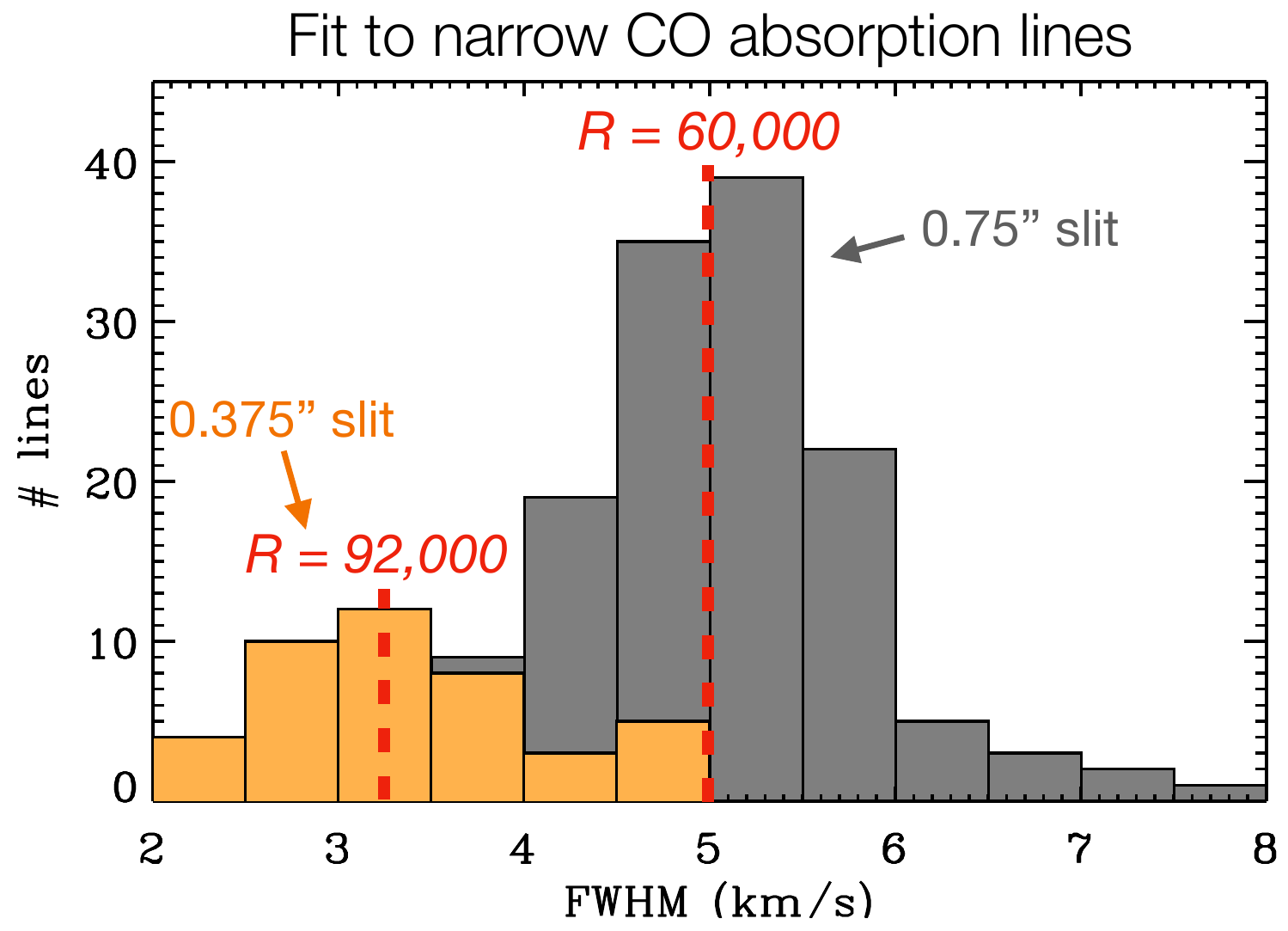} 
\caption{iSHELL resolving power with the two slits used in this survey, as estimated from Gaussian fits to narrow CO absorption lines.}
\label{fig: resolution}
\end{figure}

\section{A disk survey with IRTF-iSHELL} \label{sec: ishell}
We started this survey soon after iSHELL was deployed at IRTF in 2016 \citep{ishell}. In the first few semesters we led exploratory programs targeting a few bright disks, followed soon after by a survey first dedicated to disks around intermediate-mass stars \citep{abernathy21} and now expanding towards solar-mass stars down to a magnitude limit of $M \approx 5.5$. The survey is still ongoing, and the Appendix reports observations done in 2016--2021 under multiple programs with PIs A.Banzatti or S.Brittain. 

\begin{figure*}
\centering
\includegraphics[width=1\textwidth]{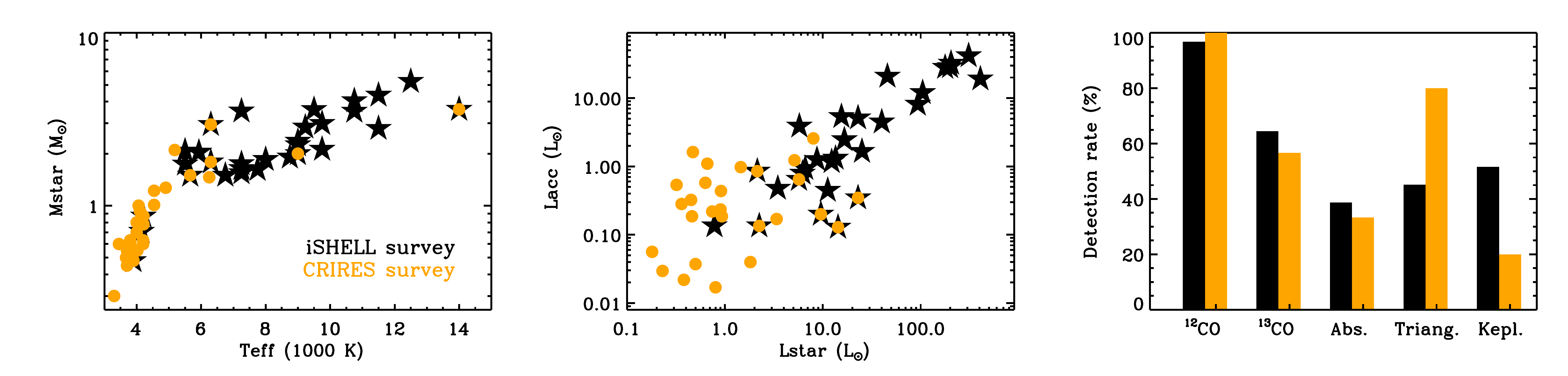} 
\caption{Sample comparison between this iSHELL survey and a previous survey done at similar resolving power with VLT-CRIRES \citep[][see Section \ref{sec: sample}]{pont11,brown13}. The third plot to the right reports detection rates for \ce{^{12}CO} $v=1-0$ emission, \ce{^{13}CO} $v=1-0$ emission, CO lines in absorption, ``triangular" line shapes, and Keplerian line shapes (see Section \ref{sec: line_shapes} for line shapes definitions).}
\label{fig: surveys_comparison}
\end{figure*}

For the observations, we followed strategies based on over two decades of experience observing CO spectra in the near-infrared. A very important aspect for high-quality, high-resolution spectroscopy at these wavelengths is to minimize the impact of telluric absorption. To do so, we planned each observation to maximize the Doppler shift between the targeted emission lines and telluric lines, and we observed a bright telluric standard near each target typically with an airmass difference below 0.1. In some cases, it is possible to achieve a total Doppler shift that brings the telluric entirely outside of the targeted disk emission line. More often, and always in case of broad CO lines and/or small stellar RVs, the telluric will carve a gap into one side of the observed $v=1-0$ CO emission lines with $J < 15$ (see more below under line stacking).

The observing strategy we have been following in this survey is to obtain high-quality $M$-band spectra with nominal S/N~$\approx100$ per pixel at 4.9$\mu$m, to study low-contrast CO emission down to 5--10~\% on the continuum (as observed to be typical for Herbig~Ae/Be disks). We have therefore used the 0.375" slit for the brightest sources ($M<4$~mag), and the 0.75" slit for fainter ones, in an effort to keep the integration time per spectrum below 1 hour. Some spectra have been observed with two slit positions to enable measurement of the spectro-astrometric signal, which is however not analysed in this first paper. Observing settings for each spectrum are reported in the Appendix.
Since January 2020, we have also used the automated Quicklook data reduction tool provided by IRTF that runs during the observations; this tool has been extremely helpful to assess the quality and S/N of the data and make strategic decisions in real-time to maximize the quality of the survey.

We homogeneously reduced all the spectra with Spextool v5.0.3 \citep{spextool}, which has been provided by IRTF specifically for iSHELL data. Spextool is written in IDL and performs all the standard reduction steps through a convenient interface that allows users to interact with the data at every step: flat-fielding, wavelength calibration, spectral extraction of individual nods, combination of nodded spectra, and telluric correction using a standard star. iSHELL's 15"-long slits provide good sky background subtraction with no loss of observing efficiency by nodding point sources up and down the slit. For telluric correction, Spextool does not apply any airmass correction factor at the time of division of the science and standard spectra; in a few cases, telluric residuals are therefore visible in the low-$J$ CO lines in the reduced spectra. The data reduction process is described in detail in the iSHELL reduction manual\footnote{http://irtfweb.ifa.hawaii.edu/research/dr\_resources/}. 
All the reduced spectra have been corrected for the barycentric velocity at the time of observations and are shown in this paper in the heliocentric velocity frame. 
The spectra have not been flux-calibrated, a process that is generally very uncertain due to flux losses through the narrow slits, but we report WISE W2 flux measurement in Table \ref{tab: sample} so that readers can flux calibrate the line fluxes reported in this work.
All the reduced, science-ready spectra are available for interactive visualization through the web interface of a new database dedicated to spectra of protoplanetary disks, \textit{SpExoDisks}, at \url{www.spexodisks.com} \citep[][and Wheeler et al., in prep.]{perez21_spexodisks}.

As iSHELL is a new instrument whose characterization was still ongoing over the past 5 years, we took the opportunity of this survey to measure its resolving power. We used point sources observed in the $M$ band by measuring the FWHM of the narrowest \ce{^{12}CO} and \ce{^{13}CO} absorption lines detected using the two slits, assuming that the narrowest of these lines must be unresolved. By taking what is currently available in this survey, three disk spectra with absorption lines were observed with the 0.375" slit, for a total of 42 measured absorption lines, and nine disks with the 0.75" slit, for a total of 141 absorption lines.
Figure \ref{fig: resolution} shows the distributions of fit results, which indicate a resolution of $\approx 3.3$~km/s for the 0.375" slit (R~$\approx 92,000$) and $\approx 5$~km/s for the 0.75" slit (R~$\approx 60,000$). These correspond or are slightly higher than what measured in unresolved lamp lines at blaze center, R~$=88,100 \pm 2,000$ (0.375" slit) and R~$=53,300 \pm 300$ (0.75" slit), an effect that could be due to point sources slightly underfilling the larger slit; more details on the instrument and its performance are reported in \citet{ishell22}.

\begin{deluxetable*}{l c c c c c c c c c c c c c c}
\tabletypesize{\footnotesize}
\tablewidth{0pt}
\tablecaption{\label{tab: sample} iSHELL sample properties as collected from the literature.}
\tablehead{ Name & Dist & $T_{\rm{eff}}$ & $M_{\star}$ & log $L_{\star}$ & log $L_{\rm{acc}}$ & $n_{13-30}$ & $R_{\rm{NIR}}$ & $R_{\rm{subl}}$ & W2 flux & RV & err & Incl & err & refs \\ 
 & (pc) & (K) & ($M_{\odot}$) & ($L_{\odot}$) & ($L_{\odot}$) &  & (au) & (au) & (Jy) & (km/s) &  & (deg) & }
\tablecolumns{15}
\startdata
51Oph & 125. & 9500. & 3.60 & 2.25 & 1.45 & -- & 0.50 & 0.49 & -- & -20.0 & 3.0 & 63.0 & 1.0 & \textit{r2, rv1, i2} \\
ABAur & 155. & 9000. & 2.36 & 1.66 & 1.32 & 0.81 & 0.32 & 0.34 & 36.7 & 15.3 & 0.0 & 11.5 & 10.1 & \textit{r1, rv3, i1} \\
AS205N & 127. & 4210. & 0.87 & 0.33 & -0.07 & -0.19 & 0.16 & 0.14 & 6.8 & -7.1 & 0.6 & 20.0 & 3.0 & \textit{r2, rv2, i4} \\
CITau & 158. & 4150. & 0.71 & -0.11 & -0.87 & -0.16 & 0.11 & 0.12 & 0.7 & 19.9 & 0.8 & 50.0 & 0.1 & \textit{r3, rv2, i3} \\
CQTau & 149. & 6750. & 1.50 & 0.82 & -0.03 & 0.78 & 0.21 & 0.18 & 2.9 & 17.6 & 0.8 & 35.0 & 1.0 & \textit{r3, rv3, i7} \\
Elias20 & 138. & 3900. & 0.48 & 0.35 & -0.87 & -0.34 & -- & 0.14 & 1.6 & -3.3 & 0.6 & 49.0 & 1.0 & \textit{--, rv5, i4} \\
HD141569 & 111. & 9750. & 2.12 & 1.40 & 0.22 & 1.30 & 0.03 & 0.29 & 0.6 & -6.0 & 5.0 & 58.5 & 2.0 & \textit{r1, rv1, i2} \\
HD142666 & 146. & 7250. & 1.75 & 1.13 & 0.12 & -0.49 & 0.11 & 0.23 & 3.0 & -7.0 & 2.7 & 55.9 & 2.8 & \textit{r2, rv1, i2} \\
HD143006 & 166. & 5500. & 1.74 & 0.54 & -0.32 & 1.25 & 0.08 & 0.15 & 1.4 & -1.5 & 0.2 & 27.1 & 3.8 & \textit{r1, rv2, i1} \\
HD145718 & 152. & 7750. & 1.62 & 1.05 & -0.35 & -- & 0.70 & 0.22 & -- & -3.6 & 2.3 & 71.9 & 1.2 & \textit{r2, rv1, i2} \\
HD150193 & 150. & 9000. & 2.25 & 1.36 & 0.71 & -0.48 & 0.39 & 0.28 & 6.5 & -4.9 & 3.9 & 47.2 & 2.3 & \textit{r2, rv1, i2} \\
HD163296 & 101. & 8750. & 1.91 & 1.19 & 0.73 & -0.50 & 0.30 & 0.24 & 11.6 & -9.0 & 6.0 & 50.2 & 0.7 & \textit{r2, rv1, i2} \\
HD169142 & 114. & 7250. & 1.55 & 0.76 & 0.59 & 1.47 & 0.33 & 0.18 & 0.9 & -3.0 & 2.0 & 21.6 & 10.7 & \textit{r2, rv1, i1} \\
HD179218 & 258. & 9750. & 2.99 & 2.02 & 1.08 & 0.13 & 2.19 & 0.42 & 4.8 & 15.1 & 2.3 & 53.8 & 6.4 & \textit{r2, rv1, i2} \\
HD190073 & 872. & 9230. & 2.85 & 1.92 & 2.31 & -1.28 & 1.78 & 0.40 & 5.6 & 0.2 & 0.1 & 21.6 & 1.6 & \textit{r2, rv1, i2} \\
HD259431 & 640. & 12500. & 5.24 & 2.91 & 2.43 & 0.55 & 0.32 & 0.85 & 9.6 & 26.0 & 8.0 & 24.5 & 8.9 & \textit{r2, rv1, i1} \\
HD35929 & 377. & 7250. & 3.53 & 1.97 & 0.91 & -1.96 & 0.21 & 0.41 & 1.1 & 21.1 & 1.8 & 31.8 & 16.2 & \textit{r1, rv1, i1} \\
HD36917 & 445. & 11500. & 4.36 & 2.61 & 1.28 & -0.18 & 0.12 & 0.64 & 2.7 & 26.3 & 3.6 & 89.4 & 19.3 & \textit{r1, rv1, i1} \\
HD37806 & 397. & 10750. & 3.52 & 2.30 & 1.46 & -1.14 & 0.81 & 0.50 & 6.7 & 47.0 & 21.0 & 60.0 & 2.6 & \textit{r2, rv1, i2} \\
HD58647 & 302. & 10750. & 4.05 & 2.49 & 1.62 & -1.91 & 0.60 & 0.57 & 10.0 & 10.0 & 12.0 & 63.3 & 0.6 & \textit{r2, rv6, i2} \\
IRS48 & 125. & 9000. & 2.00 & 1.16 & -0.89 & 1.33 & -- & 0.24 & -- & -5.7 & 0.0 & 42.0 & 6.0 & \textit{--, rv3, i9} \\
LkHa330 & 308. & 6300. & 2.95 & 1.36 & -0.46 & 1.90 & -- & 0.28 & 1.4 & 19.5 & 0.5 & 12.0 & 2.0 & \textit{--, rv2, i8} \\
MWC297 & 408. & 26000. & 19.98 & 4.78 & 4.08 & -- & 0.93 & 2.68 & 950.9 & -6.5 & 2.0 & 27.1 & 1.3 & \textit{r1, rv6, i1} \\
MWC480 & 155. & 8000. & 1.85 & 1.22 & 0.39 & -0.79 & 0.30 & 0.25 & 4.6 & 12.9 & 3.7 & 36.5 & 0.1 & \textit{r1, rv1, i3} \\
MWC758 & 155. & 7250. & 1.64 & 0.94 & 0.10 & 0.82 & 0.21 & 0.20 & 4.9 & 17.8 & 3.7 & 40.0 & 1.0 & \textit{r1, rv1, i6} \\
RNO90 & 117. & 5660. & 1.50 & 0.76 & -0.19 & -0.52 & -- & 0.17 & 3.9 & -10.8 & 1.1 & 53.0 & 1.0 & \textit{--, rv2, i3} \\
RYTau & 444. & 5930. & 2.04 & 1.09 & 0.07 & -0.19 & 0.34 & 0.23 & -- & 16.5 & 2.4 & 65.0 & 0.1 & \textit{r2, rv4, i3} \\
SR21 & 138. & 6300. & 1.79 & 0.98 & -0.70 & 2.01 & -- & 0.21 & 1.1 & -7.1 & 3.6 & 15.0 & 1.0 & \textit{--, rv2, i6} \\
SUAur & 158. & 5500. & 2.07 & 0.79 & -0.10 & 0.85 & 0.25 & 0.18 & 2.8 & 16.0 & 3.1 & 51.0 & 1.0 & \textit{r4, rv4, i5} \\
V892Tau & 117. & 11500. & 2.80 & 1.60 & 0.65 & -0.13 & -- & 0.33 & -- & 16.1 & 0.0 & 54.5 & 1.0 & \textit{--, rv3, i3} \\
VVSer & 403. & 14000. & 3.61 & 2.31 & -- & -1.24 & 0.43 & 0.51 & 3.8 & -8.0 & 0.0 & 58.7 & 0.7 & \textit{r1, rv3, i1} \\
\enddata
\tablecomments{
References -- Distances are from \citet{bj18}; stellar and accretion properties are from \citet{guzmandiaz21,wichittanakom20,herczeg14,salyk13,dsharp,fang18,simon16}; the infrared index $n_{13-30}$ is measured from \textit{Spitzer}-IRS spectra as in \citet{banz20}; $R_{\rm{subl}}$ is estimated as explained in Appendix \ref{app: dust_radii}; the W2 flux from WISE is taken from the AllWISE Data Release \citep{allwise}; references for $R_{\rm{NIR}}$, the stellar RV, and the disk inclination are as follows: \textit{r1},\textit{i1}: \citet{lazareff17}; \textit{r2},\textit{i2}: \citet{gravity19,gravity21_TTs,gravity21_51oph,ganci21}; \textit{r3}: \citet{eisner04,eisner07}; \textit{r4}: \citet{labdon19} -- \textit{rv1}: \citet{alecian13,dunkin97}; \textit{rv2}: \citet{fang18,banz19}; \textit{rv3}: \citet{liu11,vdmarel16,wolfer21}; \textit{rv4}: \citet{hartmann86}; \textit{rv5}: \citet{sullivan19_rv}; \textit{rv6}: \citet{acke08,brittain07} -- \textit{i3}: \citet{long18,long21} and Long, priv. comm.; \textit{i4}: \citet{huang18,kurtovic18}; \textit{i5}: \citet{labdon19}; \textit{i6}: \citet{pinilla18,tripathi17}; \textit{i7}: \citet{margi19}; \textit{i8}: \citet{pont11}; \textit{i9}: \citet{brown12}.
}
\end{deluxetable*}

\subsection{Sample} \label{sec: sample}
In starting this survey, we intended to gather a homogeneous overview of $M$-band CO spectra from protoplanetary disks over a wide range in conditions, including in stellar masses, accretion rates, and disks with and without inner dust cavities.
Given the sensitivity limits imposed by IRTF's 3.2-m telescope aperture, this survey has so far mostly focused on brighter protoplanetary disks around stars with $T_{\rm{eff}} \gtrsim 5500$~K (Figure \ref{fig: surveys_comparison}), i.e. some T~Tauri stars but mostly Herbig Ae/Be stars. In this work, we broadly refer to Herbig Ae/Be stars as those with $T_{\rm{eff}} \gtrsim 6500$~K and T~Tauri stars as those with lower temperatures, without including a distinction for the so-called intermediate-mass T~Tauri stars \citep[IMTTs,][]{calvet04}. These are massive ($M_{\star} \gtrsim 1.5 M_{\odot}$) but lower-temperature ($T_{\rm{eff}} \lesssim 7000$~K) precursors of the Herbig Ae/Be stars \citep[for recent lists of these, see][]{villebrun19,valegard21}, and some of these have often been included without distinction in studies of Herbig Ae/Be stars. In this iSHELL survey, stars that could be classified as IMTTs based on their mass and temperature are up to 30\% of the sample: CQ~Tau, HD~35929, HD~142666, HD~143006, HD~169142, LkHa~330, MWC~758, RNO~90, RY~Tau, SU~Aur, SR~21.

The total sample currently comprises 31 individual stars (see Table \ref{tab: sample}), including 5 disks from the ALMA DSHARP program \citep{dsharp} --~AS~205~N, Elias~20 (VSSG~1), HD~143006, HD~142666, HD~163296~-- and 9 disks that will be observed with JWST GTO and Cycle 1 GO programs: the 5 DSHARP disks plus CI~Tau, HD~35929, SR~21, RNO~90. Sample properties are compared in Figure \ref{fig: surveys_comparison} to the sample obtained at a similar resolving power of R~$\approx$~94,000 (0.2" slit) in a 24-nights Large Program with VLT-CRIRES in 2007-2010 \citep{pont11,brown13}. In the rest of this paper, we will refer to this previous dataset simply as ``the CRIRES survey''; we have included in this work the sub-sample used in \citet{bp15} excluding low S/N spectra and targets with unknown or very uncertain star/disk properties to reach a similar-sized sample to the iSHELL survey so far (30 spectra), and re-analyzed them together with the iSHELL spectra for a total sample of 61 spectra (54 individual stars, with 7 stars in common between the two samples) as described in Section \ref{sec: analysis}. The 2.5 times larger aperture of the VLT mirrors allowed the CRIRES survey to focus on solar- and low-mass stars in T~Tauri systems (Figure \ref{fig: surveys_comparison}). The combination of different samples from these two surveys can be very powerful in providing a global, unified picture of CO in inner disks, as it will be discussed below.

\begin{figure*}
\centering
\includegraphics[width=0.9\textwidth]{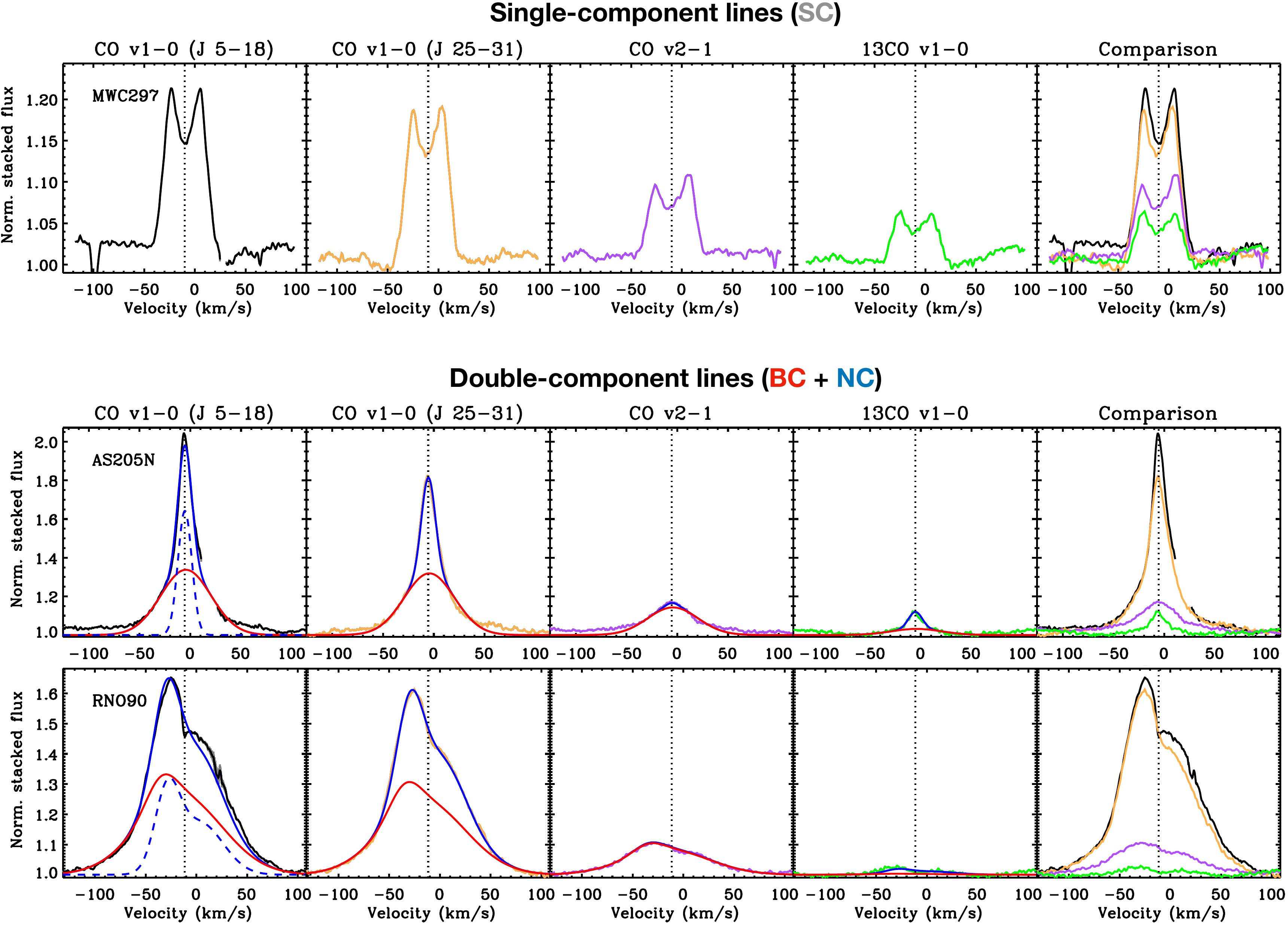} 
\caption{Examples of stacked lines, and of line decomposition into BC and NC where present. The vertical dotted line marks the line center for reference. Top: example of a spectrum with a single-component, double-peak line shape. Middle and bottom: examples of triangular lines with two CO components, BC marked in red and NC marked in blue (the dashed line shows NC on its own, the solid line shows NC as summed to BC). AS~205~N is an example where the two components can be well described with Gaussian functions, while RNO~90 an example where the $v=2-1$ line profile is used to define BC \citep[as in][]{bp15}.}
\label{fig: line_stacks}
\end{figure*}

\section{Spectral analysis procedure} \label{sec: analysis}
For this survey, we have developed a spectral analysis pipeline to systematically extract a standard set of spectral products from all spectra: \textit{i)} stacked line profiles for \ce{^{12}CO} and \ce{^{13}CO} in vibrational levels up to $v=2$ where detected, \textit{ii)} decomposed broad and narrow emission components where present, \textit{iii)} line flux measurements for transitions up to $v=3$, \textit{iv)} line centroids and FWHM for all emission and absorption components, \textit{v)} rotation diagrams for all emission and absorption components. The pipeline has been carefully developed to be flexible and handle the richness and diversity of CO disk spectra, where the wide range in line widths and excitation makes it challenging to develop a standard process that works for all. 
We describe in the following the main steps of this pipeline.

First, the telluric-corrected, normalized spectrum is rectified by removal of residual broad curvatures across echelle orders and of the HI emission lines, to produce a flat continuum across the entire spectrum. HI emission lines are fitted with up to two Gaussian functions and subtracted, and the continuum across the spectrum is instead convolved with a low-frequency filter and then subtracted. To identify and isolate pixels for the continuum determination, the pipeline uses HITRAN line lists \citep{hitran12} to locate and temporarily ignore any CO vibrational bands that are detected. After continuum rectification, every following step in the pipeline assumes a flat continuum normalized at 1 near each CO line. 

Second, spectral lines are systematically stacked to increase S/N and provide the highest quality line profile to study gas kinematics (line shape, centroid, and width). The pipeline stacks lines from the $v=1$ and $v=2$ levels of \ce{^{12}CO}, and the \ce{^{13}CO} $v=1-0$ lines. In each case, lines are stacked by weighted average, and avoiding line blends with nearby transitions. Line blends are identified by using the measured full-width-at-zero-intensity FWZI: when two detected lines are closer than FWZI/2, they are excluded from the line stacking process.
In the case of \ce{^{12}CO} $v=1-0$ lines, two line stacks are performed, one for $J$-levels between 5 and 18 (in both P and R branches), and one for $J$-levels between 25 and 31 (in the P branch). The former stack obtains the line emission shape and average strength of the low-$J$ lines, while avoiding the lowest $J$ lines that may have central absorption (see more in Section \ref{sec: absorption}). The latter obtains the line shape and strength in one of the cleanest and highest S/N parts of the spectrum, where tellurics near the \ce{^{12}CO} $v=1-0$ lines are weaker (Figure \ref{fig: spec_overview}). As a consequence, the high-$J$ stacked line profile does not suffer from gaps left from telluric correction, and aids in the characterization of the true line shapes. The \ce{^{12}CO} $v=2-1$ lines and the \ce{^{13}CO} $v=1-0$ lines are stacked between $J=3$ and $J \sim 20$. In all cases, only detected and unblended lines are included in the stacks, following the procedure explained above. This means that in the spectra with broader lines and higher excitation, i.e. producing diffuse blending across the spectrum, less lines can be used in the stacks and the kinematic characterization of CO relies more on the S/N obtained on individual lines (e.g. in HD~35929 in Figure \ref{fig: spec_overview}). Also, in spectra that have only partial coverage as was typical in the CRIRES survey, line stacks only include whichever lines are available in the $J$-level ranges noted above.

Third, line stacks are used to identify and separate broad (BC) and narrow (NC) emission components when present in a spectrum. The presence of these components has been identified in $M$-band CO spectra based on their kinematic and excitation difference \citep{goto11,bast11,bp15}. The procedure to identify and separate them is similar to what developed in \citet{bp15}, as follows. The presence of BC and NC becomes evident when the \ce{^{12}CO} $v=1-0$ and $v=2-1$ lines are compared: the $v=2-1$ lines match well the broad wings of $v=1-0$ lines, but lack extra emission at the center of the line (Figure \ref{fig: line_stacks}). In spectra with the narrowest NC (e.g. AS~205~N), the separation between BC and NC can be well described in terms of two Gaussian components. In spectra that have broader and usually more structured NC (e.g. RNO~90), the full $v=2-1$ line profile is used to define BC, and the residuals from subtraction of the $v=1-0$ and $v=2-1$ lines, after applying a flux scaling factor to match them over their wings, to define NC. Spectra where the $v=1-0$ and $v=2-1$ full line profiles match, after applying a scaling factor to account for their different strength, do not show evidence for two velocity components; this is usually the case for symmetric double-peak lines (e.g. MWC~297 in Figure \ref{fig: line_stacks}).

Fourth, the stacked lines are used to extract line fluxes across the entire spectrum from all $J$ levels. At each CO line position defined from the HITRAN line list, the stacked line profile is used as a model to fit the data and measure a line flux. This process allows to recover the flux of lines with narrow gaps like those produced by telluric absorption in the low-$J$ lines. The procedure also allows to retrieve the different fractional flux of BC and NC as a function of $J$ level, by fitting each line with a composite model that has as only free parameters the peak flux of each component. When a central absorption component is present, the composite model includes an extra narrow Gaussian line to fit and extract the absorption flux too. With these procedure, line fluxes for all emission and absorption components can be systematically and simultaneously extracted for multiple vibrational bands of \ce{^{12}CO} and \ce{^{13}CO}. For simplicity, the procedure currently ignores blended lines, as identified with the same criterium described above for the continuum flattening and line stacking procedures.

Fifth, rotation diagrams are produced using the extracted line fluxes for all emission and absorption components and for multiple vibrational bands of \ce{^{12}CO} and \ce{^{13}CO}. We produce rotation diagrams following \citet{gl99} and \citet{larsson02}, where departures from the typical linear behavior of optically thin lines in LTE become visible in curvatures that can be related to optical depth, a range in excitation temperatures, and/or non-thermal excitation of the emitting gas. Examples of curvatures observed in this survey and their interpretation in terms of a single slab of gas in LTE are shown below in Sections \ref{sec: results} and \ref{sec: discussion}; rotation diagrams for the whole sample will be included and analyzed in a forthcoming paper. 

The pipeline has been successful in processing all spectra in this survey, but it becomes increasingly ineffective when CO emission lines have FWHM larger than 100 km/s and are highly vibrationally excited, due to the widespread blending between lines. In the iSHELL sample, line stacks can only approximately be extracted for the two spectra with the broadest CO emission (FWHM~$> 180$~km/s), HD~58647 and 51~Oph, where \ce{^{12}CO} lines are excited up to $v=2$, and possibly to higher levels.

\begin{figure*}
\centering
\includegraphics[width=1\textwidth]{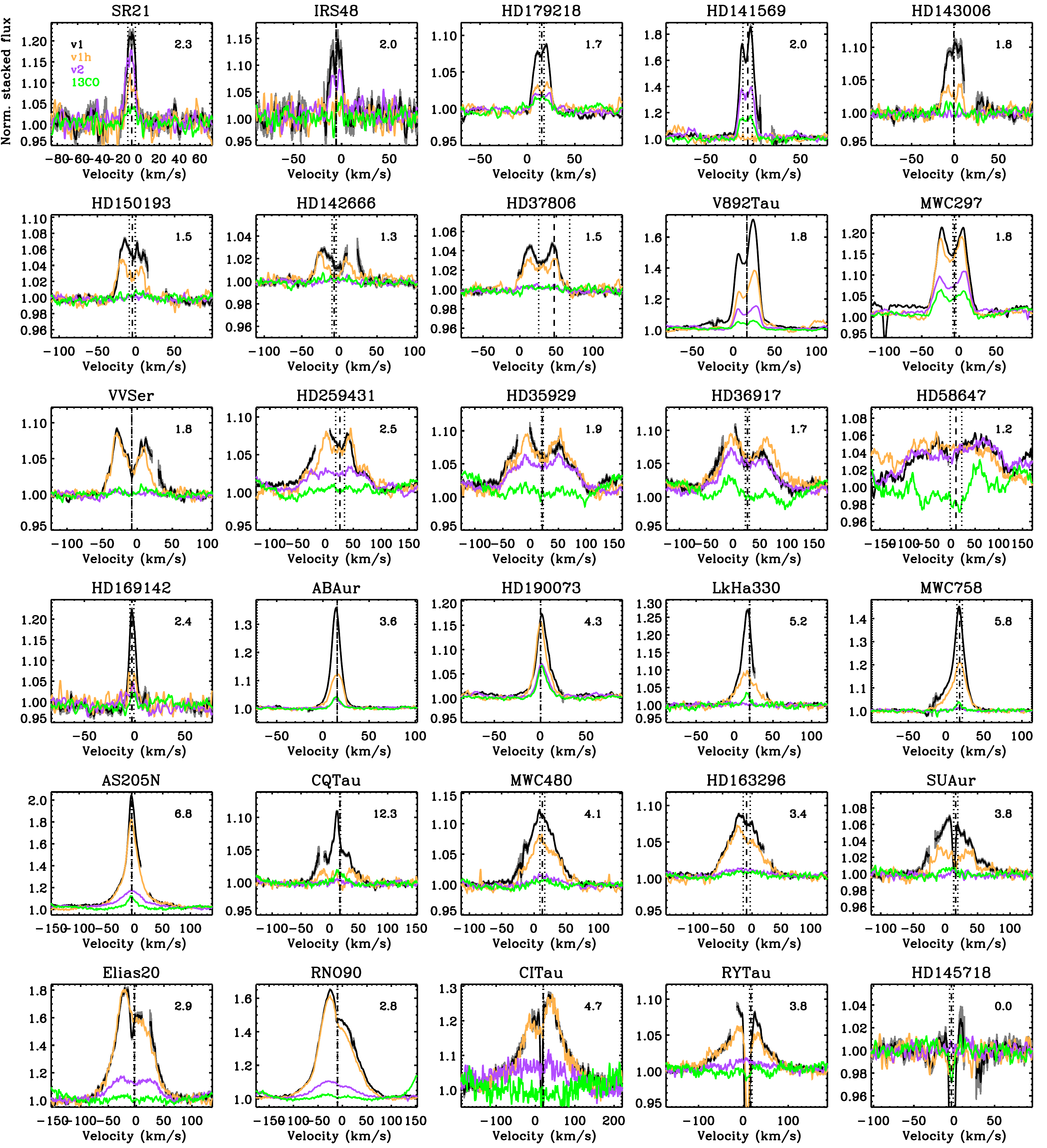}
\caption{Overview of line shapes observed in the iSHELL survey, showing line stacks for \ce{^{12}CO} $v=1-0$ $J=$~5--18 (black), $v=1-0$ $J=$~25-31 (orange), $v=2-1$ (purple), \ce{^{13}CO} $v=1-0$ (green). Double-peak lines are shown in the top half of the figure, triangular lines in the bottom half. CO is only marginally detected in emission in HD~145718, which is included at the end for reference. 51~Oph is not included because the stacked line profiles cannot be reliably extracted due to the extremely broad FWHM. The vertical dashed and dotted lines shows the stellar RV and its uncertainty. The measured line shape parameter $S$ is reported at the top right in each plot. The corresponding figure for targets in the CRIRES survey is reported in the Appendix.}
\label{fig: line_gallery}
\end{figure*}

\section{Results} \label{sec: results}
In the following subsections, we report first results from this survey in terms of line detections, shapes, velocity components, kinematics, excitation, variability, and emitting regions. A brief summary of results is provided in Section \ref{sec: res_summary}. The results will be combined and jointly discussed in Section \ref{sec: discussion}. While highlighting measurements and results from this survey, we also combine them to those from the CRIRES survey to illustrate the global picture that emerges from the synergy of the different samples (Section \ref{sec: sample} and Figure \ref{fig: surveys_comparison}).

\subsection{Line detections}
Line stacks are shown for the entire iSHELL sample in Figure \ref{fig: line_gallery}. \ce{^{12}CO} $v=1-0$ emission lines are detected in the entire sample, but only marginally in HD~145718. The $v=2-1$ emission lines are detected in 18/31 objects, $v=3-2$ emission lines in 12/31, and higher vibrational levels only in 4/31 objects. \ce{^{13}CO} is detected in emission in 19/31 objects, and for now we have not yet checked for any detections in \ce{C^{17}O} or \ce{C^{18}O}.
Absorption components are detected in 12/31 objects and only in the $v=1-0$ lines, with a lower detection rate in \ce{^{13}CO} (9/31 objects). These detection rates generally agree with previous work, which similarly found at least the $v=1-0$ CO lines to be almost ubiquitously detected in disks.
Detection rates are compared in Figure \ref{fig: surveys_comparison} to those from the CRIRES sample included in this work. Detection rates between the two surveys are strikingly different in the observed line shapes, which will be described in the next section.
Besides CO, HI is detected in nearly all sources, as reported in the Appendix, and \ce{H2O} is detected in one disk only, the disk of AS~205~N. About 30 \ce{H2O} lines are detected in the spectrum of AS~205~N, some of which are blended with CO emission. The water spectrum is included in Figure \ref{fig: spec_overview} and will be analyzed in a future paper.

\begin{figure*}
\centering
\includegraphics[width=0.99\textwidth]{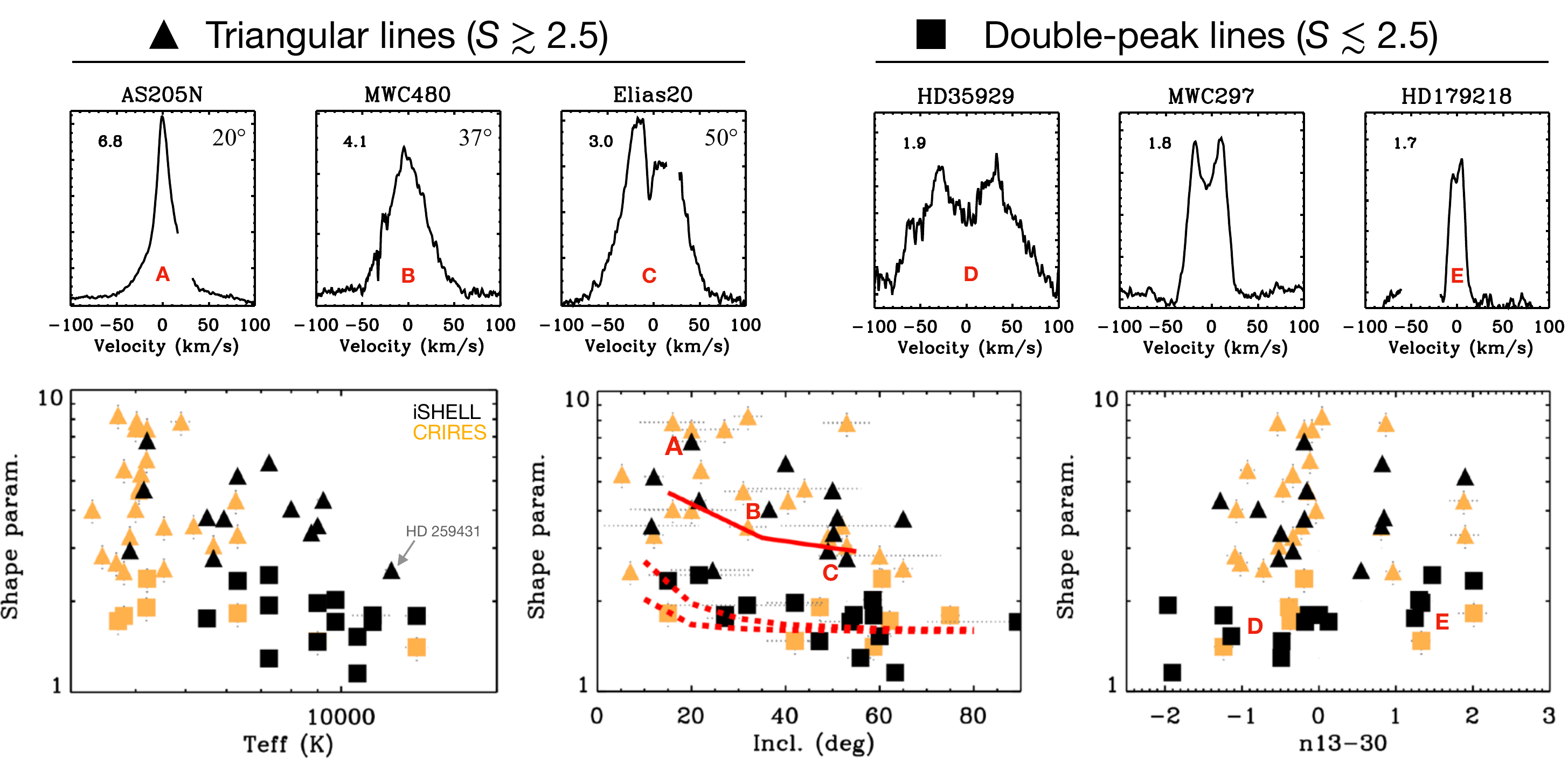} 
\caption{CO line shape parameter $S$ as a function of $T_{\rm{eff}}$, disk inclination, and infrared index for the iSHELL (black) and CRIRES (orange) surveys combined. \textit{Top row}: Representative line shapes from Figure \ref{fig: line_gallery} are separated into ``triangular'' to the left ($S \gtrsim 2.5$, triangle symbols) and double-peaked to the right ($S \lesssim 2.5$, square symbols); the measured shape parameter $S$ (Section \ref{sec: line_shapes}) is included in each plot at the top left. Line profiles are labeled with letters to identify different regions in the parameter space below.
\textit{Bottom row}: Triangular lines are more frequent in T~Tauri systems while double-peak lines in Herbig AeBe disks (left plot). Both line types are observed at all disk inclinations in both disks with and without an inner cavity (middle and right plots). Geometric effects for gas in Keplerian rotation \citep[dashed lines,][]{bast11} and self-absorption from a wind \citep[solid line,][]{pont11} are shown for reference in the middle plot.}
\label{fig: shape_param}
\end{figure*}

\subsection{Emission line shapes} \label{sec: line_shapes}
Figure \ref{fig: shape_param} shows a gallery of selected line profiles, separated into two groups based on the line shape (see also Figure \ref{fig: line_gallery}). Visually, the separation is between profiles that clearly exhibit two peaks and usually steep line wings, as expected for a ring of gas in Keplerian rotation in a disk, and those that instead show a more triangular shape with a narrower line center (often with a single peak) and broader line wings. These two types of line shapes have been identified in previous work and quantitatively described using a line profile parameter that measures the ratio of line width between the peak and base of the line. \citet{bast11} defined a parameter called P$_{10}$ using the line width at 10\% and 90\% of the line peak flux; here, we avoid line asymmetries that skew the parameter (e.g. in RNO~90, Elias~20, and V892~Tau) and use the full line width at 10\% and 75\% of the peak flux, defining the line shape parameter as $S$ = FW10\% / FW75\%. This parameter is larger for triangular line shapes, and smaller for a double-peak profile with steep wings.

\begin{figure*}[ht]
\centering
\includegraphics[width=1\textwidth]{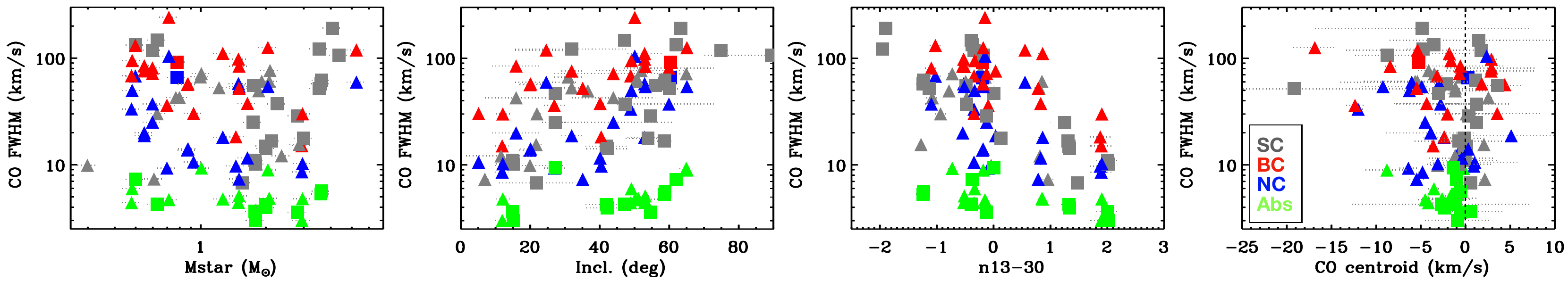} 
\caption{CO line kinematics for different components (marked in different colors as shown in the legend, with absorption components in green), by combining the iSHELL and CRIRES surveys. Symbols reflect the full line shape parameter $S$ as in Figure \ref{fig: shape_param}. The FWHM of individual emission components overall increases with disk inclination, and decreases with stellar mass and infrared index. CO line centroids predominantly show small blueshifts, in both emission and absorption components. }
\label{fig: kinem_corr}
\end{figure*}

The line shape parameter $S$ is shown in Figure \ref{fig: shape_param} by combining measurements from iSHELL and CRIRES. With the above definition for the line shape parameter, the two types of line shapes separate roughly at $S \approx 2.5$ (Figure \ref{fig: line_gallery}). While the scatter is large, one clear trend that emerges from the combination of the two surveys is that triangular lines dominate in disks around T~Tauri stars ($T_{\rm{eff}} < 7000$~K) while double-peak lines dominate in hotter, Herbig AeBe stars. It is worth noting that HD~259431, one of the hottest stars in this sample, has a border-line $S$ parameter due to a peculiar double- double-peak line structure that will be discussed below (Section \ref{sec: discussion}).

Though both line types are found at any disk inclination, the data suggest a broad trend with viewing angle where $S$ decreases with higher disk inclination. By taking the median of the whole sample in three disk inclination bins, $S=4.3$ for incl~$< 30^{\circ}$, $S=3.3$ for inclinations between $30^{\circ}$ and $60^{\circ}$, and $S=2.0$ for incl~$> 60^{\circ}$.
By focusing on triangular lines, this trend is more pronounced (illustrated with three examples labeled A, B, and C in Figure \ref{fig: shape_param}): lines that show a narrow single central peak and the largest values of $S$ are seen under low inclination angles (e.g. DR~Tau and AS~205~N, with inclinations $\lesssim 30$~deg), while broader and more structured triangular shapes are viewed under larger inclination angles (e.g. Elias~20, RNO~90, GQ~Lup, RY~Tau with inclinations $> 40$~deg and as high as $\approx 65$~deg). 

This global trend is compared in the figure to geometric effects expected from two types of gas velocity structures that have been previously considered to model the CO line profiles. The dashed curves show gas in Keplerian rotation around a star with M$_{\star} =$~0.75 and 3 M$_{\odot}$ (upper and lower curve respectively), assuming a power-law brightness profile $I \propto R^{-\alpha}$ as in \citet{bast11} with index $\alpha = 0.3$, R$_{\rm{in}}=0.1$~au, and R$_{\rm{out}}=10$~au. The shape parameter $S$ shows a dependence on viewing angle only at incl~$< 30^{\circ}$, as already shown in \citet{bast11} where models around the parameters used here have also been explored (with R$_{\rm{in}}$ as small as $0.04$~au and R$_{\rm{out}}$ as large as $100$~au, giving results that cluster around the dashed curves shown here in Figure \ref{fig: shape_param}).

A second trend is marked with a solid line that overlaps with larger values of $S$; this trend is taken from Figure 12 in \citet{pont11} showing the expected CO line shape in a disk + wind model seen under viewing angles of $15^{\circ}$, $35^{\circ}$, and $55^{\circ}$ (as explored in that work). The change in line shape in this case comes from self-absorption from the slow part of a disk wind which is intercepted at large viewing angles \citep[see][]{kurosawa06}, and will be further discussed in Section \ref{sec: discussion}. 
If these geometric effects are in place, the two types of line shapes should overlap at $S\approx$~2.5--3 as observed, i.e. around the maximum of a Keplerian model and minimum of a disk wind model. In the rest of this paper, for the sake of keeping track of the global line shape when displaying other measured properties, we therefore use 2.5 as an approximate separation to mark differently the CO line shape as observed in different objects, even if a more careful separation should account for the disk inclination.

Figure \ref{fig: shape_param} also shows that the presence of an inner dust cavity \citep[from the infrared index $n_{13-30} > 0$, see][]{brown07,furlan09,banz20} does not necessarily imply a double-peaked profile, nor disks with a double-peaked CO line always have a cavity. Triangular lines are however more frequent in full disks, and there are two distinct groups of double-peak lines: one with inner cavities, and one in full disks (points D and E in the figure).

\subsection{Emission line components} \label{sec: components}
Spectra with two emission components have been identified in previous work by comparison of the $v=1-0$ and $v=2-1$ lines shapes: a broad component ``BC" dominating the line wings and a narrow component ``NC" dominating the line center \citep{bast11,bp15}. These components are commonly found in CO spectra with a triangular line shape, and present a different vibrational excitation where BC dominates the flux of $v=2-1$ (i.e. BC is vibrationally hotter) while NC dominates in the $v=1-0$ lines (i.e. vibrationally colder, see Figure \ref{fig: line_stacks}). Previous work also found that, in T~Tauri disks, \ce{^{13}CO} lines are typically narrower than \ce{^{12}CO} lines \citep{bast11,brown13}, and match the shape of the NC component \citep{bp15}. Double-peak lines, instead, usually display a single component ``SC", i.e. lines from different isotopologues and vibrational levels all have similar shape and width (see example in Figure \ref{fig: line_stacks}).
The identification and decomposition into BC and NC is ambiguous or harder to obtain in a few cases; in this analysis we report the case of AB~Aur, where the high-$J$ lines are slightly broader than the low-$J$ but the very narrow line profile leaves the kinematic structure somewhat ambiguous, and CW~Tau, where instead the atypically asymmetric shape of $v=2-1$ did not allow the pipeline to correctly separate the components and we therefore consider it a SC type in this work \citep[the spectrum was instead separated into BC and NC in previous work, see][]{bp15}.

\subsection{Emission line kinematics} \label{sec: res_kinem}
CO line kinematics have been found in previous work to reflect both projection effects and the formation of an inner dust cavity, with line FWHM larger for higher disk inclinations \citep[e.g.][]{blake04} and FWHM smaller for larger $n_{13-30}$ \citep[e.g.][]{salyk11}. Figure \ref{fig: kinem_corr} shows these global trends by combining results from the iSHELL and CRIRES surveys (the trends are visible in each emission component, BC, NC, and SC). These trends are consistent with Keplerian motion imprinting CO line widths, with broader lines when the emission comes from smaller stellocentric distances, and with inner gas depletion removing CO emission at high velocities \citep[e.g.][]{bp15}. 
In addition, there is a global trend with stellar mass/luminosity and the emission line FWHM (Figure \ref{fig: kinem_corr} shows the trend with stellar mass), where lower-mass stars tend to have broader CO lines (a median of $\approx 40$~km/s at masses $< 1 M_{\odot}$) and higher-mass stars narrower CO lines (a median of $\approx 20$~km/s at masses of 1--3~$M_{\odot}$). Previous work explained this trend as a luminosity effect, where the typical range of gas temperatures that excites CO moves to larger disk radii (and therefore a narrower line FWHM) as a function of stellar luminosity \citep{salyk11}. For the first time, this survey now shows that the trend breaks at masses $> 3 M_{\odot}$, where CO is always broad (FWHM 50--130~km/s); in this sample, these stars are: 51~Oph, HD~259431, HD~35929, HD~36917, HD~37806, HD~58647, VV~Ser, and MWC~297. MWC~297 is the most massive star in this sample with a mass of nearly 20~$M_{\odot}$ \citep{guzmandiaz21}, and is not included in Figure \ref{fig: kinem_corr} for better visualization of the rest of the sample.
It should also be mentioned that, in each two-dimensional plane, all these trends present significant scatter that is partly due to the multi-dimensional dependencies of CO emission, as show by the multiple panels in the figure. This multi-dimensionality is one aspect where future work should invest to better characterize the dependencies of CO emission on star/disk properties.

Another key kinematic feature is the Doppler shift in line centroids. Measurements of stellar RVs are needed to measure any velocity shifts in the observed CO lines in reference to their system. RV measurements for pre-main sequence stars are notoriously challenging due to high veiling, line broadening, and/or the number of stellar absorption lines that are or can be observed. Measured RVs for T~Tauri stars can easily differ by a few km/s in different works, suggesting that the real uncertainties are at this level, beyond what reported in nominal error-bars. 
RV measurements for Herbig AeBe stars are even more challenging due to the fewer photospheric lines available, which are also much broader then those observed in T~Tauris, resulting in uncertain RV measurements that sometimes differ by a few tens of km/s in different works using different photospheric lines. In this work, for five Herbig stars we have therefore adopted velocities from millimeter observations of cold CO gas that more reliably measure the systemic velocity (see references in Table \ref{tab: sample}). The case of HD~37806 shows a large shift in the CO line centroid from the stellar frame, but the only RV measurement that we have found in the literature is very uncertain \citep[$47 \pm 21$~km/s,][]{alecian13}. 

Figure \ref{fig: kinem_corr}, right panel, shows the CO line centroids in the stellar reference frame (i.e. after subtraction of the stellar RV in each source), demonstrating that both absorption and emission lines are predominantly blue-shifted. A similar result was reported for the CRIRES survey from the line asymmetries, which were found to be mostly on the blue side of CO lines \citep[Figure 6 in][]{brown13}. The centroid blue-shifts, although mostly confined within -5~km/s and therefore individually consistent with or close to the combined uncertainty from wavelength calibration and stellar RVs, are clearly systematic across the sample and cannot be explained with random noise. We run a Kolmogorov–Smirnov test to compare the observed distribution of CO centroids (100 datapoints in total, combining all components, with mean of -1.6~km/s and standard deviation of the mean of 0.4~km/s) to a normal distribution with same sample size and same standard deviation as the distribution of the data (3.4~km/s), but centered at 0~km/s. We repeat the KS test 10,000 times by randomly populating the normal distribution centered on 0 for comparison to the observed distribution, and measure the distribution of 10,000 KS probability values. The median probability of this distribution is 0.008 demonstrating a highly significant deviation from a sample randomly centered at zero, and only 8\% of the realizations in the tail of the distribution have probabilities larger than 0.1. 
A tendency for broader lines to show larger blue-shifts is also tentatively present in the data (Figure \ref{fig: kinem_corr}, right panel), and it is reminiscent of a trend observed in forbidden oxygen [OI] emission that traces slow inner disk winds \citep[][see more in Section \ref{sec: discussion}]{banz19,weber20}.

\begin{figure*}
\centering
\includegraphics[width=1\textwidth]{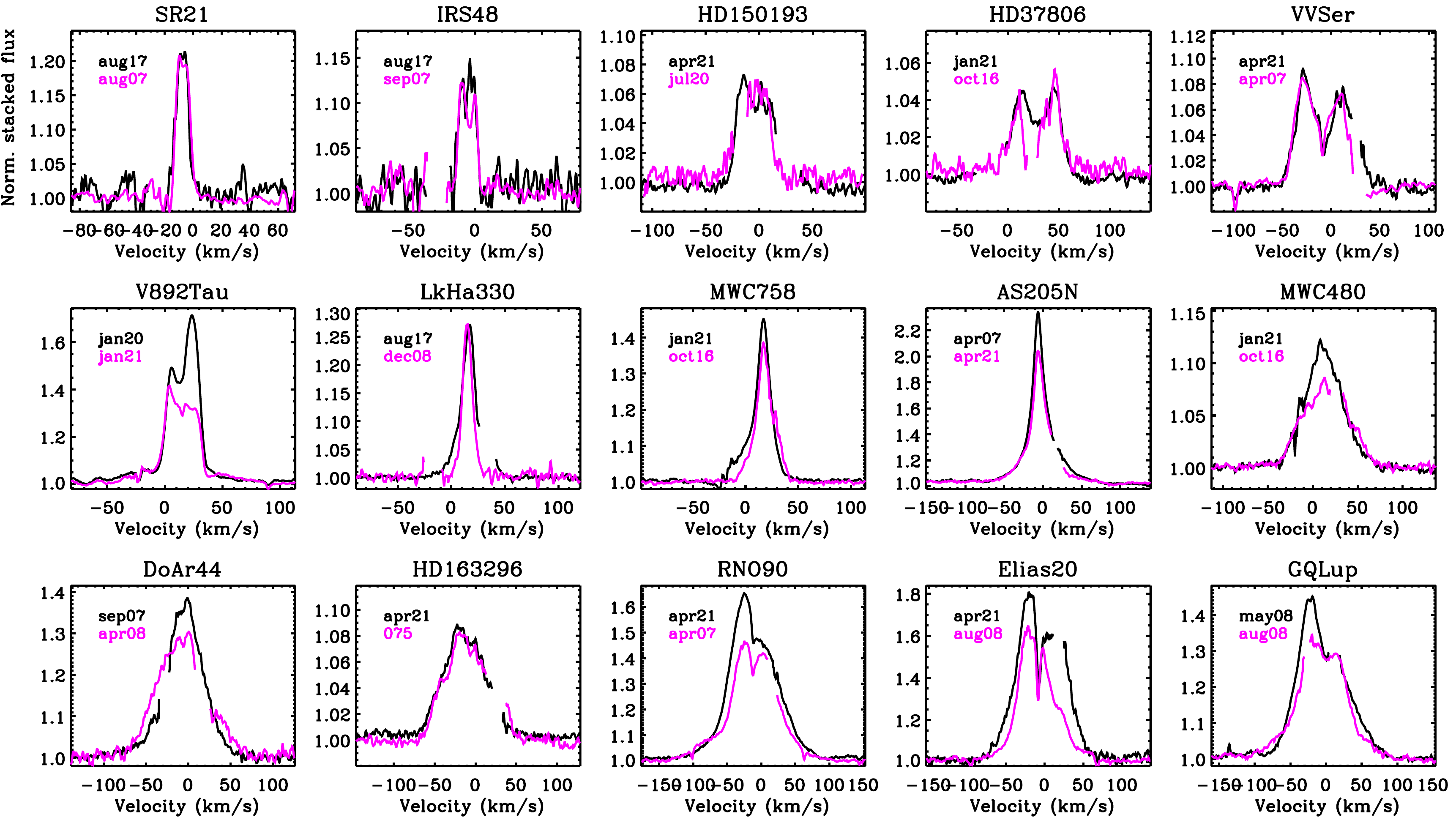} 
\caption{Overview of kinematic variability observed in CO $v=1-0$ emission lines. Epochs are written in each plot in the format mmmyy (three letters for the month, two numbers for the year). Spectra from 2007--2008 have been taken with CRIRES \citep{pont11,brown13}, those from 2016-2021 with iSHELL (this work). Double-peak line shapes are shown at the top of the figure, followed by triangular lines below. In the case of HD~163296, the second spectrum is taken 2 days later with the 0.75" slit (Appendix \ref{app: logs}); this target is included in this figure for reference to kinematic variability observed before in \citet{heinbert16_var}.}
\label{fig: line_variability}
\end{figure*}

\begin{figure*}
\centering
\includegraphics[width=1\textwidth]{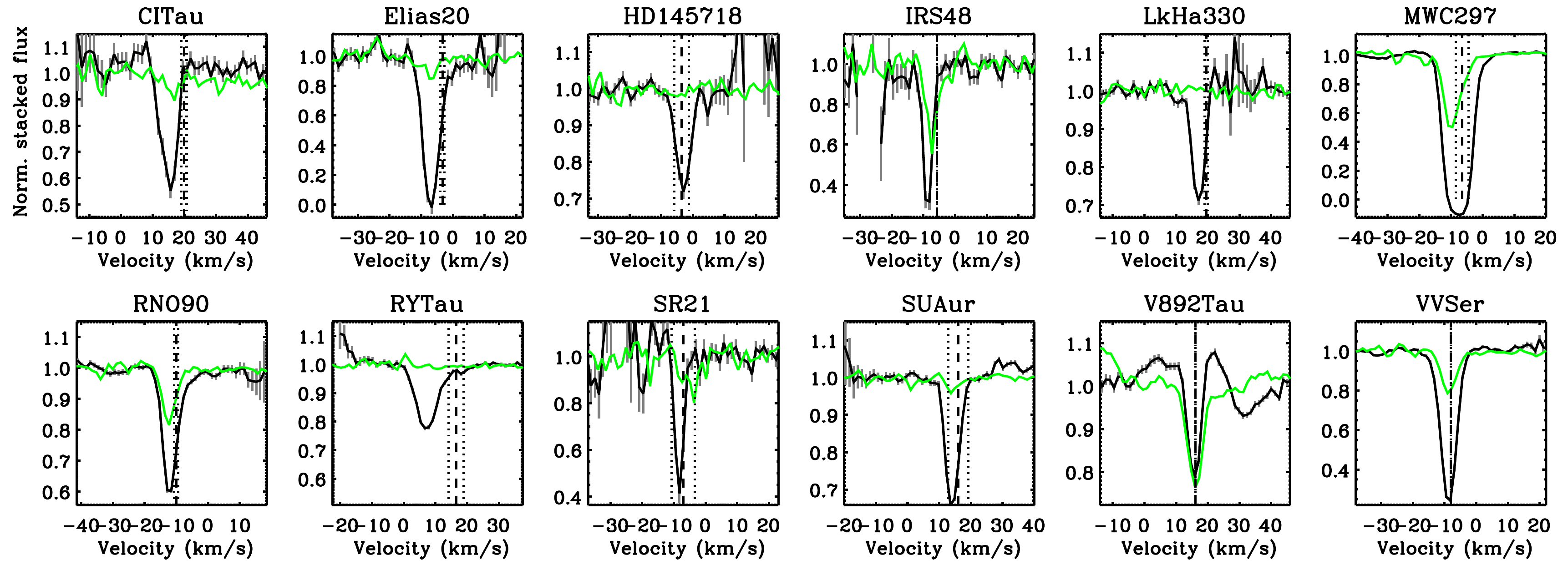} 
\caption{Overview of $v=1-0$, $J=1$ absorption lines (black: \ce{^{12}CO}, green: \ce{^{13}CO}). The spectra are shown after subtraction of CO emission (some residuals from imperfect subtraction are visible in V892~Tau). The vertical dashed and dotted lines shows the stellar RV and its uncertainty.}
\label{fig: line_gallery_abs}
\end{figure*}

\subsection{Emission line variability}
Early evidence for kinematic variability in $M$-band CO spectra was found in a sample of five disks, where two of them showed a strong change between the red and blue side of the line profile in different epochs \citep{najita03}. Within the CRIRES survey, line variability was checked over a timescale of two years in a sample of 11 disks, and found it only in two disks: RU~Lup in terms of a simple line flux change, and in VV~CrA as the disappearance of outflow absorption \citep{brown13}.
Stronger variability, both in line flux and kinematics, was found in a system that had an accretion outburst, EX~Lup, where the BC strongly increased compared to the NC and showed time-dependent shifts in the line asymmetry between the blue and red peaks of the line \citep{goto11,banz15}, suggesting an orbiting hot spot in the inner accretion region of the disk. 
Kinematic variability has also been detected by comparison of three NIRSPEC and CRIRES spectra taken over a decade (2001--2012) in the CO spectrum of HD~163296, and was attributed to a variable sub-Keplerian component in a wind \citep{heinbert16_var}.
Another type of kinematic variability has been detected in HD~100546, where time-dependent shifts in a CO line asymmetry over 14 years (2003--2017) have suggested the presence of a hot spot orbiting the star near the dust inner cavity rim, possibly connected to a forming planet \citep{brittain13,brittain14,brittain19}.
With this new iSHELL survey, it is now possible to investigate line variability over timescales as long as 14 years (though not obtained in any systematic way), and for a sample of at least 15 disks.

Figure \ref{fig: line_variability} shows an overview of variability observed in CO $v=1-0$ emission lines. We check for variability in the gas line kinematics only (not in the line flux, since spectra have not been flux calibrated) using the stacked line profiles, and find that double-peaked profiles can be kinematically stable over timescales between 1 and 14 years, though only six disks of this type are available with multiple epochs and the S/N of one epoch is lower than in the other one in most cases. The most clear case of kinematic stability is the spectrum in VV~Ser, which was found very stable also on timescales of 1~yr \citep{pont11}. A notable example of a double-peak line that strongly varies is in V892~Tau, which hosts a circumbinary disk around two Herbig stars \citep[e.g.][]{long21}. $M$-band CO emission is observed in this system for the first time, finding that it shows strong variability in a red-shifted asymmetry over one year. This target is being followed up with another epoch in 2022, and this peculiar case of kinematic variability will be studied in a future paper.

Triangular lines, on the contrary, generally show kinematic variability with evident changes in line shape and/or width on any timescale. A group of CO lines that stands out in this context is that observed in RNO~90, Elias~20, and GQ~Lup, which all show an almost identical line structure where strong blue-shifted asymmetries appear and disappear (bottom right of Figure \ref{fig: line_variability}). This variability shows a remarkable similarity across three disks that share a high disk inclination (50--60~deg), and is observed on timescales as short as 3 months and as long as 14 years, suggesting that it could be periodic or at least recurring. This variability does not seem to be related to different slit widths possibly filtering out extended emission, because in GQ~Lup it is observed using the same slit and same position angles.
The nature and origin of this variability will be the study of future work.

\subsection{Absorption components} \label{sec: absorption}
As observed in the CRIRES survey before, CO absorption spectra from disks often show low rotational excitation with deep lines for the lowest $J$s and rapidly decreasing strength up to at most $J = 7$ beyond which no lines are detected, a structure that can be produced by a low rotational temperature. 
In the iSHELL sample, high $J$-levels are detected in absorption up to $J =$~20--40 only in four disks: CI~Tau, HD~145718, RY~Tau, SU~Aur. In the CRIRES sample included in this work, high $J$ levels are detected in CW~Tau and IQ~Tau only. These six disks are all highly inclined with inclinations between 50 and 65 deg, and they have a triangular line shape (except for IQ~Tau that is double-peaked, and HD~145718 where the emission line shape cannot be characterized).
Absorption line FWHMs span from the resolving power limits (3 and 5~km/s depending on the slit used) up to 9--10~km/s in RY~Tau.
Absorption components are observed both in double-peak emission line spectra (7/20 disks) and in triangular lines (8/33 disks), and in both types they tend to be observed in disks with $incl > 40$~deg, at least in this sample. A tentative trend of increasing absorption FWHM for increasing disk inclination might be present at $incl > 50$~deg (Figure \ref{fig: kinem_corr}).
The absorption component in MWC~297 shows up as an outlier in this figure, with a FWHM of 10~km/s at an inclination of $\approx 30^{\circ}$; this peculiarity is further described in the Appendix.

\begin{figure*}[ht!]
\centering
\includegraphics[width=0.95\textwidth]{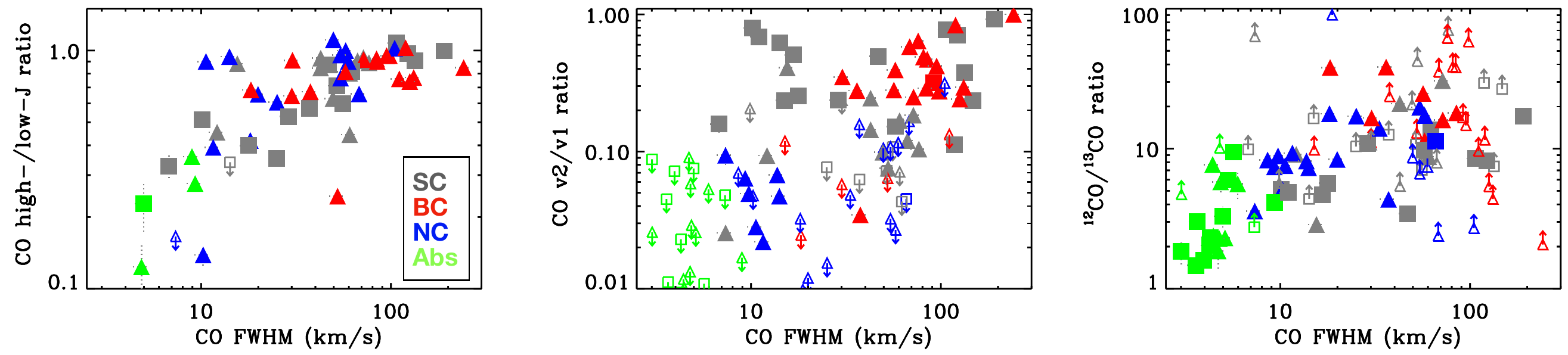} 
\caption{CO line excitation for different components, by combining the iSHELL and CRIRES surveys. Symbols reflect the full line shape parameter $S$ as in Figure \ref{fig: shape_param}. The ratio of high-$J$ to low-$J$ $v=1-0$ lines, a proxy for the rotational excitation, increases with line FWHM (left). Absorption components generally have values $< 0.1$, and are not included for better visualization of the trend. The $v2/v1$ ratio, a proxy for vibrational excitation, shows grouping of datapoints according to their FWHM as will be discussed in Section \ref{sec: discussion} (middle). The \ce{^{12}CO}/\ce{^{13}CO} ratio, a proxy for column density, increases with line FWHM. In these three excitation tracers, absorption components, especially those measured in highly inclined disks, show measurements that overlap with those obtained for the emission components, especially in NC. These three excitation tracers show no global relation to the stellar or accretion luminosities (Appendix \ref{app: co_excit}).}
\label{fig: excit_corr}
\end{figure*}

\begin{figure*}[ht]
\centering
\includegraphics[width=0.95\textwidth]{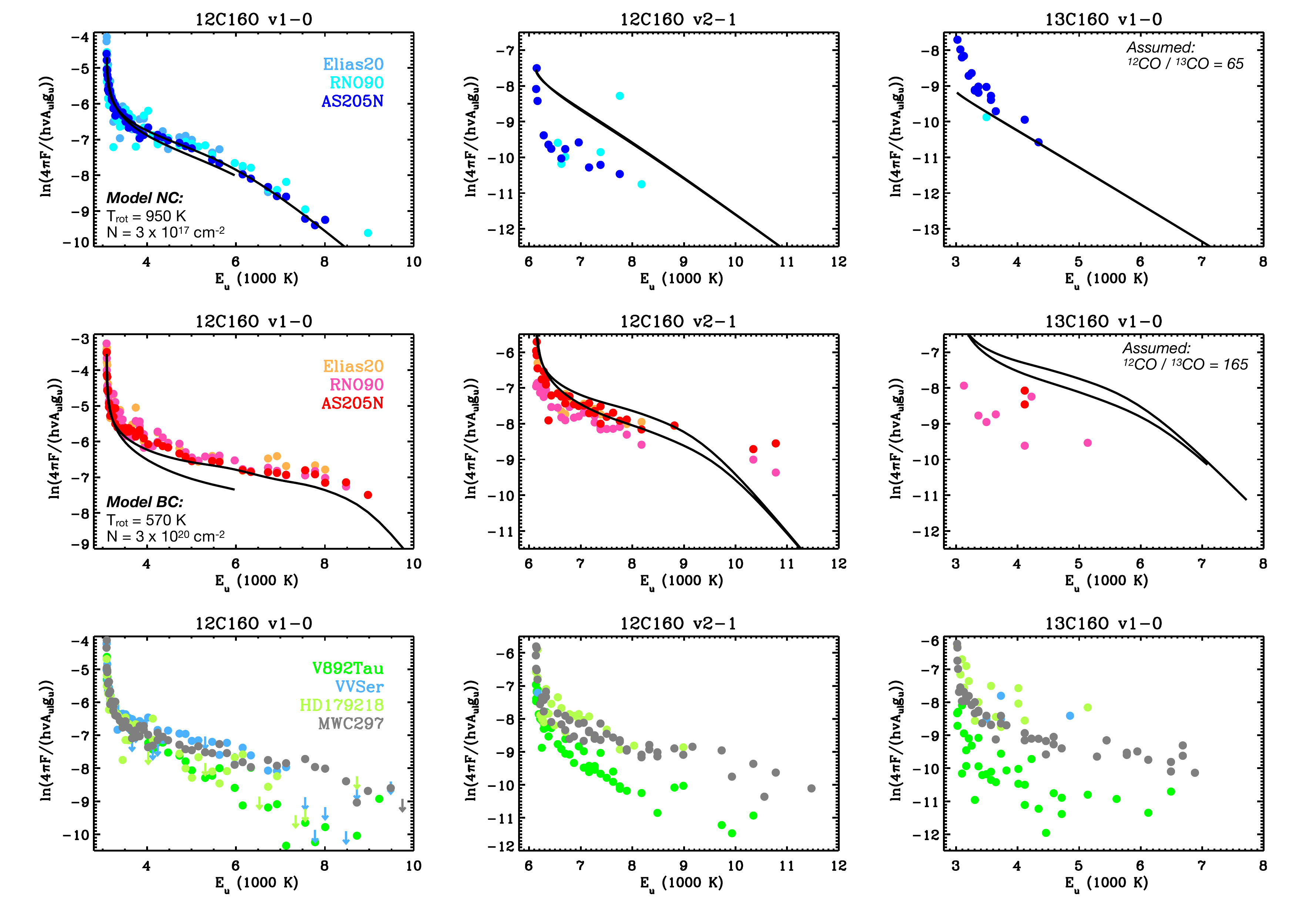} 
\caption{Different excitation of emission components. \textit{Top}: examples of NC in triangular lines. \textit{Middle}: examples of BC in triangular lines. \textit{Bottom}: examples of SC in double-peak lines. Line fluxes are not flux calibrated, therefore the rotation diagrams are vertically scaled to match the $v=1-0$ lines in individual objects to better highlight the similarity or difference in curvature, which reflect the excitation conditions. LTE slab model fits to the $v=1-0$ lines are overplotted in black for NC and BC, showing their prediction for the $v=2-1$ and \ce{^{13}CO} lines in the middle and right panels (see Section \ref{sec: res_excit}). Upper limits are not plotted for the $v=2-1$ and \ce{^{13}CO} lines for a better visualization of detections.}
\label{fig: rotdiagr_gallery}
\end{figure*}

\subsection{Line excitation} \label{sec: res_excit}
\subsubsection{Line ratios}
The spectral completeness of iSHELL spectra provides an unprecedented view of the different excitation conditions of multiple CO components in disks. In this work, we systematically collect some line ratios to provide a global overview of diverse excitation conditions and trends: the high-$J$ ($J = 25$--31 ) to low-$J$ ($J = 5$--18) line flux ratio for $v=1-0$ lines (a proxy for the rotational excitation), the $v=2-1$ to $v=1-0$ line flux ratio (a proxy for the vibrational excitation), and the \ce{^{12}CO}/\ce{^{13}CO} ratio (a proxy for column density). To measure these ratios, we use the line stacks presented above in Section \ref{sec: analysis} (Figure \ref{fig: line_stacks}). These ratios are a convenient empirical proxy that we use to highlight broad trends in CO excitation across large samples, but should not be expected to fully capture the complexity and diversity of CO spectra. To illustrate that, in the next Section we will also show some examples of rotation diagrams to motivate a more comprehensive study of excitation conditions in future work.

These three tracers of excitation and physical conditions show strong dependency on the line width, with some global trends that encompass all lines regardless of type of profile from the broadest lines down to the absorption components detected in inclined disks (Figure \ref{fig: excit_corr}). These global trends suggest that, on top of viewing projection effects (Figure \ref{fig: kinem_corr}), the range of line widths measured in different line types or velocity components (BC vs NC) reflects different gas locations in the disk, scanning disk radii from the innermost regions inside dust sublimation to disk regions out to possibly tens of au (the absorption components).
The CO rotational excitation, in particular, has not been studied systematically in the CRIRES survey, due to the lack of spectral coverage of the high-$J$ lines in most of the sample (Appendix \ref{app: crires_lines}). Its proxy shows a strong trend with line width, with BC having the highest excitation and narrow components (whether in emission or absorption) the lowest.
The \ce{^{12}CO}/\ce{^{13}CO} ratio also shows a positive trend with FWHM. These first two trends are evident even in single spectra, e.g. AS~205~N, by comparison of BC (strong in the high-$J$s but weak in \ce{^{13}CO}) and NC (weaker in the high-$J$s but stronger in \ce{^{13}CO}, Figure \ref{fig: line_stacks}). 

The vibrational excitation, instead, shows a more complex behavior. There is a global trend where $v2/v1$ increases with FWHM, and this trend is visible even in single spectra, e.g. AS~205~N, by comparison of BC (vibrationally hotter) and NC (vibrationally colder, see Figure \ref{fig: line_stacks}). However, a high vibrational excitation is also found in a group of narrow double-peak lines, those observed in disks that have large inner dust cavities (top left side of the middle panel in Fig.\ref{fig: excit_corr}). Parts of these trends have already been seen and discussed in previous work \citep{bp15,banz18}, and will be further discussed in Section \ref{sec: discussion}.
These three excitation tracers do not show any obvious global trends with stellar and accretion luminosity (see Appendix \ref{app: co_excit}), suggesting that CO excitation reflects a range of local disk conditions (especially the level of mixing with dust, as will be discussed below) rather than simply the stellar or accretion spectrum.

\subsubsection{Rotation Diagrams}
These line flux ratios are simple proxies of temperature and density only in LTE conditions, while CO emission in disks has been often found to have non-LTE excitation \citep[e.g.][]{thi13}. Different gas temperatures and densities, together with different types of excitation, can produce distinct curvatures in rotation diagrams \citep{gl99}. While a systematic rotation diagram analysis for the full sample is left to future work, we include here some representative examples to illustrate the unprecedented coverage of iSHELL spectra and its potential in revealing different excitation in different types of lines. For reference, previous work covered lines with upper level energy $E_{up}$ up to 5000~K (6000~K or 7500~K in a few cases only, and with large gaps of 500--1000~K in between) and only in a smaller sample, by performing multiple observations with different spectral settings \citep[e.g.][]{najita03,blake04,salyk11,bast11,brown13,sanchez21}. iSHELL spectra now systematically cover $v=1-0$ lines up to $E_{up}$ = 10,000~K, revealing the full shape of curvatures that reflect the combination of line opacities, temperature, and type of excitation \citep{gl99}. 

We include in Figure \ref{fig: rotdiagr_gallery} some examples to illustrate the excitation of CO in spectra with different line shapes and components. A similar curvature is visible in most cases at $E_{up} <$~4,000~K, which can be explained with optically thick emission. However, iSHELL spectra now reveal very distinct curvatures at $E_{up} >$~5,000~K that were never fully observed before. We briefly describe them here, using a slab model of gas in LTE as guidance \citep{banz12}. With the model, we fit in each case only the $v=1-0$ lines, and we report in the figure how that model compares with the $v=2-1$. Also, we report how that model compares to \ce{^{13}CO} lines, by assuming a \ce{^{12}CO}/\ce{^{13}CO} column density ratio of 65 as observed in the ISM, and up to 165 as measured in some disks \citep{smith15}. 
Narrow components NC in triangular lines show a steeper rotational curvature than the broad components BC in the same disks. This different curvature is consistent with a higher rotational temperature T$_{rot}$ in NC than in BC ($\approx950$~K vs $\approx550$~K), and a much lower column density in NC than in BC ($\approx 10^{17}$~cm$^{-2}$ vs $\approx 10^{20}$~cm$^{-2}$). This range of column density corresponds in the models to median line opacities $\tau \approx 0.1$ (optically thin) in NC and $\tau \approx 20$ (moderately optically thick) in BC.
Together with a different rotational excitation, NC and BC also show a very different vibrational excitation, as shown above in this section using the stacked line ratios. NC $v=2-1$ lines are sub-thermally excited compared to T$_{rot}$ estimated for the $v=1-0$ lines, suggesting conditions where T$_{vib} <<$~T$_{rot}$. BC, instead, are only slightly sub-thermally excited, consistent with T$_{vib} \approx$~T$_{rot}$. Another very clear difference is in the \ce{^{12}CO}/\ce{^{13}CO}, which could be lower than the ISM value of 65 in NC, and higher than 165 in BC.
Double-peak lines, instead, seem to comprise the full range of excitation conditions found in NC and BC, consistent with a range of rotational temperatures and column densities. One notable difference is that these lines also show cases with T$_{vib} >$~T$_{rot}$ (e.g. in V892~Tau). Rotation diagrams for the full sample will be included in a future paper.

A simple slab model in LTE provides a remarkable success in fitting the full curvature in the rotation diagram of $v=1-0$ lines, suggesting that the emission could be rotationally thermalized. This was already discussed before based on fit results from narrower spectral ranges \citep[e.g.][]{brittain07,thi13,vdplas15}, suggesting that UV and IR pumping could dominate the vibrational excitation but not affect much the rotational temperature due to collision rates that are a factor 10--100 larger, and Einstein-$A$ coefficients that are orders of magnitude lower, than for the vibrational transitions.
In some cases (the BC and some double-peak lines), the $v=1-0$ model almost matches even the $v=2-1$ line curvature and $v2/v1$ ratio, suggesting the emission could be close to being thermalized even vibrationally due to a large gas density \citep{bosman19}.
Nonetheless, deviations from LTE excitation are very evident in the vibrational excitation of NC components as well as in several spectra with SC lines, suggesting that non-LTE excitation could be common. A comprehensive analysis of CO excitation is left to future work.

The \ce{^{13}CO} lines, too, show deviations from the simple prediction from the slab model even assuming a range in \ce{^{12}CO}/\ce{^{13}CO} ratios as observed in previous work \citep{smith15}. The mismatch could be due to different emitting regions, or non-LTE vibrational excitation. The similar line shapes between \ce{^{12}CO} and \ce{^{13}CO} are consistent with a similar radial emitting region, but modeling is needed to determine if a different combination of emitting radii and depths into the disk surface could explain the emission from the two isotopologues without producing noticeable line kinematic differences.

\begin{figure*}
\centering
\includegraphics[width=0.8\textwidth]{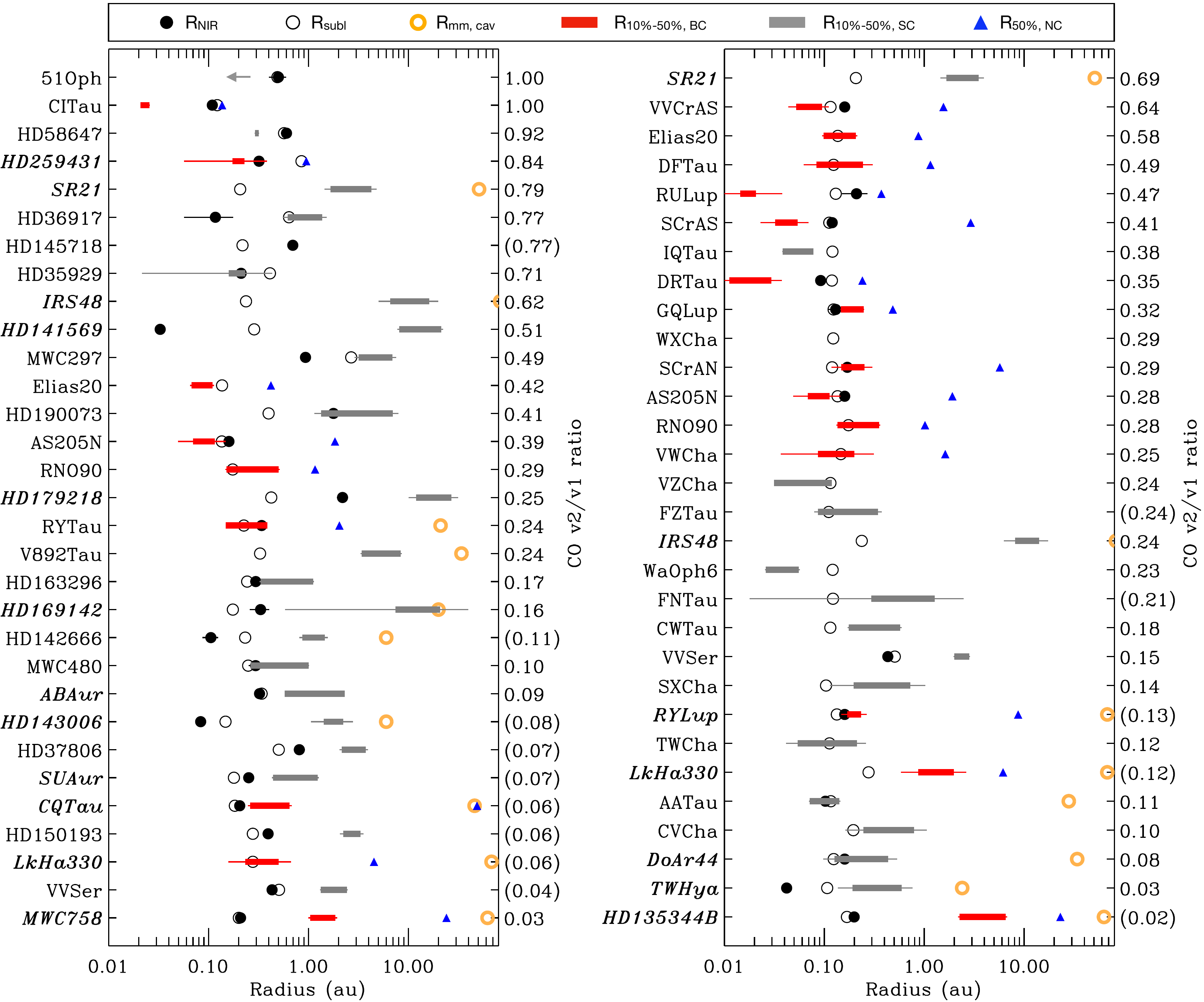} 
\caption{
Comparison between inner dust and CO gas radii for the iSHELL (left) and CRIRES (right) surveys. The samples are ordered from bottom to top by the measured CO $v2/v1$ ratio (right axis with upper limits in parentheses, displaying the measured value for BC in triangular lines). Names of disks with evidence for an inner dust cavity from the infrared index $n_{13-30} > 0$ are marked in bold italics. $R_{\rm{NIR}}$ is the measured inner dust radius from $H$- or $K$-band interferometry with the VLTI \citep{lazareff17,gravity19,gravity21_TTs}. $R_{\rm{subl}}$ is an empirical estimate of $R_{\rm{NIR}}$ as a function of the stellar luminosity based on data collected in \citet{marcosarenal21}, see Appendix \ref{app: dust_radii}. $R_{\rm{10\%}}$ and $R_{\rm{50\%}}$ are estimates of the CO emitting radius from the HW10\% and HWHM as explained in Section \ref{sec: emit_radii}. Error-bars (thin lines) are dominated by the uncertainty in disk inclinations. $R_{\rm{mm, cav}}$ report the approximate size of an inner dust cavity from spatially-resolved millimeter interferometry \citep{pinilla18,huang18,francis20}.}

\label{fig: radii_compar}
\end{figure*}

\begin{figure*}
\centering
\includegraphics[width=1\textwidth]{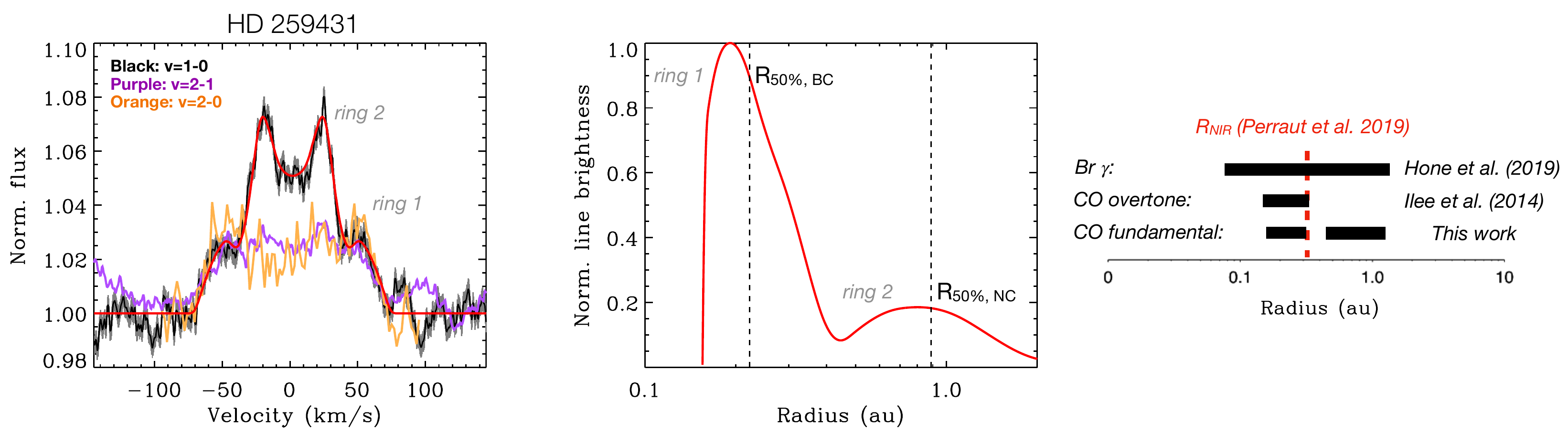} 
\caption{Two Keplerian rings of CO emission observed in HD~259431. A model fit to the high-$J$ stacked line profile (which do not suffer from telluric absorption) is shown in red, using the inversion method described in \citet{bosman21}. Estimates of the emitting radius of CO from the measured HWHM, i.e. the $R_{\rm{50\%}}$ values plotted in Figure \ref{fig: radii_compar}, are shown for comparison in the middle plot. This disk also shows a Keplerian ring of CO overtone emission \citep{ilee14} that corresponds to the inner ring observed in fundamental emission in this work. To illustrate that, in the left plot we include the line profile of $v=2-0$ emission as recently observed with IGRINS \citep{grant21}.}
\label{fig: kepl_rings}
\end{figure*}

\subsection{Line emitting radii} \label{sec: emit_radii}
If the observed gas line profiles are broadened by Keplerian rotation in a disk, a simple empirical estimate for the radial location of the emitting gas can be derived from the measured line broadening $\Delta v$ and Kepler's law as $R =  GM_{\star} ~ (sin(i) / \Delta v)^2$, where $i$ is the viewing angle (i.e. the disk inclination). As a Doppler broadening representative for the radial distribution of the emission we take the half-width-at-half-maximum (HWHM), and call this radius $R_{\rm{50\%}}$ or more generically $R_{\rm{CO}}$ as in \citet{bp15}. We also take the half-width-at-10\% of the line peak (HW10\%) as an estimate of an inner radius $R_{\rm{10\%}}$ for the emitting CO gas. These radial estimates are shown in Figure \ref{fig: radii_compar} for each disk as a bar that extends from $R_{\rm{10\%}}$ out to $R_{\rm{50\%}}$, to indicate the inner to intermediate radial extent of the emission. The largest source of uncertainty in these estimates is typically the disk inclination, producing the large error bars visible in the figure for some objects. A shift in the estimated emitting region may also be produced by a different disk inclination, in cases where inner and outer disk are misaligned. While these are and should be taken just as approximate radial estimates, they give a convenient tracer of the inner distribution of the gas emission that can be compared across the whole sample. For the NC, in Figure \ref{fig: radii_compar} we only mark $R_{\rm{50\%}}$, and discuss further in Section \ref{sec: discussion} whether a Keplerian interpretation of the line profile is appropriate.

For comparison to the emitting gas radial estimates, in Figure \ref{fig: radii_compar} we also include measured radii for rings of inner disk dust $R_{\rm{NIR}}$ where available from $H$- or $K$-band interferometry with PIONIER and GRAVITY on the VLTI \citep{lazareff17,gravity19,gravity21_TTs}, which trace the innermost hot dust and are typically interpreted as tracing the distance from the star where dust sublimates, i.e. a tracer of the inner dust disk rim. 
These radii are complemented with an estimate of the inner dust rim radius as a function of the stellar luminosity (that we call $R_{\rm{subl}}$) following recent work \citep[][see Appendix \ref{app: dust_radii}]{marcosarenal21}, to obtain a $R_{\rm{subl}}$ estimate for all targets in the sample, even those that have not been spatially resolved with NIR interferometry yet. These $R_{\rm{subl}}$ estimates based on the stellar luminosity are generally close to the measured $R_{\rm{NIR}}$, where available, as they should, because $R_{\rm{subl}}$ is estimated from the measured relation between $R_{\rm{NIR}}$ and $L_{\star}$. There are a few cases where they are different by a factor of $\approx$~5--10 (HD~141569, HD~179218, HD~36917), which is consistent with the presence of some outliers (see Figure \ref{fig: Rdust_empirical} in the Appendix). 

\begin{figure*}
\centering
\includegraphics[width=1\textwidth]{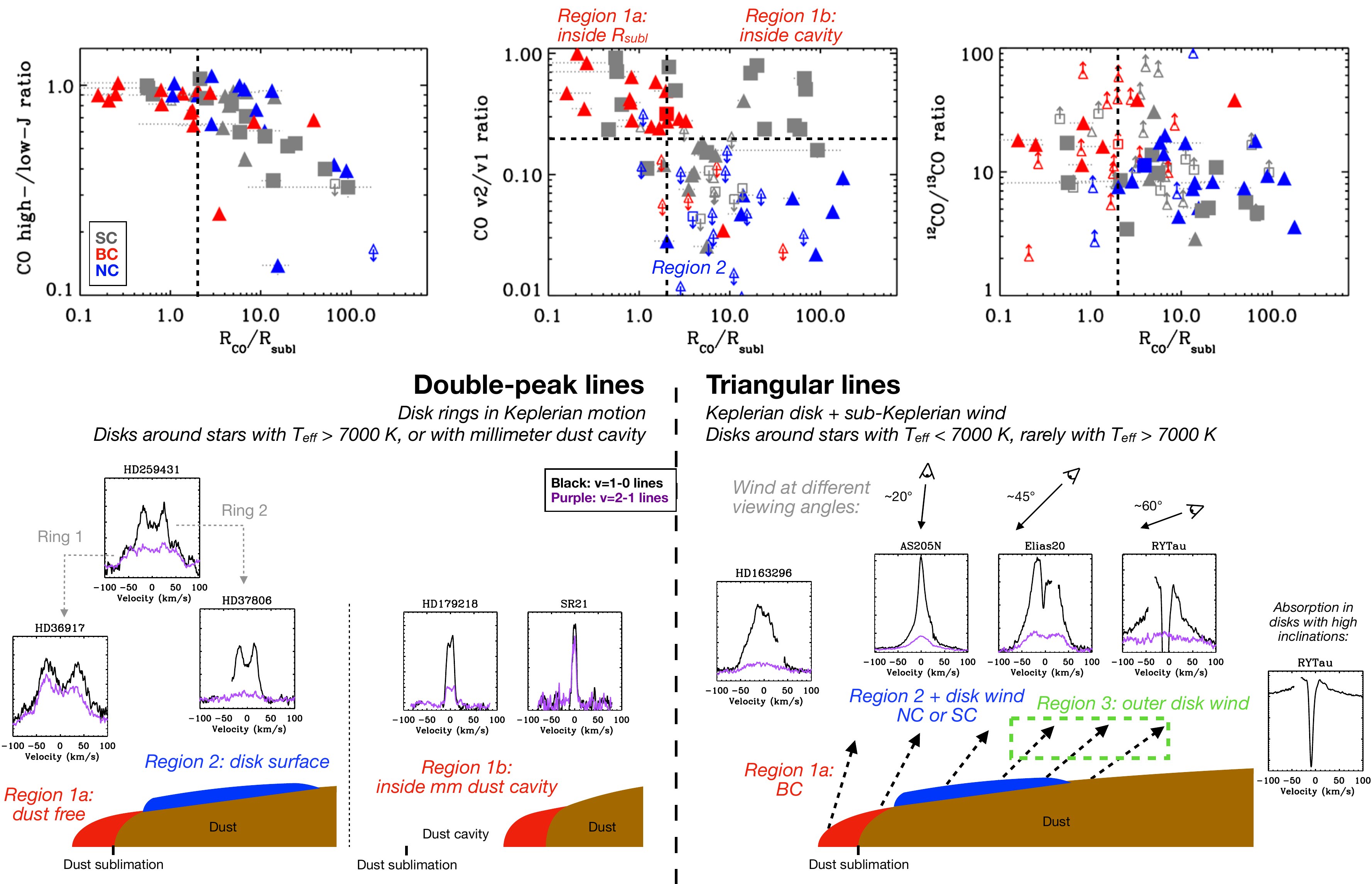} 
\caption{Overview of line shape types and their different emitting regions based on the observed line kinematics and excitation, as discussed in Section \ref{sec: discussion} (see also Figure \ref{fig: classification}). The three plots at the top of the figure are the same as in Figure \ref{fig: excit_corr} but converting the CO FWHM into a representative emitting radius $R_{\rm{CO}}$ and normalizing that to the dust sublimation radius $R_{\rm{subl}}$ in each disk (see Section \ref{sec: emit_radii}). }
\label{fig: discussion}
\end{figure*}

In Figure \ref{fig: radii_compar}, the sample is ordered by the measured CO $v2/v1$ flux ratio for an easier comparison to what will be shown in Figure \ref{fig: discussion}. When $v2/v1$ ratio is lower than $\approx 0.2$, CO generally emits from radii larger than the inner dust radius, i.e. from a surface layer above the dusty disk. Instead, when $v2/v1 > 0.2$, CO generally emits from a region inside and overlapping to the inner dust radius, i.e. a region that should be partly dust-free and partly overlap with the inner dust rim. A very interesting case in this context is the spectrum observed in HD~259431, where two distinct double-peak and symmetric components in ro-vibrational CO are detected for the first time in a disk, to our knowledge (Figure \ref{fig: kepl_rings}). The inner ring (ring 1 in the figure) is vibrationally hot ($v2/v1 \approx 0.8$) while the outer ring is vibrationally cold ($v2/v1 < 0.1$). For reference, the figure compares the radial distribution of the emission from the model to the simple estimates of $R_{\rm{50\%}}$ from the measured HWHM. A comparison to spatially-resolved measurements of the inner location of hot dust in this disk by \citet{gravity19} demonstrates that these two rings separate at the onset of the innermost dust surviving close to the star ($R_{\rm{NIR}}$ in the figure). The innermost ring also coincides with CO fundamental emission observed by \citet{ilee14}, once scaled using the new stellar mass and disk inclinations used here (see Table \ref{tab: sample}), and the inner emitting region of Brackett $\gamma$ emission as observed by \citet{hone19}. In general, disks seem to have only either one or the other CO emitting region, though hints for a similar double-ring structure are visible in HD~35929 too. A similar structure has been seen before only in HD~101412 but in optical [OI] emission, and not in fundamental or overtone CO emission \citep[Figure 10 in][]{vdplas15}; even in that case, the broad inner ring of CO coincides in both fundamental and overtone CO emission.

As found in previous work and shown in Figure \ref{fig: excit_corr}, the other situation where large $v2/v1 > 0.2$ are measured is that of disks with inner dust cavities that extend beyond the sublimation radius (typically called ``transitional" disks) and are usually attributed to dynamical interaction with companions and/or disk dispersal by winds \citep[e.g.][]{espaillat14,alexander14}. In these disks (marked with boldface names in Figure \ref{fig: radii_compar}), CO emits from larger radii than $R_{\rm{NIR}}$. The radial size of the inner dust cavity as measured with millimeter interferometry, tracing the mm-cm size dust grains, is always larger than both $R_{\rm{NIR}}$ and $R_{\rm{CO}}$ in these disks, demonstrating that the observed NIR dust and CO gas are located inside the millimeter cavity. This stratified radial structure with CO surviving inside the dust cavity has been spatially resolved in a few disks with spectro-astrometry of fundamental CO lines \citep{pont08} and ALMA imaging of millimeter CO lines \citep{vdmarel16}.

\subsection{Summary of results} \label{sec: res_summary}
We conclude with a short summary of the main results described above in this section. All results and trends refer to what found from the combined iSHELL+CRIRES sample (Section \ref{sec: sample}).

\paragraph{Double-peak emission lines} 
A double-peak line profile, which is the typical shape expected for gas in Keplerian rotation in a disk, is detected in $M$-band CO emission for the first time at a high rate in the iSHELL survey ($\approx50\%$ of the sample), providing the most striking difference in detection rates in comparison to the previous CRIRES survey (Figure \ref{fig: surveys_comparison}). Double-peak CO lines (corresponding to a shape parameter $S $ = FW10\% / FW75\% $\lesssim 2.5$, see Section \ref{sec: line_shapes}) are observed at any disk inclination, in disks with and without inner dust cavities, but are more frequent in stars with $T_{\rm{eff}} > 7000$~K.

\paragraph{Triangular emission lines} 
Triangular lines are distinguished by narrower line centers and broader lines wings ($S \gtrsim 2.5$), and they are more frequent in stars with $T_{\rm{eff}} < 7000$~K and in disks without an inner dust cavity. They typically show two kinematic components with different FWHM and excitation (BC and NC, Section \ref{sec: components}), and they show kinematic variability over any timescale explored so far (Figure \ref{fig: line_variability}). They show a single narrow peak when observed at low inclinations ($< 30$~deg) and broader, more structured line centers when observed at high inclinations ($> 45$~deg), often with an asymmetry on the blue side of the line and narrow, blue-shifted central absorption. 

\paragraph{Absorption lines}
Absorption lines are detected in both double-peak and triangular emission lines, but they tend to be observed in disks with high inclinations ($incl > 40$~deg). They are narrow ($< 10$~km/s, down to the spectral resolution limits of 3--5~km/s) and more often rotationally ``cold" (excited up to $J < 7$). They are typically blue-shifted in triangular lines, and less so in double-peak lines (Figure \ref{fig: line_gallery_abs}). Broader (up to $10$~km/s) and more rotationally excited absorption (up to $J =20$--40) tends to be observed in spectra with triangular emission lines and highly-inclined disks, with $incl > 50$~deg.

\paragraph{Trends in line FWHM and centroids}
CO emission line FWHMs are as narrow as $< 10$~km/s and as broad as 200~km/s and show the following trends (in any line type and any kinematic component, Figure \ref{fig: kinem_corr}): they decrease with stellar mass between 0.5 and 3~$M_{\odot}$, they increase with disk inclination, and decrease in disks with an inner dust cavity (Section \ref{sec: res_kinem}). A new finding from the iSHELL survey is that in disks around stars of $>$~3~$M_{\odot}$, CO seems to break the overall trend and always have FWHM~$\gtrsim 50$~km/s (Figure \ref{fig: kinem_corr}). Line centroids, in reference to the stellar RVs, are predominantly blue-shifted by a few up to $-10$~km/s, especially in triangular lines.

\paragraph{Trends in line excitation}
CO excitation is studied from three line flux ratios: the high-$J$ to low-$J$ ratio of $v=1-0$ lines (a proxy for rotational excitation), the $v=2-1$ to $v=1-0$ ratio (a proxy for vibrational excitation), and the \ce{^{12}CO}/\ce{^{13}CO} ratio (a proxy for column density). All three ratios increase with line FWHM, encompassing all line types and components (Figure \ref{fig: excit_corr}), while they do not show strong global trends with stellar or accretion luminosity. Additionally, the vibrational ratio shows a group of narrow double-peak lines with high vibrational ratios associated to disks with an inner dust cavity.

\section{Discussion} \label{sec: discussion}
By combining this new iSHELL survey, which has so far mostly focused on disks around Herbig Ae/Be stars, to a previous survey done with CRIRES, which focused instead on T~Tauri stars (Figure \ref{fig: surveys_comparison}), in Section \ref{sec: results} we have provided an overview of the current status of high-resolution $M$-band spectroscopy of protoplanetary disks in terms of the main global trends that have so far been found in the data. While the CO spectra show a wide and complex diversity in line profiles and excitation conditions, we have attempted to provide a high-level overview by identifying a minimum number of line types that can be described in terms of different velocity structures for the emitting gas. By building on and extending the analysis from previous work, two fundamental types of line profiles have been identified from their shape as ``triangular" or double-peaked, which can be quantified in terms of a line shape parameter that measures the ratio of the line widths at the base and peak of the line (Section \ref{sec: line_shapes} and Figure \ref{fig: shape_param}). These two line types have measured properties that largely overlap under several parameters (disk inclination, infrared index, CO FWHM, CO excitation), suggesting that the origin of their different velocity structure should not be primarily related to a different geometric orientation, the presence/absence of a dust cavity, or different gas excitation conditions. The only property that shows some segregation between the two line types is their stellar properties, where triangular lines dominate the sample at $T_{\rm{eff}} \lesssim 7000$~K while double-peak lines dominate the sample at $T_{\rm{eff}} \gtrsim 7000$~K (Figure \ref{fig: shape_param}). 

\begin{figure*}
\centering
\includegraphics[width=1\textwidth]{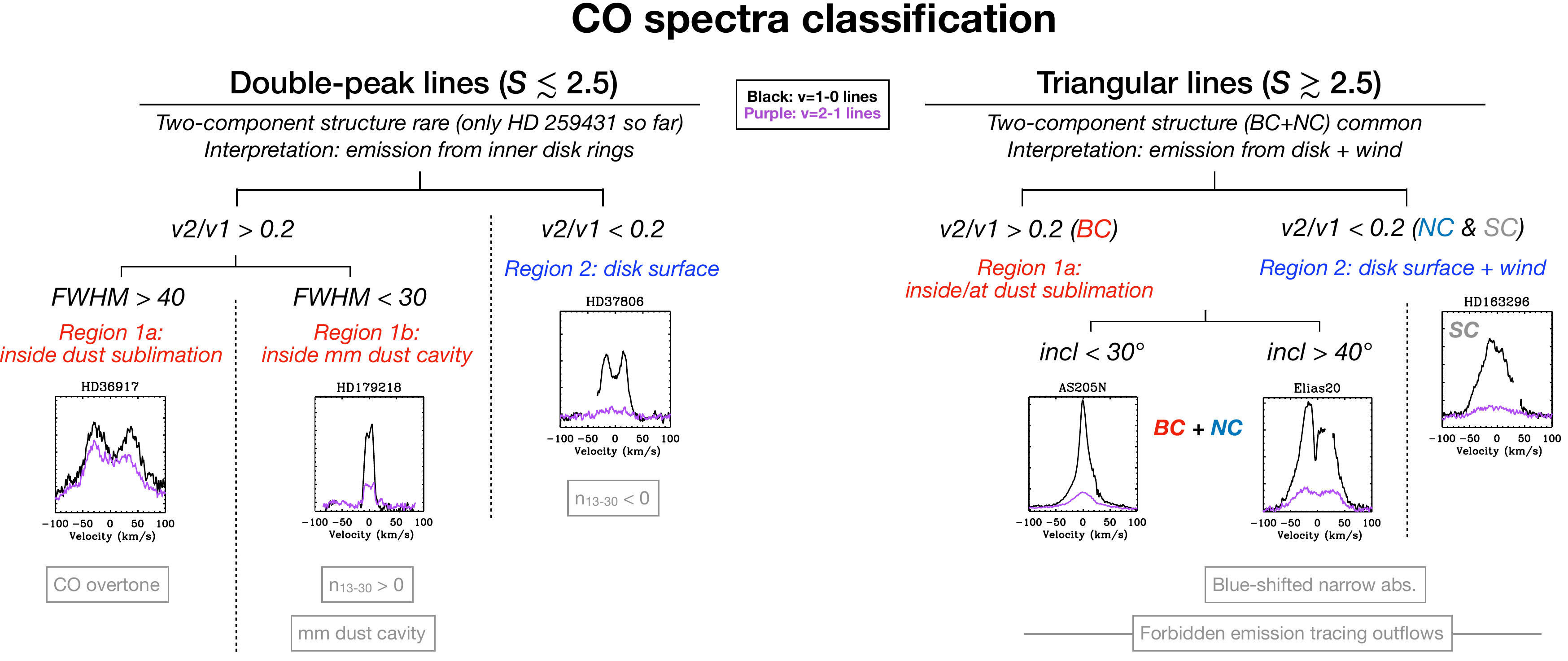} 
\caption{Classification of CO lines based on the observed properties and interpretation discussed in this work (see also Figure \ref{fig: discussion}). For the decomposition of triangular lines into BC and NC, see examples in Figure \ref{fig: line_stacks}. Additional distinctive tracers discussed in the text are reported in grey boxes below each line type.
}
\label{fig: classification}
\end{figure*}

In this section, we combine the measured properties described in Section \ref{sec: results}, including both kinematics and excitation of the observed CO gas, and discuss them jointly to describe a unified picture of the regions and conditions that different line types and velocity components may trace in disks. We focus this discussion on the two main scenarios that have been considered in the literature so far to explain the observed properties of CO $M$-band emission: gas in Keplerian rotation in the disk and the low-velocity part of an inner disk wind \citep[e.g.][]{najita03,pont11,woitke16}. 
In Section \ref{sec: conclusions}, we will include an outlook to how these scenarios can be extended and further investigated in the next decade, to observe key signatures that can distinguish them.
The classification and interpretation of CO line spectra are visually summarized in Figures \ref{fig: discussion} and \ref{fig: classification}. At the cost of some repetition, we have preferred to include both figures to better highlight multiple aspects. The first figure reports the excitation tracers and trends from Figure \ref{fig: excit_corr} in terms of their $R_{\rm{CO}}$ in reference to $R_{\rm{subl}}$ in each disk (as defined above in Section \ref{sec: emit_radii}), in reference to a disk cartoon to visualize the relative location of CO gas and dust emission in different inner disk regions. The second figure reports a more schematic classification of CO lines based on the observed properties and the interpretation discussed in this work, to aid the classification of future spectra.

\subsection{Double-peak lines: disk rings of CO gas with $v2/v1$ reflecting the radial disk region and gas-to-dust ratio} \label{sec: disc_dp_lines}
A double-peak line shape is the easiest case to explain kinematically, as it naturally results from a relatively narrow ring of gas in Keplerian rotation around the star (see e.g. example above in Figure \ref{fig: kepl_rings}). This iSHELL survey is the first that observes a high frequency ($\sim50\%$) of double-peak profiles in $M$-band CO emission, and as discussed above this seems to be due to the sample focus on Herbig Ae/Be systems (Section \ref{sec: results}). The 16 disks in this survey with double-peak profiles show for the first time that $M$-band CO lines of this type can have FWHM as large as $\approx 200$~km/s and as narrow as 9~km/s (Figures \ref{fig: line_gallery} and \ref{fig: kinem_corr}), spanning the entire range of CO line widths measured in disks in general (and of BC and NC components combined, from the triangular lines). This type of line shape is therefore not specific to a single disk location but can trace different radii in different disks, and it can be observed at any disk inclination.

Three groups can be identified from their $R_{\rm{CO}}$ in reference to $R_{\rm{subl}}$ and their CO vibrational excitation in Figure \ref{fig: discussion}, groups that include both triangular and double-peak lines but that we will discuss first here focusing on the double-peak lines. Two of these groups show the same $v2/v1$ excitation but very different radial location, and we call these ``region 1a/b". The broadest lines emit from $R_{\rm{CO}} \lesssim $~$R_{\rm{subl}}$ (region 1a in the figure) and have the highest vibrational and rotational excitation observed in CO spectra. The narrowest lines emit from well beyond the sublimation radius (region 1b, with $R_{\rm{CO}}$/$R_{\rm{subl}} \approx$~10-100) and have high vibrational but low rotational excitation. While disks in region 1a all have low $n_{13-30} < 0$ (and as low as $-2$), suggesting the absence of large inner disk cavities and a very flat disk surface or small disk size (to keep the flux at 30~$\mu$m low), disks in region 1b distinguish themselves by their large $n_{13-30} > 0$, clearly indicative of large inner dust cavities \citep{brown07,furlan09,banz20}.
Intermediate lines emit from a region right beyond the sublimation radius ($R_{\rm{CO}}$/$R_{\rm{subl}} \approx$~2--10) and still show a high to intermediate rotational excitation, but a large drop in vibrational excitation ($v2/v1 < 0.2$ and as low as 0.01, region 2). 

\citet{bosman19} performed a dedicated thermo-chemical modeling of line kinematics and excitation for CO gas in disks around intermediate-mass stars, and explained the different vibrational excitation $v2/v1$ of regions 2 and 1b as a consequence of dust depletion: region 2 has typical conditions for an inner disk surface with gas-to-dust ratios of $\approx100$ and $T_{\rm{gas}}\approx$~$T_{\rm{dust}}$. Region 1b, instead, has gas-to-dust ratios $> 10,000$ and large columns of gas ($> 10^{18}$~cm$^{-2}$) increasing the $v2/v1$ ratio. The modeling showed that when the gas temperature drops below 1000~K, the $v=2$ level quickly becomes depopulated in both LTE and non-LTE excitation, needing a large column of CO to re-populate at the observed level ($v2/v1 > 0.2$).
The detection of lines from levels up to at least $v=6$ in some of these disks (MWC~297, V892~Tau, HD~141569, HD~179218 in the iSHELL sample, plus HD~100546 and HD~97048 observed before by \citet{vdplas15}) additionally suggests UV fluorescence to populate $v > 3$ in some disks with dust-depleted inner cavities \citep{brittain07,vdplas15}. The general paucity of line detections from $v > 3$ even in disks around Herbig Ae/Be stars suggests that UV pumping is uncommon and is not simply a function of the stellar/accretion spectrum, but requires specific conditions in terms of inner disk structure and gas-to-dust ratio.

The analysis in \citet{bosman19} did not include any spectra from region 1a, but still predicted what should have been observed: by emitting from a dust-free and hot region within dust sublimation, any CO emission should be highly vibrationally excited in the $v=2$ level and possibly trace large columns of gas. In this work, that prediction is confirmed by spectrally resolving for the first time a highly-vibrationally excited fundamental CO spectrum in five of these disks (51~Oph, HD~35929, HD~36917, HD~58647, and HD~259431; CO emission from this inner dust-free region will be further discussed below in Section \ref{sec: dust-free}). As shown above in Section \ref{sec: emit_radii}, the case of HD~259431 is particularly remarkable because its two Keplerian components correspond in excitation to region 1a (ring \#1, broader CO and high vibrational excitation) and region 2 (ring \#2, narrower CO and low vibrational excitation), providing one example where both are detected in the same disk demonstrating how $v2/v1$ reflects the relative location of $R_{\rm{CO}}$ in reference to $R_{\rm{subl}}$.

To summarize, there is substantial evidence that double-peak CO lines can be well explained with disk rings of CO gas in Keplerian rotation around the star, with their vibrational excitation in the first two levels ($v2/v1$) that closely reflects the gas-to-dust ratio in three distinct disk regions \citep{bosman19}. High $v2/v1$ ratios can be produced by large CO columns found in strongly dust-depleted regions, either inside dust sublimation (region 1a) or inside an inner dust cavity (region 1b); low $v2/v1 < 0.2$, instead, trace a dust-rich inner disk surface (region 2).

\subsection{Triangular lines: CO gas in a low-velocity region of inner disk winds} \label{sec: disc_tr_lines}
Despite their strikingly different line velocity structure, especially for shape parameters as high as 6--10 as observed in e.g. AS~205~N, triangular lines could still be produced simply by gas in Keplerian rotation in a disk. The general condition for that to happen is that the disk surface is warm out to much larger radii than the narrower radial emitting region that produces a double-peak line, so that the line center is filled in by lower-velocity gas emission from larger disk radii. Detailed thermo-chemical models of inner disk emission, in fact, do produce triangular lines with two distinct velocity components emitting from the inner rim and part of a flared disk atmosphere \citep[e.g.][]{woitke16,heinbert16,bosman19}, components that could naturally explain the broad+narrow (BC+NC) velocity structure commonly observed in disks around T~Tauri stars. 

Challenges to this scenario have come from the gas excitation, the extent of the emitting radial region, and the spectro-astrometric signal. Model explorations by \citet{woitke16} have found that the narrow component requires specific geometric conditions (large disk flaring and/or scale height) and should be rotationally cold, populating transitions only up to $J < 15$. This is in conflict with NC that is strong up to $J > 31$ in most disks where it is present (Figure \ref{fig: excit_corr}). 
In terms of emitting region, assuming a single power-law temperature profile for the inner disk gas, previous work found that the NC emission should be radially extended out to 10--30~au, which has been excluded at least in some disks with CRIRES data before \citep{bast11,brown13}.
Moreover, in a few disks where high enough S/N was achieved (primarily AS~205~N and RU~Lup), the compact and highly asymmetric shape of the spectro-astrometric signal measured in their NC components showed that the gas velocity needs a sub-Keplerian outflowing component to fill in the line center \citep{pont11}.
Previous work therefore proposed that lines with narrow centers and broad wings are inconsistent with only Keplerian rotation, and suggested the possibility that they include a slow disk wind \citep{pont11,bast11,brown13}. Based on the scarcity of double-peak profiles, and the similarity in line shapes observed across the CRIRES survey, \citet{brown13} further proposed that a wind origin might be common for $M$-band CO emission as observed in T~Tauri disks in general. We expand on this discussion in this work by combining, in addition to the line shapes, the unprecedented view on CO excitation provided by the large spectral coverage of iSHELL spectra. 
 
There are multiple lines of evidence suggesting that all triangular lines, and not only those with single narrow peaks and the largest values of $S$, may trace the same gaseous structure as seen under different viewing angles. 
It is known that line widths overall increase with inclination angle (Figure \ref{fig: kinem_corr}), a projection effect that is expected when gas is in rotation in a disk. Moreover, this work now finds that the shape parameter $S$ changes with inclination angle too (Figure \ref{fig: shape_param}) suggesting that geometric effects are affecting line shapes, with line centers that become broader relative to their wings at higher disk inclinations. As shown in Section \ref{sec: line_shapes}, this broad trend is expected both from gas in Keplerian rotation and from gas in a slow disk wind that turns into absorption at the line center at high disk inclinations, with the difference that gas in Keplerian rotation should only affect the line shape at the lowest inclinations and only produce $S \lesssim 2.5$, according to models explored in previous work (Section \ref{sec: line_shapes}). 
Another aspect now also emerges from considering the excitation of CO lines. Triangular lines share the common presence of two emission components with distinct width and excitation (Section \ref{sec: results}), something that is instead extremely rare in double-peak lines (the only case so far is HD~259431, found in this work).
The rotation diagrams of triangular lines show nearly identical shapes in the $v=1$ and $v=2$ lines (Figure \ref{fig: rotdiagr_gallery}) in both NC and BC regardless of their FWHM, from the narrower lines seen at low inclinations (e.g. AS~295~N) to the broader lines seen at higher inclinations (e.g. RNO~90). 
Rather than requiring a wind origin only for spectra with the largest $S$ values as seen at low disk inclinations, these common properties could instead be indicative of a scenario where all triangular lines trace a same gas structure as viewed under different angles in different disks. 

While not conclusive yet, nor necessarily the unique solution in all cases, the scenario that a triangular line shape may in general include a slow wind component, rather than emission from an extended disk atmosphere, finds support in multiple lines of evidence in addition to the ones mentioned above, of which probably the strongest is the asymmetric spectro-astrometric signal measured in a few disks \citep{pont11}.
Triangular lines show a large fraction of line asymmetries, and mostly on the blue side of the line \citep[see Figure 6 in][]{brown13}, and they frequently show small blueshifts (Figure \ref{fig: kinem_corr} and Section \ref{sec: res_kinem}), all indicative of outflowing gas.
They also show complex gas kinematic variability (Figure \ref{fig: line_variability}), which is yet to be understood but could be due to variability in the mass loss into the wind \citep[see e.g.][]{ellerbroek14,heinbert16_var}, especially for those objects that show the strong and variable blue asymmetry (Elias~20, GQ~Lup, RNO~90).

In a scenario where a triangular line shape in general indicates CO emission from the disk surface and the low-velocity part of a wind, the gas would still maintain most of the Keplerian motion from the disk launching region, which would explain the increase of FWHM with disk inclination, as suggested in previous work \citep{brown13}. Disk winds are expected to have complex structures where poloidal and toroidal gas velocities change across disk radii as well as along the streamlines \citep[e.g.][]{blandford82,kurosawa06}. The complex radial and vertical structure of winds is expected to produce line profiles in forbidden emission spectra often with prominent single peaks and blue asymmetries, where the wind launching region is not easily traced from the observed line velocities \citep[e.g.][]{weber20}. CO emission lines from these winds have not been modeled yet for direct comparison to observations, but recent disk wind models that include thermo-chemistry do predict that CO should be present at the low-velocity ($< 10$~km/s) base of winds \citep{wang19}, supporting the fact that any blueshifts observed in CO lines should at most be small and therefore harder to detect. 

In a disk wind scenario, it may not be correct to apply a simple Keplerian interpretation to the CO line widths in triangular lines. It is also not yet clear whether both BC and NC would trace the wind, and the same wind, but the spectro-astrometric signal supports a wind origin in a few disks at least for the central part of the line, corresponding to NC \citep{pont11}. A Keplerian interpretation is thus most probably more correct for BC than for NC, which should be the component that includes sub-Keplerian wind emission. We therefore include in Figure \ref{fig: discussion} an estimate of $R_{\rm{CO}}$ as calculated for the double-peak lines in the previous section, with the caveat that it may not be fully correct for NC in a wind scenario. From these $R_{\rm{CO}}$ estimates, the emitting radii of BC are found to mostly trace the dust-free region inside dust sublimation and at the dust inner rim (region 1a), with a few exceptions, while the NC may trace just above a dust-rich disk surface beyond dust sublimation (region 2, in a few cases possibly extending out to $> 10$~au). 

In support of the interpretation of different emitting radii/regions comes again the CO excitation: BCs are typically highly vibrationally excited, in a similar way to Keplerian lines that emit from region 1a, while NCs show a low vibrational excitation similarly to Keplerian lines from region 2.
In potential conflict to this interpretation is the rotational temperature found from LTE fits to the iSHELL spectra in Figure \ref{fig: rotdiagr_gallery}, where BC is found to be rotationally colder than NC. This could mean that even the rotational excitation is not in LTE, or perhaps that by being excited in an inner dust-free region, BC is colder due to the absence of dust \citep[using Equation 15 for  from][and adopting $T_{\rm{eff}}$ = 4000~K, a stellar radius of 2 solar radii, the gas temperature at 0.1~au in the dust-free inner region is $\approx500$~K, i.e. lower than the 1500~K at the dust sublimation radius]{dullemond10}. These two different excitation scenarios should be investigated in detail in future work.

\begin{figure}
\centering
\includegraphics[width=0.45\textwidth]{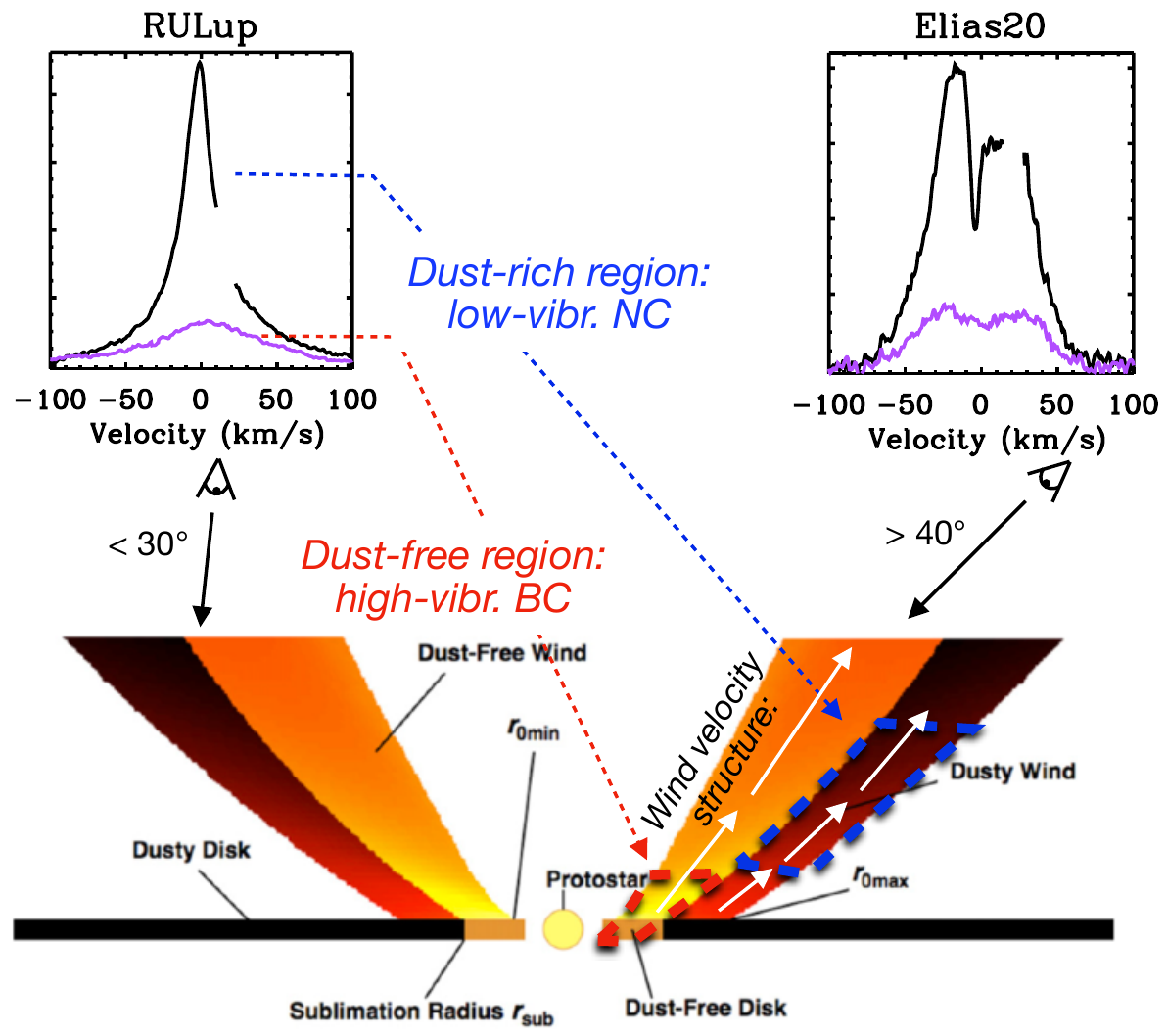}
\caption{Potential interpretation of BC and NC in triangular lines within the context of a multi-shell wind launched near the dust sublimation radius. The cartoon of the wind physical structure is adopted from \citet{banskonigl12}, and the velocity structure from what typically expected in magneto-centrifugal winds, where the poloidal velocity increases along each streamline and decreases at larger launching radii \citep[][]{blandford82}. The figure is just for illustration of a different velocity and CO excitation in different regions of the wind, and it is not to scale.}
\label{fig: windlayers_discussion}
\end{figure}

If the observed lines are still mostly broadened by Keplerian rotation and the $R_{\rm{CO}}$ estimates are approximately correct, the emitting radii of both BC and NC (well inside of 3~au in most disks) would point, to our current knowledge, to an MHD wind because photoevaporative winds can only be launched from beyond the gravitational radius for thermal escape \citep[see e.g. the similar discussion of forbidden line tracers in][]{simon16,banz19}, at least in its inner region traced by $M$-band CO emission. Recent wind models are exploring the combination of MHD and photoevaporative winds, showing that they can be launched in a radially stratified structure in the same disk \citep[e.g.][]{bai16,wang19}.
A layered wind structure can also include a gradient in gas-to-dust ratio that may affect the excitation of CO in a similar way to what discussed above for the double-peak lines. In fact, magneto-centrifugal disk winds launched at the sublimation radius have long been proposed to lift dust within them to explain the unusually high NIR excess observed in some Herbig Ae systems, including AB~Aur and HD~163296: such winds are proposed to have dust-free and dust-rich layers in an onion-shell type of structure \citep{banskonigl12}. A recent example is the disk of SU~Aur (which is observed in the iSHELL survey with a triangular line shape, see Figure \ref{fig: line_gallery}), where a dusty wind launched at the disk inner dust rim, very close to where CO emission is observed, has been proposed to explain both the observed NIR excess in the SED and the NIR visibilities \citep{labdon19}. 

Future work should explore whether an onion-like stratified structure with a gradient in gas-to-dust ratios could give rise to the two components observed in triangular lines, their different velocity structure, and their different vibrational excitation: BC tracing an inner dust-free layer closer to the disk surface, i.e. vibrationally more excited and dominated by Keplerian broadening, and NC an outer dust-rich layer just above the disk surface, i.e. vibrationally less excited and moving at sub-Keplerian speed. For the sake of visualization of this tentative scenario to be tested in the future, we illustrate it in Figure \ref{fig: windlayers_discussion}, where we include the CRIRES spectrum of RU~Lup and the iSHELL spectrum of Elias~20 for reference of triangular CO lines at different inclinations. Recent spectro-astrometry of forbidden emission in RU~Lup found a similar stratification where the wind velocity decreases with distance from the star, suggesting that the emission traces gas from different streamlines possibly at different scale heights \citep[Figure 10 in][]{whelan21}, a structure that is expected in magneto-centrifugal disk winds and that could apply similarly to the BC and NC observed in $M$-band CO spectra (Figure \ref{fig: windlayers_discussion}).

\subsection{CO absorption in highly inclined disks: intercepting gas columns in a disk wind}
CO absorption spectra detected in the CRIRES survey have been interpreted by \citet{brown13} as foreground gas absorption when unresolved and showing low rotational excitation (FWHM $<$~5~km/s and only $J<8$ detected), and as outer disk absorption when partially or fully resolved and showing higher rotational excitation \citep[FWHM $=$~5--10~km/s and higher $J$-levels detected, see also][]{rettig06}. \citet{brown13} also found a trend where the rotational excitation of absorption components increases with FWHM, qualitatively consistent with the low-excitation tail of the trend shown above in Figure \ref{fig: excit_corr}.

\begin{figure}
\centering
\includegraphics[width=0.45\textwidth]{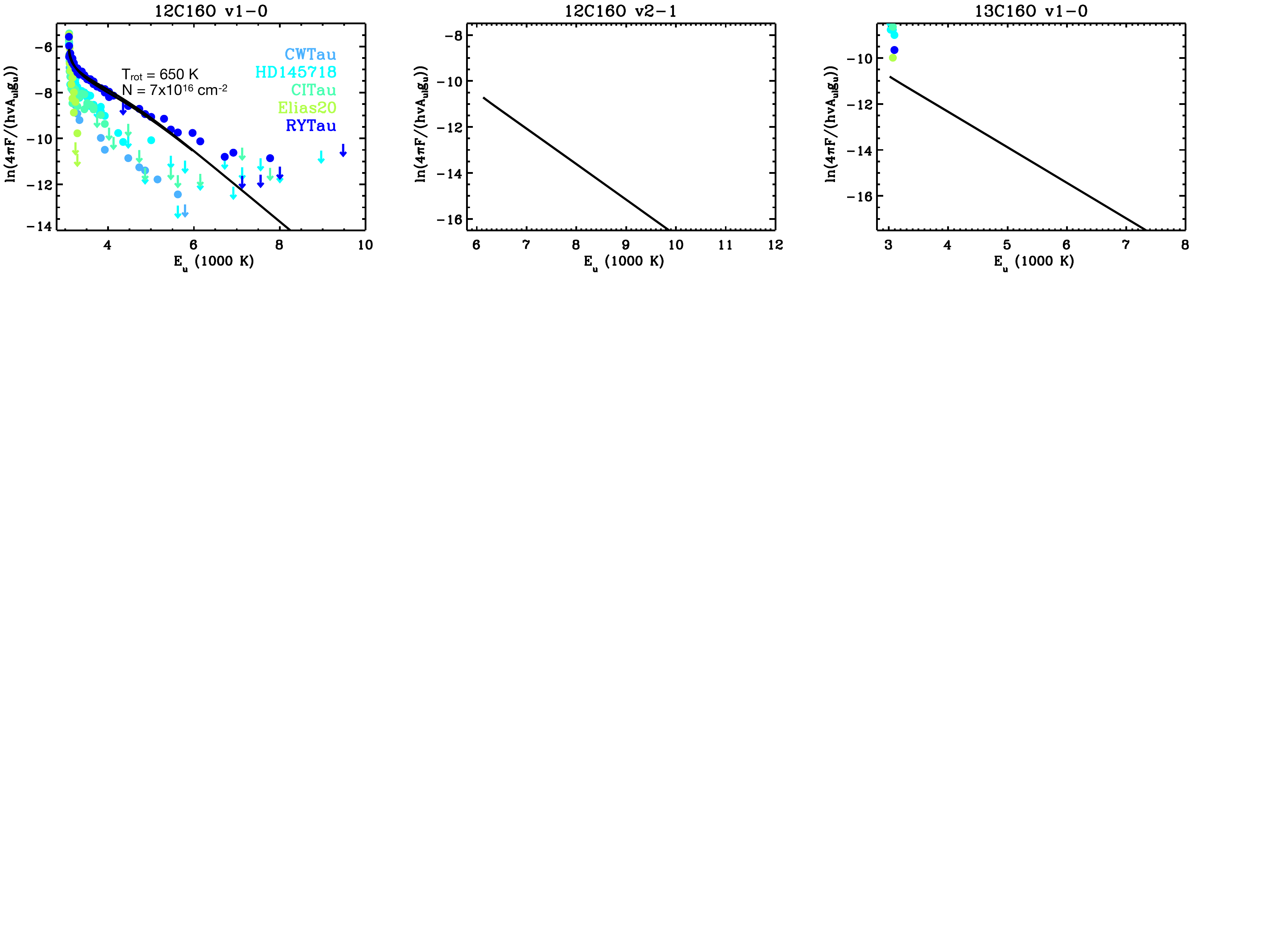} 
\caption{Rotation diagrams of absorption components observed in highly-inclined disks (incl = 50--65~deg), including an LTE model fitted to the highest rotationally-excited absorption spectrum observed in RY~Tau, which almost approaches the excitation of the emission NC components (Figure \ref{fig: rotdiagr_gallery}). Elias~20 is included as an example of the lower rotational excitation in spectra where absorption is detected only up to $J < 8$, showing a much steeper curve.}
\label{fig: RD_absorpt}
\end{figure}

In this work, we combine multiple observed properties to improve our understanding of the nature of these narrow absorption components. 
First, the three line excitation tracers in absorption components follow global trends shown by the emission components (Figure \ref{fig: excit_corr}). Absorption components are rotationally and vibrationally cold and have the lowest \ce{^{12}CO}/\ce{^{13}CO} ratios, with values that overlap with those measured in NC in all three tracers. This suggests that the gas traced by NC and absorption components shares at least in part similar excitation conditions. We do not include absorption components in Figure \ref{fig: discussion} because these are the least likely to be broadened by Keplerian rotation, especially in a disk wind scenario, but from the measured FWHM the nominal corresponding disk radii would overlap with the outer regions of NC emission. Moreover, the rotational excitation of the absorption components in the most inclined disks shows a curvature and properties that approach what observed in the emission NC components (Figures \ref{fig: RD_absorpt} and \ref{fig: rotdiagr_gallery}).

Second, the prevalence of blue-shifts in absorption lines (Figure \ref{fig: kinem_corr} and \ref{fig: line_gallery_abs}) suggests that these may be tracing parts of a slow disk wind. A viewing geometry that would be favorable for this scenario is high disk inclinations, where the line of sight is grazing the disk surface and the absorption is tracing a large column of CO gas that extends across disk radii. This scenario was tested using the ray-tracer RADLite in \citet{pont11}, finding that the same slow disk wind that produces the NC emission components would turn into a central narrow ($\approx 5$~km/s) absorption component when observed at inclinations higher than 40~deg \citep[Figure 12 in][]{pont11}. Section \ref{sec: line_shapes} shows how this scenario could at least in part reproduce the values and trend observed in triangular line in general.
It is important to note that winds that turn into a low-velocity, narrow absorption on top of wind emission have already been explored before in HI tracers of MHD accretion-driven disk winds \citep[e.g.][]{kurosawa06}.

In at least one source observed at high inclination, CW~Tau, the spectro-astrometric signal is highly asymmetric and matches the velocity width of the narrow central absorption component \citep[Figure 9 in][]{pont11}, further supporting a wind origin. CW~Tau has long been known for its jet and wind observed and spatially-resolved in multiple gas tracers \citep[e.g.][]{hirth94_cwtau,dougados00}, it has a triangular CO line shape similar to that observed in Elias~20, GQ~Lup, and RNO~90, and its absorption component is similarly highly rotationally excited (at least up to $J=32$, the coverage of the CRIRES spectrum) as the absorption spectrum observed in RY~Tau (Figure \ref{fig: RD_absorpt}), which also shares its high disk inclination of $65$~deg. RY~Tau too is known for its large scale jet suggesting a magneto-centrifugal inner disk wind launched within 1~au \citep{coffey15,garufi19}, which is the region where $M$-band CO is observed and could explain in a single scenario both emission and absorption components in this object and possibly apply to inner disks more in general.

It should be noted that these low-velocity absorption components are distinct from the faster outflows (blueshifts of 30--100~km/s) observed in $M$-band CO spectra in more embedded YSOs \citep{thi10,herczeg11} and in a few highly-accreting objects \citep{brittain07b,goto11,brown13,banz15}. These faster outflows show broader lines with FWHM~$>30$~km/s, are found to be variable and sometimes episodic in connection to accretion outbursts, sometimes show multiple clumps at different velocities, and have a high rotational excitation suggesting temperatures in excess of $\approx 1000$~K. Considered together, these properties suggest that these absorption components with larger blue-shifts might trace shocked gas in clumps along the outflow \citep{thi10,brown13}.

\subsection{Why a $T_{\rm{eff}}$ dependence for line shapes?}
Previous work mostly identified the difference in CO spectra as observed from disks of T~Tauri and Herbig stars in terms of excitation, in relation to the different strength of UV radiation \citep[e.g.][]{brown13,vdplas15}, or in terms of the observed FWHM, in relation to a gas emitting area that moves outward with higher luminosity or with the formation of an inner dust cavity \citep{salyk11}. However, Section \ref{sec: results} shows that triangular and double-peak lines overlap very well in $v2/v1$, once BC and NC are distinguished, as well as in the observed range of FWHM that translates into radial emitting regions across 0.1--10~au, still with generally smaller FWHM in disks around more massive stars (up to 3~$M_{\odot}$) and disk with inner dust cavities (Figure \ref{fig: kinem_corr}). We also discussed that the vibrational excitation in the first two levels primarily reflects, rather than the irradiation spectrum (Appendix \ref{app: co_excit}), the level of dust depletion in three different radial regions \citep[Sections \ref{sec: emit_radii} and \ref{sec: disc_dp_lines}, and][]{bosman19}. This work for the first time focuses on a property that shows a primary dependence on stellar properties and that could be related to a fundamental difference in the velocity structure of the emitting gas, as discussed above in this section: the CO line shape as double-peaked or triangular (Figure \ref{fig: shape_param}).

The properties of the line shape parameter described in Section \ref{sec: line_shapes} suggest that the nature of line shapes as double-peaked rather than triangular, while both types are affected by geometric factors, is not mainly determined by a specific viewing angle nor the presence of an inner cavity in the disk, but seems to be mostly determined by their central stars. This suggests that something in (or some combination of) the stellar structure, luminosity, irradiation spectrum, and magnetic field in stars with different mass (and age) affect the inner disk and any gaseous structure that is traced in CO emission.

In a disk wind interpretation for the triangular lines, this line shape difference could be naturally explained if T~Tauri disks typically host inner disk winds, while Herbig disks typically do not. This general scenario has in fact been largely supported in previous work from multiple tracers. T~Tauri disks typically have evidence for outflows and/or inner winds with a range of blueshift velocities in forbidden emission \citep[e.g.][]{hartigan95,banz19}. Instead, Herbigs disks are well known to very rarely show high-velocity outflows \citep[e.g.][]{corcoran97}, and even forbidden emission gives evidence for being dominated by Keplerian motion in the disk rather than a slow disk wind as in T~Tauri disks \citep{acke05}.

From a theoretical point of view, disks around both T~Tauri and Herbig stars are expected to launch inner disk winds, but with some differences due to the different stellar structure and its evolution. \citet{kunitomo20,kunitomo21} propose that due to the change in stellar structure from convective to radiative early in the evolution of a 3~$M_{\odot}$ star, the decrease in magnetic field would rapidly cease any MHD winds by a few Myr, leaving disk evolution to photoevaporative winds at larger disk radii. Instead, a solar-mass star would have high X-ray and magnetic field up to 10~Myr or longer, supporting MHD and photoevaporative winds throuhgout the disk lifetime. Stellar magnetic field measurements in pre-main-sequence stars do support a strong stellar mass dependence in the strength and evolution of magnetic fields, with weaker fields below $\approx200$~G in Herbig Ae/Be stars compared to kG in T~Tauri stars \citep[e.g.][]{alecian13,villebrun19,jarvinen19}. In a scenario where CO traces an inner disk wind that is MHD in nature, the internal stellar structure and duration of magnetic fields could be related to the $T_{\rm{eff}}$ dependence of line shapes, where solar-mass, convective stars typically have triangular lines tracing an inner MHD wind supported by their stronger magnetic field, while (older) Herbig Ae/Be stars with their radiative structure do not support inner MHD winds and typically have Keplerian lines tracing the disk only.
A possibility to consider is that inner winds may be present in Herbigs but simply not visible in CO due to dissociation of the molecule; however, this scenario would still not explain the rare evidence for outflows from forbidden emission tracers (see above in this section).
It should also be noted that the origin of a magnetic field that supports inner MHD disk winds is still unclear and could be related to the ambient molecular cloud rather than convection inside the star or a different combination of the two in different systems \citep[e.g.][]{ferreira97,ferreira06}, and MHD winds can be radially stratified with the stellar magnetic field dominating the launching region inside $\approx0.5$~au \citep[e.g.][]{zanni13}, the region where most CO BC lines are emitting from (Figure \ref{fig: radii_compar}).

If a triangular CO line shape does in fact reflect the presence of an inner disk wind, a phenomenon that seems to be typically absent in disks around stars of $T_{\rm{eff}} > 7000$~K, why do some Herbig Ae stars do have triangular lines? In this sample, these stars are only four: AB~Aur, HD~163296, HD~190073, MWC~480. AB~Aur has previous suggestions for an MHD wind from H$\alpha$ and modeling of the high NIR excess \citep{banskonigl12,perraut16}. HD~190073 is the youngest known magnetic HAeBe star \citep{alecian13}, which could perhaps sustain an inner MHD wind, but it is also the farthest away source in this sample ($\approx900$~pc) so that its disk is not well known.
As for the disks in MWC~480 and HD~163296, which are instead very well studied, it is very remarkable to note that in addition to being known for their flat geometry from the SED and spatially-resolved gas observations \citep{law21_maps,zhang21_maps}, therefore not supporting the scenario to have an unusually extended emitting disk surface area, they are among the very few Herbig Ae/Be stars known to launch a large-scale jet \citep[e.g.][]{grady10,ellerbroek14}. 

\begin{figure}
\centering
\includegraphics[width=0.45\textwidth]{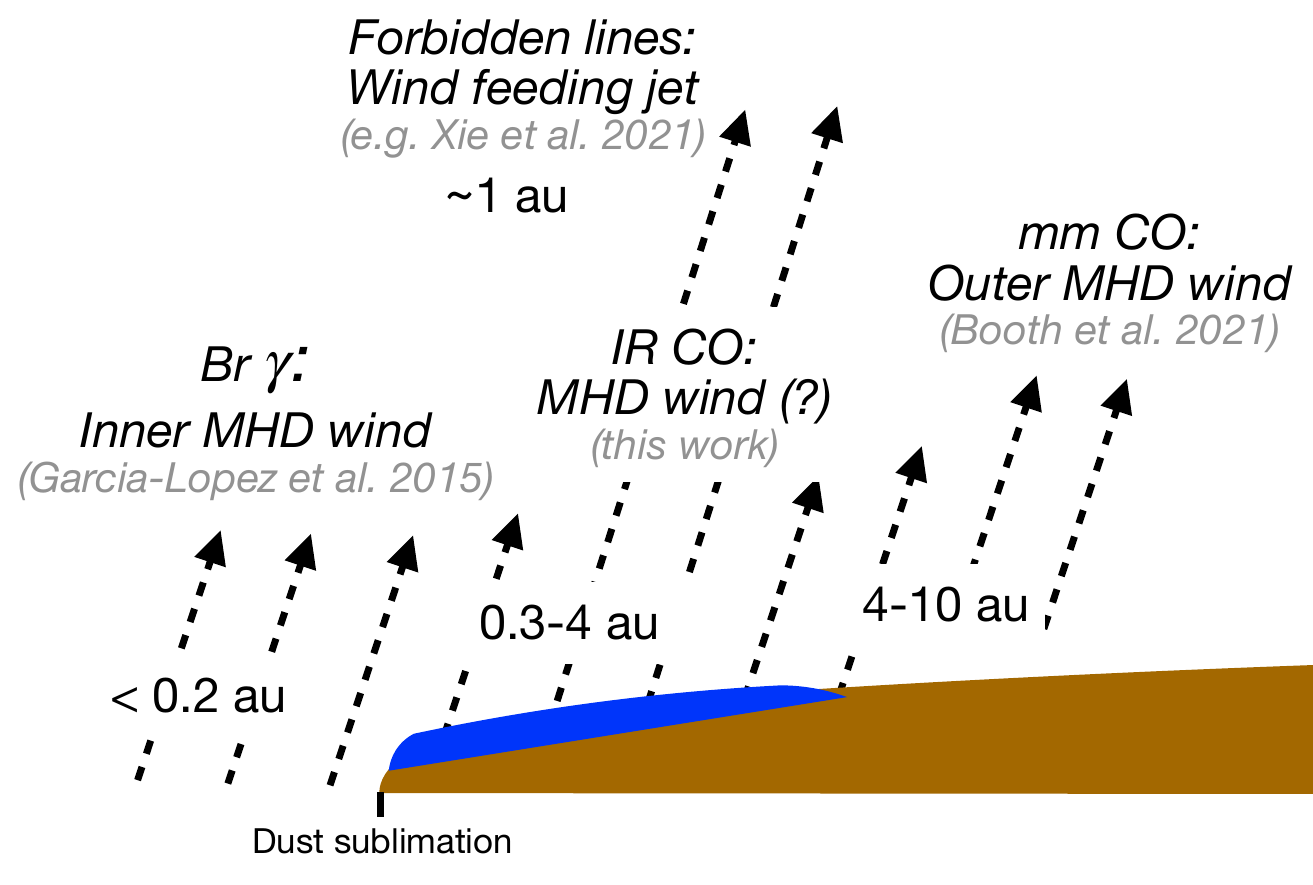} 
\caption{Multiple wind components proposed to explain emission from different gas tracers as observed in HD~163296.}
\label{fig: HD163296}
\end{figure}

HD~163296, in particular, has been extensively studied at multiple wavelengths and shows multiple evidence for an outflow and wind from imaging and spectroscopy \citep[e.g.][]{klaassen13}. A recent ALMA study reveals in great detail the kinematic structure of a large-scale molecular disk wind traced in optically thick CO $J=2-1$ line, a wind that is most likely MHD-driven and is launched around 4~au \citep{booth21}, overlapping to where $M$-band CO emission is observed, whose kinematic structure and variability has already been proposed to trace a disk wind \citep{heinbert16_var}.
With all these tracers, the disk of HD~163296 gives probably one of the best examples to date of a radially layered outflow structure observed from inside dust sublimation \citep[an inner wind traced with Br$\gamma$ emission from 0.04--0.16~au,][]{garcialopez15} out to launching regions around 1~au \citep[for the bipolar jet HH 409,][]{ellerbroek14,xie21_hd163296} and at 4--10~au \citep[for the extended molecular outflow observed with ALMA,][]{booth21}, where the change in magnetic lever arm as a function of radius could be due to a change in angular momentum extraction efficiency \citep[e.g.][]{ferreira97,bethune17}. By tracing CO gas at 0.3--3.6~au (from the FW10\% out to the FW75\% of the CO line), $M$-band CO emission stands right in between a wind traced in Br$\gamma$ and the wind traced with ALMA around 4--10~au, supporting a picture where different tracers probe different parts of a radially-extended outflow process (Figure \ref{fig: HD163296}).

\begin{figure*}
\centering
\includegraphics[width=1\textwidth]{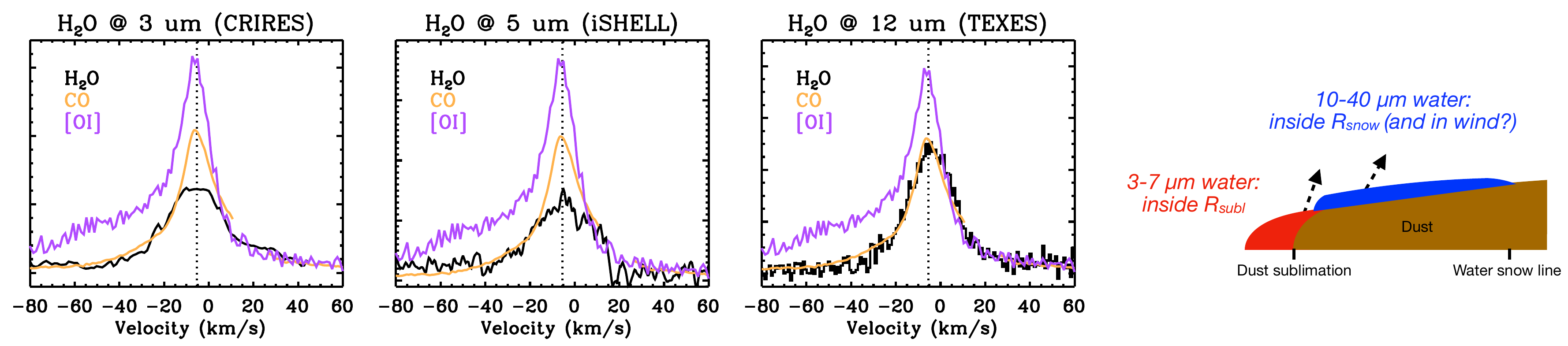} 
\caption{Water line profiles observed in AS~205~N, as compared to the CO line profile (this work) and to the [OI] line profile \citep{banz19}. The three emission lines are scaled to highlight where line kinematics are similar between water and CO, whether over the wings only or at the line center too. The first plot shows water emission as observed with CRIRES near $3 \mu$m \citep{banz17}. The second plot shows a stack of 6 clean water lines observed in the iSHELL spectrum (Figure \ref{fig: spec_overview}). The third shows water emission near $12 \mu$m as observed with TEXES \citep{najita18}.}
\label{fig: AS205_water}
\end{figure*}

\subsection{On molecular gas emission from the inner dust-free disk region} \label{sec: dust-free}
Just a decade ago, \citet{dullemond10} called it a ``mystery why disks do not show stronger emission from molecules in the dust-free inner disk", given that molecules are regularly observed in mid-infrared spectra above the dusty disk surface \citep[e.g.][]{cn11,salyk11_spitz} and that theoretically the inner disk gas could still be molecule-rich if large gas columns are present \citep{bruderer13}. Observations of the CO first overtone emission, tracing hot CO gas within the dust inner rim, have in fact historically yield low detections rates of only 7--20\% \citep{carr89,ilee14}. However, CO overtone emission requires both a higher temperature and a higher column than fundamental emission to be detectable, therefore the low detection rates may not necessarily imply that the inner dust-free region is molecule-free. \citet{ilee18} in fact suggested that, in disks around Herbig Ae/Be stars, the low detection rate of CO overtone can be explained not by the absence of CO but by a line-to-continuum contrast that favors CO detection only in a relatively narrow range of high accretion rates around $10^{-5} M_{\odot}$/yr.

Our view of the innermost disk region is now being greatly improved by new NIR interferometry observations, especially with GRAVITY, that are spatially resolving the inner dust rim in several disks, and narrow rings of CO overtone emission well within dust sublimation in at least a few disks \citep{gravity19,gravity20,gravity21_51oph}. In particular, in 51~Oph, \citet{gravity21_51oph} estimates that the observed CO overtone is both hot (1900--2800~K) and with a very large column density ($10^{21-22}$~cm$^{-2}$). As said above, in this disk we are not able to extract a clean line profile due to uncommonly large CO line widths and high vibrational excitation, but assuming a FWHM of at least $\approx200$~km/s, similar to the case of HD~58647, an outer limit to CO fundamental emission can be placed at $< 0.25$~au, well inside the measured $R_{\rm{NIR}}$ and corresponding to the emitting region of CO overtone \citep[0.1~au,][]{gravity21_51oph}. With the iSHELL spectra from this survey, it is now possible for the first time to test how commonly the radial region of CO overtone and CO fundamental emission match in a sample of 5 disks where both are observed: 51~Oph, HD~35929, HD~36917, HD~58647, and HD~259431. In at least 4 out of 5, CO fundamental emission comes from within the measured $R_{\rm{NIR}}$ (Figure \ref{fig: radii_compar}), corresponding to the region of CO overtone \citep{ilee14}, and we have already highlighted the example of HD~259431 where the two CO bands match in tracing the inner Keplerian ring in Figure \ref{fig: kepl_rings}. With the only other case found in previous work, HD~101412 \citep{vdplas15}, there are to date 6 disks where CO overtone and CO fundamental emission match. Remarkably, the iSHELL survey clearly shows for the first time a discontinuity in the broad trend between CO FWHM and stellar mass (Figure \ref{fig: kinem_corr}), where CO emission in disks around stars of mass $> 3 M_{\odot}$ seems to transition to a different regime and mostly probe the disk region inside dust sublimation.

With this iSHELL survey, and using new measurements of $R_{\rm{NIR}}$ from GRAVITY \citep{gravity19,gravity21_TTs}, it is now also possible to test how commonly CO fundamental emission traces the region inside the dust sublimation radius in disks in general. In disks that have triangular lines, the BC seems to regularly trace this inner dust-free region (Figures \ref{fig: radii_compar} and \ref{fig: discussion}) even if CO overtone is not commonly detected in disks, suggesting that, as expected, fundamental CO lines are more easily excited down to lower temperatures and/or columns than what required to populate CO overtone emission. Although historically it is CO overtone that has been used to detect molecular gas within dust sublimation \citep[e.g.][]{ilee14}, this analysis suggests that CO fundamental emission may be a better tracer to use in future work, with the caveat that CO fundamental can also emit from larger disk radii and therefore that different emitting regions need to be distinguished using the observed line kinematics. It is interesting to note that the column density from a simple LTE fit to the BC in fundamental CO emission in some disks indicates very high columns of gas (see the case of AS~205~N in Figure \ref{fig: rotdiagr_gallery}), similar to the high columns estimated in 51~Oph from CO overtone. The lower temperature of $\approx 500$~K estimated for BC, if real and not simply an artifact of the LTE assumption, would also be consistent with the non-detection of CO overtone, which instead requires higher temperatures of $> 1000$~K to be significantly excited \citep[e.g.][]{ilee18}.

\subsection{The chemical structure of inner disks and winds}
Figure \ref{fig: AS205_water} shows the velocity profile of \ce{H2O} emission as detected in AS~205~N (Fig. \ref{fig: spec_overview}) in comparison to the CO line profile, and \ce{H2O} emission detected near 3~$\mu$m and 12~$\mu$m from other works \citep{banz17,najita18}. Water and CO line profiles at infrared wavelengths have been found in previous work to be similar, suggesting similar emitting regions \citep{pont10,banz17,salyk19}. The present work now combines for the first time spectrally-resolved water line profiles between 3 and 12~$\mu$m, and compares them to CO and optical forbidden oxygen [OI] emission, a key tracer of winds and jets \citep[e.g.][]{edwards87,hartigan95,banz19}. The hotter water lines at near-IR wavelengths are dominated by BC, while the colder lines at mid-IR wavelengths are dominated by NC. The narrow line center has a similar shape in [OI] too, possibly implying that all three species trace the same wind, although different parts of it; [OI], in fact, is excited also in a faster part of the outflow, as indicated by the blue-shifted wing at $< -30$~km/s, where CO and water are clearly absent. If CO in this disk traces a low-velocity region of a disk wind, as discussed above, the same region of the wind could be water-rich. Multi-wavelength thermo-chemical modeling of carbon and oxygen gas tracers has the potential to study gradients in the physical and chemical structure (temperature, density, and C/O ratios) of inner disk surfaces and winds. The great advantage offered by the detection of water lines near 5~$\mu$m, presented in this work, is the possibility of a simultaneous analysis of CO and \ce{H2O} that does not suffer from the large uncertainties introduced by flux calibration of spectra at different wavelengths, and from the variability of spectra taken at different epochs (see again Figure \ref{fig: line_variability}).

\section{Summary and main conclusions} \label{sec: conclusions}
In this work, we have presented an overview and first results from an ongoing $M$-band spectroscopic survey of planet-forming disks performed with iSHELL on IRTF, using two slits that provide resolving power R~$\approx$~60,000--92,000 (5 and 3.3~km/s). The spectral completeness of iSHELL data is unprecedented at these wavelengths and resolution, covering 4.52--5.24~$\mu$m in one shot with only narrow gaps between the echelle orders (Figure \ref{fig: spec_overview}). The spectra cover $> 50$ lines from the R and P branches of \ce{^{12}CO} and \ce{^{13}CO} for each of multiple vibrational levels. This coverage provides an unprecedented view on CO excitation in inner disks, that we have started to extract and analyze in this work in terms of three empirical tracers (Section \ref{sec: res_excit}): the high-$J$ to low-$J$ line flux ratio of $v=1-0$ lines (a proxy for rotational excitation), the $v=2-1$ to $v=1-0$ line flux ratio (a proxy for vibrational excitation), and the \ce{^{12}CO}/\ce{^{13}CO} ratio (a proxy for column density).
Some of the most notable results of this survey are:
\begin{itemize}
    \item the largest spectral coverage ever obtained in CO fundamental emission, for a total of $>100$ line fluxes that can be extracted from each spectrum;
    \item first detection of a large fraction of double-peak lines in $M$-band CO emission (16/31 disks in the current sample);
    \item first observation of $M$-band CO at resolution R~$> 50,000$ in: CQ~Tau, V892~Tau, HD~143006, HD~35929, AB~Aur, MWC~480, MWC~758, MWC~297, RY~Tau, SU~Aur, CI~Tau, 51~Oph, HD~36917, HD~58647, and HD~259431;
    \item first detection of two separate Keplerian rings of CO emission in a disk at $<2$~au (in HD~259431);
    \item first detection of common CO kinematic variability over timescales between 1 and 14 years;
    \item first detection of \ce{H2O} lines near 5~$\mu$m in a disk (in AS~205~N).
\end{itemize}

\subsection{Two fundamental line profiles and their interpretation}
The survey currently includes 31 disks around stars mostly within $5500 < T_{\rm{eff}} < 12,000$. We have combined it with a previous survey done with VLT-CRIRES that focused on cooler stars with $T_{\rm{eff}} < 6,500$ \citep{pont11,brown13} and found generally similar detection rates for \ce{^{12}CO} and \ce{^{13}CO}, but a striking difference in terms of line shapes. Building on previous work, we have identified and studied two fundamental types of CO line shapes to better understand their different kinematics: lines that exhibit a double-peaked structure as expected from gas in Keplerian rotation, and a ``triangular" shape that exhibit broader line wings and narrower line centers, often with a single central peak. 
We summarize the classification of CO spectra according to the measured properties of these two line types and their emission components in Figures \ref{fig: discussion} and \ref{fig: classification}.
The data show that these two types of line shapes are not distinguished by a different viewing angle, the presence of absence of an inner dust cavity, or different excitation conditions. Rather, the line shape shows a dependence on the stellar properties, with double-peak lines most common at $T_{\rm{eff}} \gtrsim 7,000$ and triangular lines most common at $T_{\rm{eff}} \lesssim 7,000$. In this first paper, we have focused on this global trend to identify and discuss a scenario that could explain it.

Double-peak line shapes are readily described with relatively narrow rings of gas in Keplerian rotation in a disk. 
By comparing their emitting radii $R_{\rm{CO}}$ with the innermost location of hot dust $R_{\rm{subl}}$, taken as a tracer of the dust sublimation radius, we have identified three emitting regions that seem to reflect a fundamental dependence of $v2/v1$ on the local dust content \citep[as previously suggested in][]{banz18}. Regions 1a and 1b both have high $v2/v1 = $~0.2--1 but a very different emitting region, 1a inside $R_{\rm{subl}}$ and 1b inside an inner dust cavity as observed in millimeter dust emission. Both regions share a high level of dust depletion in principle allowing to observe larger columns of CO gas, which has been proposed as the explanation for the high $v2/v1$ in region 1b \citep{bosman19}. While region 1a is also highly rotationally excited, likely due to the higher temperature closer to the star, region 1b instead shows lower rotational excitation that has been attributed to non-LTE conditions and UV fluorescence exciting higher vibrational levels up to $v=6$ in some disks \citep{brittain07,vdplas15}.
An intermediate region outside $R_{\rm{subl}}$, but still relatively close to the inner dust rim ($R_{\rm{CO}}$/$R_{\rm{subl}} \approx $~2--10), shows a large drop in values of $v2/v1 = $~0.01--0.2, which has been interpreted as tracing typical conditions found in a dust-rich inner disk surface \citep[e.g.][]{bosman19}. 

Moving now to triangular lines, their emitting radius (by assuming Keplerian broadening, which should be valid at least for the broad component BC) and excitation follow the same broad trends as found in the double-peak lines, suggesting that all CO lines may trace a similar range of disk regions, gas-to-dust ratios, and excitation conditions.
However, the different shape of triangular lines points to a different kinematic structure that has been previously interpreted either as gas in Keplerian rotation over a radially extended disk surface, or as a Keplerian disk + sub-Keplerian low-velocity wind filling in at the line center. While a single interpretation does not necessarily applies to all disks with this line shape, in this work we have extensively discussed the wind scenario and find that, unlike an extended disk surface, it could simultaneously explain several aspects observed in the data: 
1) the asymmetric spectro-astrometric signal detected in at least a few disks so far, 2) the large fraction of blue-shifted emission line asymmetries and centroids, 3) the large fraction of kinematic variability, and especially those cases where variability is on the blue side of the lines, 4) line shapes that change with viewing angle, in terms of a wind that turns into absorption at the line center, 5) blue-shifted narrow absorption lines detected in disks at high inclinations, and, fundamentally, 6) the dependence of line kinematics on $T_{\rm{eff}}$, where T~Tauri disks typically have inner disk winds as observed from optical forbidden line emission at 0.05--5~au (similar to what estimated for BC and NC in CO), while Herbig disks very rarely have winds and jets. The few cases where a triangular CO line shape is observed in Herbig disks also support the wind scenario: most notably, MWC~480 and HD~163296 are among the very few Herbig Ae stars where large-scale outflows are observed. In HD~163296, in particular, there is evidence from multiple tracers, including new spatially resolved ALMA data \citep{booth21}, that an inner MHD disk wind is launched as close as $< 0.2$~au and out to 4--10~au, overlapping to the emitting region of CO fundamental emission (Figure \ref{fig: HD163296}).

\subsection{Future work}
Moving forward on the interpretation of line kinematics, future work should invest on spectro-astrometry and spatial resolution. Spectro-astrometry at high enough S/N could clarify if triangular lines in general do in fact include a wind component, including the broader lines observed at higher disk inclinations $> 40$~deg. We note that the spectro-astrometric signal so far has been found consistent with a Keplerian motion in some of these sources (RNO~90, GQ~Lup), but closer inspection shows what model fits need improvement and that there could be asymmetries that require higher S/N than what achieved before \citep[Figure 3 in][]{pont11}. 
On the spatial resolution side, the difference between a larger emitting disk surface and a compact emission could be readily visible and distinguished with future ELT-METIS imaging, as demonstrated for disks around Herbig Ae/Be stars in \citet{bosman19}.
Confirming with spectro-astrometry and/or high-resolution imaging a wind origin for CO emission in the disks of AB~Aur, MWC~480, and HD~163296 would be very important, and provide what seems to be the most direct tracer so far of an inner disk wind observed at its launching region in Herbig Ae disks.

Other kinematically complex structures should also be investigated in future work, including a warped inner ring or a vortex that could also plausibly produce complex line kinematics that deviate from a simple Keplerian profile. These structures have been proposed to explain inner disk misalignment and large-scale shadows projected to the outer disk \citep[e.g.][]{benisty17,min17}, which could be produced from dynamical interactions with giant protoplanets forming in the inner disk \citep[as proposed e.g. in CI~Tau,][]{jkrull16}. 
The combination of systematic model simulations of these different scenarios and high spectral/spatial-resolution observations will likely be required for a comprehensive analysis of a range of plausible kinematic structures. 
Study of the timescales and kinematics of CO line variability will also play a key role in distinguishing different scenarios. Line asymmetries that vary on orbital timescales corresponding to the line velocity are consistent with point sources orbiting the star \citep{brittain19}, while line variability could just be stochastic, and/or correlated to accretion variability, if it traces an inner disk wind.

The complex and multi-component kinematic structure of $M$-band CO spectra also poses non-trivial challenges for the analysis of future JWST-MIRI and NIRSpec spectra, whose medium resolving power (R~$\approx 3000$, or 100~km/s) will blend together the flux from multiple emission and absorption components. The correct analysis of CO spectra from JWST will likely require support from ground-based higher-resolution spectrographs.
Future work should also combine the observed line kinematics to a detailed study of CO excitation and its deviations from LTE conditions, a study that the large spectral coverage of iSHELL (and of VLT-CRIRES+ by combining multiple settings) now support at an unprecedented level.

\acknowledgments
We thank I.Pascucci, E.Whelan, and M.Flock for helpful discussions on the interpretation of CO emission in terms of inner disk winds. 
We thank the referee for constructive feedback and suggestions that helped improve the paper.
This work includes data gathered at the Infrared Telescope Facility, which is operated by the University of Hawaii under contract 80HQTR19D0030 with the National Aeronautics and Space Administration.
This work is partly based on observations made with ESO telescopes at the Paranal Observatory under program 179.C-0151.
The authors wish to recognize and acknowledge the very significant cultural role and reverence that the summit of Maunakea has always had within the indigenous Hawaiian community. We are grateful to the Hawaiian community to have the opportunity to conduct observations from this mountain.

\bibliography{ishell_paper}{}
\bibliographystyle{aasjournal}

\appendix

\section{Spectral line measurements from this survey}
Tables \ref{tab: measurements} and \ref{tab: measurements_crires} report measured line properties for the CO spectra included in this paper from the iSHELL and CRIRES surveys (see Section \ref{sec: sample}), separated by line types using the shape parameter $S$ as described in Section \ref{sec: line_shapes}.

\begin{deluxetable*}{l c c c c c | c c c c c c c c}
\tabletypesize{\footnotesize}
\tablewidth{0pt}
\tablecaption{\label{tab: measurements} Measured CO line properties from the iSHELL survey.}
\tablehead{ Name & Comp & $S$ & Cen & FWHM & FW10\% & $F_{\rm{low}-J}$ & err & $F_{\rm{high}-J}$ & err & $F_{\rm{v=2-1}}$ & err & $F_{\rm{\ce{^{13}CO}}}$ & err \\
 &  &  & (km/s) & (km/s) & (km/s) &  \multicolumn{8}{c}{(erg s$^{-1}$ cm$^{-2}$)} }
\tablecolumns{14}
\startdata
HD145718 & -- & -- & -- & -- & -- & 7.97e-16 & 2.35e-15 & 0.00e+00 & 1.22e-15 & 1.53e-16 & 6.14e-16 & 0.00e+00 & 7.32e-16 \\
\multicolumn{6}{c}{\textit{-- Double-peak lines (associated to disk rings) --}} \\
HD141569 & SC & 2.0 & -6.8 & 16.7 & 26.4 & 2.68e-14 & 2.34e-16 & 2.79e-16 & 1.14e-15 & 1.35e-14 & 1.11e-16 & 5.74e-15 & 1.10e-16 \\
HD142666 & SC & 1.3 & -3.5 & 55.7 & 71.0 & 2.54e-15 & 1.15e-16 & 1.52e-15 & 4.96e-17 & 4.22e-17 & 2.74e-16 & 1.79e-16 & 3.60e-16 \\
HD143006 & SC & 1.8 & -0.3 & 25.1 & 31.1 & 5.26e-15 & 2.08e-16 & 1.85e-15 & 8.60e-17 & 0.00e+00 & 4.04e-16 & 3.10e-16 & 5.06e-16 \\
HD150193 & SC & 1.5 & -3.7 & 37.2 & 44.7 & 5.11e-15 & 1.34e-16 & 2.92e-15 & 5.86e-17 & 9.33e-17 & 3.19e-16 & 2.67e-16 & 4.06e-16 \\
HD169142 & SC & 2.4 & -2.5 & 6.8 & 11.8 & 3.31e-15 & 6.66e-17 & 1.08e-15 & 1.14e-16 & 5.25e-16 & 7.17e-17 & 1.98e-16 & 3.35e-16 \\
HD179218 & SC & 1.7 & 14.9 & 17.8 & 24.4 & 3.10e-15 & 1.74e-17 & 1.24e-15 & 2.77e-17 & 7.85e-16 & 1.53e-17 & 5.46e-16 & 1.98e-17 \\
HD35929 & SC & 1.9 & 16.3 & 122.2 & 149.6 & 1.96e-14 & 2.20e-16 & 1.90e-14 & 1.03e-16 & 1.38e-14 & 5.31e-17 & 2.40e-15 & 7.24e-17 \\
HD36917 & SC & 1.7 & 17.5 & 106.7 & 159.5 & 1.77e-14 & 2.27e-16 & 1.91e-14 & 9.70e-17 & 1.37e-14 & 5.17e-17 & 2.08e-15 & 6.19e-17 \\
HD37806 & SC & 1.5 & 27.8 & 51.8 & 67.1 & 4.24e-15 & 8.09e-17 & 3.01e-15 & 4.49e-17 & 2.77e-16 & 3.06e-16 & 1.78e-16 & 3.93e-16 \\
HD58647 & SC & 1.2 & 5.1 & 191.2 & 199.2 & 1.82e-14 & 8.27e-17 & 1.82e-14 & 3.16e-17 & 1.67e-14 & 2.64e-17 & 1.07e-15 & 3.26e-17 \\
IRS48 & SC & 2.0 & -5.9 & 14.2 & 22.1 & 3.82e-15 & 1.62e-16 & 9.68e-16 & 1.30e-15 & 2.38e-15 & 1.16e-16 & 4.24e-16 & 8.72e-16 \\
MWC297 & SC & 1.8 & -9.5 & 46.8 & 68.9 & 1.86e-14 & 1.48e-17 & 1.62e-14 & 1.25e-17 & 9.20e-15 & 6.06e-18 & 5.42e-15 & 7.52e-18 \\
SR21 & SC & 2.3 & -7.8 & 10.2 & 17.3 & 4.59e-15 & 1.07e-16 & 2.36e-15 & 1.70e-16 & 3.65e-15 & 9.13e-17 & 9.00e-16 & 8.75e-17 \\
V892Tau & SC & 1.8 & 16.4 & 28.8 & 45.5 & 3.72e-14 & 3.09e-17 & 1.97e-14 & 4.82e-17 & 8.81e-15 & 2.33e-17 & 3.43e-15 & 2.95e-17 \\
VVSer & SC & 1.8 & -6.9 & 62.2 & 84.5 & 8.72e-15 & 1.02e-16 & 7.12e-15 & 4.95e-17 & 3.06e-16 & 3.76e-16 & 6.38e-16 & 3.10e-17 \\
HD259431 & NC & 2.5 & 19.9 & 59.6 & 137.2 & 5.06e-15 & 1.32e-16 & 4.55e-15 & 6.17e-17 & 1.43e-16 & 5.87e-16 & 3.48e-16 & 6.84e-16 \\
HD259431 & BC & 2.5 & 20.6 & 119.4 & 137.2 & 7.90e-15 & 1.32e-16 & 8.11e-15 & 6.17e-17 & 6.60e-15 & 3.15e-17 & 5.21e-16 & 6.84e-16 \\
\multicolumn{6}{c}{\textit{-- Triangular lines (associated to disk + winds) --}} \\
ABAur & SC & 3.6 & 14.1 & 12.1 & 24.2 & 9.11e-15 & 1.49e-17 & 4.14e-15 & 2.69e-17 & 8.52e-16 & 1.35e-17 & 1.00e-15 & 1.84e-17 \\
HD163296 & SC & 3.4 & -14.4 & 60.6 & 111.5 & 1.10e-14 & 8.13e-17 & 8.86e-15 & 4.55e-17 & 1.85e-15 & 2.32e-17 & 1.26e-15 & 2.75e-17 \\
HD190073 & SC & 4.3 & 2.2 & 15.5 & 33.6 & 5.43e-15 & 1.25e-17 & 4.79e-15 & 1.83e-17 & 2.22e-15 & 9.46e-18 & 1.89e-15 & 1.08e-17 \\
MWC480 & SC & 4.1 & 11.7 & 49.5 & 94.2 & 1.20e-14 & 6.79e-17 & 7.53e-15 & 6.01e-17 & 1.18e-15 & 3.06e-17 & 3.38e-16 & 5.81e-16 \\
SUAur & SC & 3.8 & 10.3 & 60.7 & 100.5 & 8.06e-15 & 1.12e-16 & 3.57e-15 & 8.19e-17 & 0.00e+00 & 5.72e-16 & 1.72e-16 & 6.63e-16 \\
AS205N & NC & 6.8 & -5.1 & 14.1 & 72.2 & 1.99e-14 & 4.97e-17 & 1.88e-14 & 6.29e-17 & 9.39e-16 & 3.27e-17 & 2.72e-15 & 3.97e-17 \\
CITau & NC & 4.7 & 22.2 & 104.7 & 266.9 & 3.56e-14 & 6.30e-16 & 3.65e-14 & 4.04e-16 & 3.36e-16 & 1.12e-14 & 3.36e-16 & 1.32e-14 \\
CQTau & NC & 12.3 & 12.1 & 7.3 & 83.5 & 8.92e-16 & 1.55e-17 & 0.00e+00 & 1.48e-16 & 8.37e-17 & 1.22e-17 & 2.49e-16 & 1.54e-17 \\
Elias20 & NC & 2.9 & -9.6 & 49.8 & 121.3 & 3.94e-14 & 8.72e-16 & 4.38e-14 & 4.36e-16 & 4.33e-16 & 4.15e-15 & 1.52e-15 & 4.58e-15 \\
LkHa330 & NC & 5.2 & 16.4 & 10.3 & 45.0 & 3.38e-15 & 8.10e-17 & 4.69e-16 & 4.55e-17 & 0.00e+00 & 1.61e-16 & 6.61e-16 & 3.31e-17 \\
MWC758 & NC & 5.8 & 17.4 & 11.6 & 48.4 & 7.61e-15 & 2.18e-17 & 2.98e-15 & 3.82e-17 & 1.67e-16 & 1.99e-17 & 8.17e-16 & 3.25e-17 \\
RNO90 & NC & 2.8 & -15.0 & 54.2 & 150.8 & 3.61e-14 & 1.11e-16 & 3.46e-14 & 8.49e-17 & 1.07e-15 & 1.17e-15 & 1.83e-15 & 7.67e-17 \\
RYTau & NC & 3.8 & 7.2 & 54.3 & 200.8 & 6.10e-15 & 1.35e-16 & 4.67e-15 & 6.21e-17 & 6.04e-16 & 6.58e-16 & 0.00e+00 & 9.24e-16 \\
AS205N & BC & 6.8 & -3.6 & 57.5 & 72.2 & 4.48e-14 & 2.49e-16 & 3.65e-14 & 1.05e-16 & 1.76e-14 & 6.07e-17 & 3.93e-15 & 7.36e-17 \\
CITau & BC & 4.7 & -25.1 & 241.4 & 266.9 & 2.72e-14 & 6.30e-16 & 2.31e-14 & 4.04e-16 & 2.72e-14 & 4.45e-16 & 2.76e-16 & 1.32e-14 \\
CQTau & BC & 12.3 & 12.1 & 52.4 & 83.5 & 5.99e-15 & 1.00e-16 & 1.47e-15 & 4.80e-17 & 0.00e+00 & 3.83e-16 & 3.76e-16 & 4.63e-16 \\
Elias20 & BC & 2.9 & -4.9 & 95.0 & 121.3 & 6.71e-14 & 8.72e-16 & 6.39e-14 & 4.36e-16 & 2.82e-14 & 2.22e-16 & 6.49e-16 & 4.58e-15 \\
LkHa330 & BC & 5.2 & 17.4 & 30.1 & 45.0 & 7.81e-15 & 1.78e-16 & 5.04e-15 & 7.21e-17 & 2.18e-16 & 4.50e-16 & 0.00e+00 & 7.04e-16 \\
MWC758 & BC & 5.8 & 13.4 & 37.6 & 48.4 & 1.03e-14 & 6.49e-17 & 6.94e-15 & 4.84e-17 & 3.57e-16 & 2.54e-17 & 1.95e-17 & 4.38e-16 \\
RNO90 & BC & 2.8 & -18.6 & 84.0 & 150.8 & 6.14e-14 & 1.11e-16 & 5.67e-14 & 8.49e-17 & 1.78e-14 & 5.54e-17 & 6.23e-16 & 1.61e-15 \\
RYTau & BC & 3.8 & -0.4 & 125.7 & 200.8 & 1.10e-14 & 1.40e-16 & 8.12e-15 & 9.34e-17 & 2.64e-15 & 6.18e-17 & 0.00e+00 & 2.04e-15 \\
\multicolumn{6}{c}{\textit{-- Absorption lines --}} \\
CITau & abs & 4.7 & 15.4 & 4.7 & 266.9 & 4.08e-15 & 3.69e-16 & 3.32e-16 & 3.69e-16 & 3.32e-16 & 3.69e-16 & 2.19e-15 & 5.22e-16 \\
Elias20 & abs & 2.9 & -6.8 & 6.0 & 121.3 & 1.06e-14 & 5.62e-16 & 5.06e-16 & 5.62e-16 & 5.06e-16 & 5.62e-16 & 1.89e-15 & 4.06e-16 \\
HD145718 & abs & -- & -3.0 & 5.0 & -- & 3.36e-15 & 2.53e-16 & 7.65e-16 & 1.70e-16 & 2.28e-16 & 2.53e-16 & 1.02e-15 & 1.59e-16 \\
IRS48 & abs & 2.0 & -8.1 & 4.0 & 22.1 & 5.57e-15 & 4.01e-16 & 3.61e-16 & 4.01e-16 & 3.61e-16 & 4.01e-16 & 3.48e-15 & 4.73e-16 \\
LkHa330 & abs & 5.2 & 16.8 & 4.8 & 45.0 & 2.09e-15 & 1.22e-16 & 1.09e-16 & 1.22e-16 & 1.09e-16 & 1.22e-16 & -3.60e-17 & 2.06e-16 \\
MWC297 & abs & 1.8 & -7.9 & 9.4 & 68.9 & 1.26e-14 & 1.96e-17 & 1.76e-17 & 1.96e-17 & 1.76e-17 & 1.96e-17 & 3.06e-15 & 1.77e-17 \\
RNO90 & abs & 2.8 & -11.8 & 5.1 & 150.8 & 4.41e-15 & 1.13e-16 & 1.02e-16 & 1.13e-16 & 1.02e-16 & 1.13e-16 & 1.91e-15 & 7.56e-17 \\
RYTau & abs & 3.8 & 7.7 & 9.0 & 200.8 & 8.21e-15 & 1.38e-16 & 2.95e-15 & 1.45e-16 & 1.24e-16 & 1.38e-16 & 0.00e+00 & 0.00e+00 \\
SR21 & abs & 2.3 & -8.0 & 3.0 & 17.3 & 3.51e-15 & 3.08e-16 & 2.77e-16 & 3.08e-16 & 2.77e-16 & 3.08e-16 & 1.89e-15 & 3.40e-16 \\
SUAur & abs & 3.8 & 14.2 & 4.9 & 100.5 & 2.63e-15 & 7.62e-17 & 3.26e-16 & 7.92e-17 & 6.86e-17 & 7.62e-17 & 4.55e-16 & 6.97e-17 \\
V892Tau & abs & 1.8 & 15.5 & 3.6 & 45.5 & 2.17e-15 & 9.79e-17 & 8.81e-17 & 9.79e-17 & 8.81e-17 & 9.79e-17 & 1.48e-15 & 9.78e-17 \\
VVSer & abs & 1.8 & -8.6 & 5.7 & 84.5 & 7.93e-15 & 8.61e-17 & 7.75e-17 & 8.61e-17 & 7.75e-17 & 8.61e-17 & 8.40e-16 & 7.68e-17 \\
\enddata
\tablecomments{The second column reports the velocity component as defined in Section \ref{sec: components} (``abs" stands for absorption). The third column reports the full line shape parameter $S$ as defined in Section \ref{sec: line_shapes}. The next three columns report the line centroid in heliocentric reference (Cen, not corrected for the stellar RV), full-width-at-half-maximum (FWHM), and full-width-at-10\% (FW10\%) as measured in the \ce{^{12}CO} $v=1-0$ lines. The FW10\%, like the parameter $S$, is measured from the full line profile only, not for BC and NC separately. The line fluxes in the rest of the table are measured from the continuum-normalized spectra, therefore they can be used for line flux ratios as done in this paper, but should be flux calibrated with the W2 flux reported in Table \ref{tab: sample} before using them individually.
}
\end{deluxetable*}

\begin{deluxetable*}{l c c c c c | c c c c c c c c}
\tabletypesize{\footnotesize}
\tablewidth{0pt}
\tablecaption{\label{tab: measurements_crires} Measured CO line properties from the CRIRES sample included in this work.}
\tablehead{ Name & Comp & $S$ & Cen & FWHM & FW10\% & $F_{\rm{low}-J}$ & err & $F_{\rm{high}-J}$ & err & $F_{\rm{v=2-1}}$ & err & $F_{\rm{\ce{^{13}CO}}}$ & err \\
 &  &  & (km/s) & (km/s) & (km/s) &  \multicolumn{8}{c}{(erg s$^{-1}$ cm$^{-2}$)} }
\tablecolumns{14}
\startdata
\multicolumn{6}{c}{\textit{-- Double-peak lines (associated to disk rings) --}} \\
AATau & SC & 1.8 & 17.2 & 118.3 & 166.5 & 7.91e-14 & 5.51e-16 & 0.00e+00 & 0.00e+00 & 8.89e-15 & 2.28e-16 & 2.41e-15 & 2.68e-15 \\
IQTau & SC & 1.7 & 11.8 & 133.6 & 190.6 & 3.31e-14 & 1.11e-15 & 3.01e-14 & 2.06e-16 & 1.25e-14 & 1.33e-16 & 0.00e+00 & 4.43e-15 \\
IRS48 & SC & 1.5 & -5.5 & 15.0 & 19.8 & 2.67e-15 & 4.19e-17 & 0.00e+00 & 0.00e+00 & 6.30e-16 & 2.51e-17 & 9.88e-17 & 1.62e-16 \\
SR21 & SC & 1.8 & -7.1 & 11.0 & 15.9 & 4.37e-15 & 1.79e-17 & 0.00e+00 & 0.00e+00 & 3.01e-15 & 1.27e-17 & 9.01e-16 & 1.03e-17 \\
VVSer & SC & 1.4 & -10.2 & 57.5 & 68.3 & 8.39e-15 & 5.64e-17 & 0.00e+00 & 0.00e+00 & 1.28e-15 & 1.52e-17 & 8.60e-16 & 1.55e-17 \\
WaOph6 & SC & 1.9 & -6.1 & 147.4 & 215.6 & 4.20e-14 & 1.47e-16 & 0.00e+00 & 0.00e+00 & 9.85e-15 & 1.02e-16 & 0.00e+00 & 1.57e-15 \\
\multicolumn{6}{c}{\textit{-- Triangular lines (associated to disk + winds) --}} \\
CVCha & SC & 3.7 & 12.0 & 76.6 & 136.8 & 4.77e-14 & 1.28e-16 & 4.25e-14 & 9.70e-17 & 4.97e-15 & 5.98e-17 & 5.52e-16 & 6.79e-16 \\
CWTau & SC & 2.6 & 11.6 & 71.6 & 129.3 & 7.21e-14 & 1.31e-16 & 6.73e-14 & 9.36e-17 & 1.33e-14 & 5.86e-17 & 2.35e-15 & 4.94e-17 \\
DoAr44 & SC & 3.5 & -5.5 & 52.7 & 98.2 & 3.80e-14 & 1.15e-16 & 0.00e+00 & 0.00e+00 & 2.87e-15 & 1.03e-16 & 7.27e-16 & 9.01e-16 \\
FNTau & SC & 4.0 & 12.3 & 9.9 & 20.5 & 1.03e-14 & 3.30e-16 & 0.00e+00 & 0.00e+00 & 0.00e+00 & 2.11e-15 & 6.37e-16 & 1.81e-15 \\
FZTau & SC & 5.5 & 15.9 & 30.1 & 60.1 & 7.20e-14 & 3.26e-15 & 0.00e+00 & 0.00e+00 & 9.53e-15 & 1.70e-14 & 5.62e-15 & 6.74e-15 \\
SXCha & SC & 4.7 & -12.9 & 42.7 & 81.8 & 2.09e-14 & 7.78e-16 & 1.94e-14 & 3.42e-16 & 3.01e-15 & 1.62e-16 & 0.00e+00 & 3.86e-15 \\
TWCha & SC & 4.6 & 17.4 & 66.7 & 132.1 & 6.55e-14 & 1.22e-15 & 5.93e-14 & 5.33e-16 & 7.87e-15 & 2.79e-16 & 2.03e-16 & 8.26e-15 \\
TWHya & SC & 2.5 & 12.3 & 7.3 & 12.9 & 2.19e-14 & 5.48e-17 & 0.00e+00 & 0.00e+00 & 5.58e-16 & 4.78e-17 & 0.00e+00 & 3.45e-16 \\
VZCha & SC & 4.0 & 18.9 & 42.6 & 82.9 & 1.10e-13 & 6.22e-16 & 9.26e-14 & 2.16e-16 & 2.68e-14 & 9.80e-17 & 5.22e-15 & 1.18e-16 \\
AS205N & NC & 7.5 & -6.0 & 13.7 & 72.7 & 2.62e-14 & 9.93e-18 & 0.00e+00 & 0.00e+00 & 1.76e-15 & 1.21e-17 & 3.20e-15 & 8.83e-18 \\
DFTau & NC & 2.8 & 13.7 & 37.2 & 138.7 & 1.67e-14 & 1.22e-16 & 0.00e+00 & 0.00e+00 & 1.63e-16 & 2.63e-15 & 3.84e-15 & 6.63e-17 \\
DRTau & NC & 5.3 & 24.1 & 10.6 & 49.2 & 2.11e-14 & 2.71e-17 & 1.90e-14 & 2.38e-17 & 5.95e-16 & 1.56e-17 & 2.79e-15 & 1.65e-17 \\
Elias20 & NC & 3.3 & -15.2 & 33.3 & 99.3 & 3.90e-14 & 9.77e-17 & 0.00e+00 & 0.00e+00 & 3.29e-16 & 1.88e-15 & 2.80e-15 & 4.78e-17 \\
GQLup & NC & 2.4 & -2.6 & 65.8 & 119.8 & 2.47e-14 & 1.15e-16 & 0.00e+00 & 0.00e+00 & 2.57e-16 & 1.12e-15 & 2.19e-15 & 6.54e-17 \\
HD135344B & NC & 4.3 & 4.0 & 9.7 & 31.1 & 2.52e-15 & 3.56e-17 & 0.00e+00 & 8.24e-17 & 1.24e-16 & 5.85e-18 & 2.83e-16 & 5.03e-18 \\
LkHa330 & NC & 3.3 & 14.7 & 8.6 & 22.8 & 3.64e-15 & 3.60e-17 & 0.00e+00 & 0.00e+00 & 1.03e-16 & 2.54e-16 & 4.38e-16 & 3.06e-17 \\
RNO90 & NC & 3.1 & -13.0 & 57.8 & 158.2 & 2.76e-14 & 9.70e-17 & 2.76e-14 & 5.46e-17 & 2.67e-16 & 7.40e-16 & 1.60e-15 & 3.95e-17 \\
RULup & NC & 7.9 & -3.9 & 20.0 & 101.7 & 2.61e-14 & 9.73e-17 & 1.71e-14 & 4.62e-17 & 2.54e-16 & 3.11e-16 & 3.11e-15 & 2.02e-17 \\
RYLup & NC & 7.9 & -1.6 & 18.2 & 130.3 & 1.10e-14 & 7.26e-17 & 4.61e-15 & 5.71e-17 & 0.00e+00 & 3.56e-16 & 6.18e-16 & 2.29e-17 \\
SCrAS & NC & 10.8 & -2.8 & 12.0 & 114.7 & 1.30e-14 & 4.17e-17 & 0.00e+00 & 0.00e+00 & 0.00e+00 & 2.89e-16 & 1.06e-15 & 2.55e-17 \\
SCrAN & NC & 7.2 & -3.9 & 9.5 & 58.9 & 1.28e-14 & 2.64e-17 & 0.00e+00 & 0.00e+00 & 1.49e-15 & 2.06e-17 & 1.73e-15 & 1.35e-17 \\
VVCrAS & NC & 8.3 & -0.5 & 18.8 & 102.4 & 9.70e-15 & 2.35e-17 & 0.00e+00 & 0.00e+00 & 0.00e+00 & 9.54e-17 & 0.00e+00 & 1.07e-16 \\
VWCha & NC & 5.9 & 12.7 & 25.2 & 108.8 & 3.21e-14 & 2.08e-16 & 1.95e-14 & 7.54e-17 & 0.00e+00 & 4.93e-16 & 1.88e-15 & 9.99e-17 \\
WXCha & NC & 2.7 & 22.6 & 68.0 & 165.9 & 3.01e-14 & 1.18e-15 & 1.98e-14 & 3.82e-16 & 2.06e-16 & 4.99e-15 & 2.06e-16 & 1.26e-14 \\
AS205N & BC & 7.5 & -1.0 & 56.5 & 72.7 & 5.28e-14 & 3.63e-17 & 0.00e+00 & 0.00e+00 & 1.48e-14 & 2.17e-17 & 2.13e-15 & 1.64e-17 \\
DFTau & BC & 2.8 & 19.4 & 80.9 & 138.7 & 4.89e-14 & 1.22e-16 & 0.00e+00 & 0.00e+00 & 2.39e-14 & 1.38e-16 & 5.84e-16 & 1.27e-15 \\
DRTau & BC & 5.3 & 26.6 & 30.3 & 49.2 & 5.39e-14 & 4.15e-17 & 4.91e-14 & 3.13e-17 & 1.89e-14 & 2.19e-17 & 3.25e-15 & 2.21e-17 \\
Elias20 & BC & 3.3 & -6.4 & 68.4 & 99.3 & 3.13e-14 & 9.77e-17 & 0.00e+00 & 0.00e+00 & 1.82e-14 & 1.01e-16 & 3.75e-16 & 8.86e-16 \\
GQLup & BC & 2.4 & -8.1 & 91.8 & 119.8 & 2.19e-14 & 1.15e-16 & 0.00e+00 & 0.00e+00 & 6.93e-15 & 5.55e-17 & 2.07e-16 & 1.31e-15 \\
HD135344B & BC & 4.3 & 0.7 & 18.3 & 31.1 & 3.83e-15 & 5.04e-17 & 2.62e-15 & 1.88e-17 & 0.00e+00 & 9.39e-17 & 1.01e-16 & 7.72e-18 \\
LkHa330 & BC & 3.3 & 15.9 & 15.1 & 22.8 & 4.11e-15 & 1.10e-16 & 0.00e+00 & 0.00e+00 & 2.20e-16 & 4.89e-16 & 0.00e+00 & 4.19e-16 \\
RNO90 & BC & 3.1 & -7.2 & 98.0 & 158.2 & 4.81e-14 & 9.70e-17 & 4.58e-14 & 5.46e-17 & 1.33e-14 & 3.53e-17 & 4.65e-16 & 8.29e-16 \\
RULup & BC & 7.9 & -0.5 & 84.7 & 101.7 & 4.70e-14 & 1.01e-16 & 4.23e-14 & 6.18e-17 & 2.21e-14 & 3.42e-17 & 2.60e-15 & 3.20e-17 \\
RYLup & BC & 7.9 & -1.0 & 110.7 & 130.3 & 8.80e-15 & 8.90e-17 & 6.70e-15 & 1.02e-16 & 0.00e+00 & 1.17e-15 & 3.49e-16 & 9.16e-16 \\
SCrAS & BC & 10.8 & -6.4 & 88.5 & 114.7 & 2.99e-14 & 8.92e-17 & 0.00e+00 & 0.00e+00 & 1.24e-14 & 9.41e-17 & 0.00e+00 & 9.45e-16 \\
SCrAN & BC & 7.2 & -9.3 & 45.1 & 58.9 & 2.78e-14 & 4.76e-17 & 0.00e+00 & 0.00e+00 & 8.15e-15 & 4.62e-17 & 2.06e-16 & 4.10e-16 \\
VVCrAS & BC & 8.3 & -2.8 & 76.0 & 102.4 & 1.70e-14 & 4.13e-17 & 0.00e+00 & 0.00e+00 & 1.09e-14 & 1.66e-17 & 0.00e+00 & 2.76e-16 \\
VWCha & BC & 5.9 & 16.5 & 71.8 & 108.8 & 8.18e-14 & 2.20e-16 & 7.49e-14 & 1.05e-16 & 2.04e-14 & 6.43e-17 & 5.10e-15 & 1.52e-16 \\
WXCha & BC & 2.7 & 21.6 & 131.8 & 165.9 & 5.52e-14 & 1.18e-15 & 4.26e-14 & 3.82e-16 & 1.62e-14 & 2.46e-16 & 4.22e-16 & 1.26e-14 \\
\multicolumn{6}{c}{\textit{-- Absorption lines --}} \\
CWTau & abs & 2.6 & 14.8 & 9.3 & 129.3 & 2.22e-14 & 1.01e-16 & 6.14e-15 & 1.10e-16 & 9.05e-17 & 1.01e-16 & 0.00e+00 & 0.00e+00 \\
Elias20 & abs & 3.3 & -7.4 & 4.4 & 99.3 & 9.18e-15 & 1.08e-16 & 9.68e-17 & 1.08e-16 & 9.68e-17 & 1.08e-16 & 1.19e-15 & 1.09e-16 \\
IQTau & abs & 1.7 & 14.3 & 7.3 & 190.6 & 9.59e-15 & 6.01e-16 & 3.69e-15 & 3.94e-16 & 5.41e-16 & 6.01e-16 & 1.93e-15 & 2.61e-15 \\
IRS48 & abs & 1.5 & -6.8 & 4.2 & 19.8 & 5.77e-15 & 5.42e-17 & 4.88e-17 & 5.42e-17 & 4.88e-17 & 5.42e-17 & 2.83e-15 & 6.75e-17 \\
LkHa330 & abs & 3.3 & 18.7 & 3.0 & 22.8 & 1.77e-15 & 4.58e-17 & 4.12e-17 & 4.58e-17 & 4.12e-17 & 4.58e-17 & 8.31e-17 & 3.76e-16 \\
RNO90 & abs & 3.1 & -12.5 & 4.4 & 158.2 & 4.28e-15 & 4.36e-17 & 3.92e-17 & 4.36e-17 & 3.92e-17 & 4.36e-17 & 1.80e-15 & 4.15e-17 \\
RYLup & abs & 7.9 & -0.9 & 4.8 & 130.3 & 6.86e-15 & 9.13e-17 & 8.22e-17 & 9.13e-17 & 8.22e-17 & 9.13e-17 & 3.01e-15 & 7.79e-17 \\
SR21 & abs & 1.8 & -6.4 & 3.7 & 15.9 & 4.30e-15 & 4.81e-17 & 4.32e-17 & 4.81e-17 & 4.32e-17 & 4.81e-17 & 1.43e-15 & 5.80e-17 \\
VVSer & abs & 1.4 & -8.9 & 5.3 & 68.3 & 7.12e-15 & 2.45e-17 & 2.20e-17 & 2.45e-17 & 2.20e-17 & 2.45e-17 & 1.20e-15 & 3.54e-17 \\
WaOph6 & abs & 1.9 & -10.3 & 4.3 & 215.6 & 2.43e-15 & 5.58e-17 & 5.02e-17 & 5.58e-17 & 5.02e-17 & 5.58e-17 & 1.05e-15 & 6.36e-17 \\
\enddata
\tablecomments{See notes in Table \ref{tab: measurements}.}
\end{deluxetable*}

\section{Gallery of iSHELL spectra}
\begin{figure*}
\centering
\includegraphics[width=0.9\textwidth]{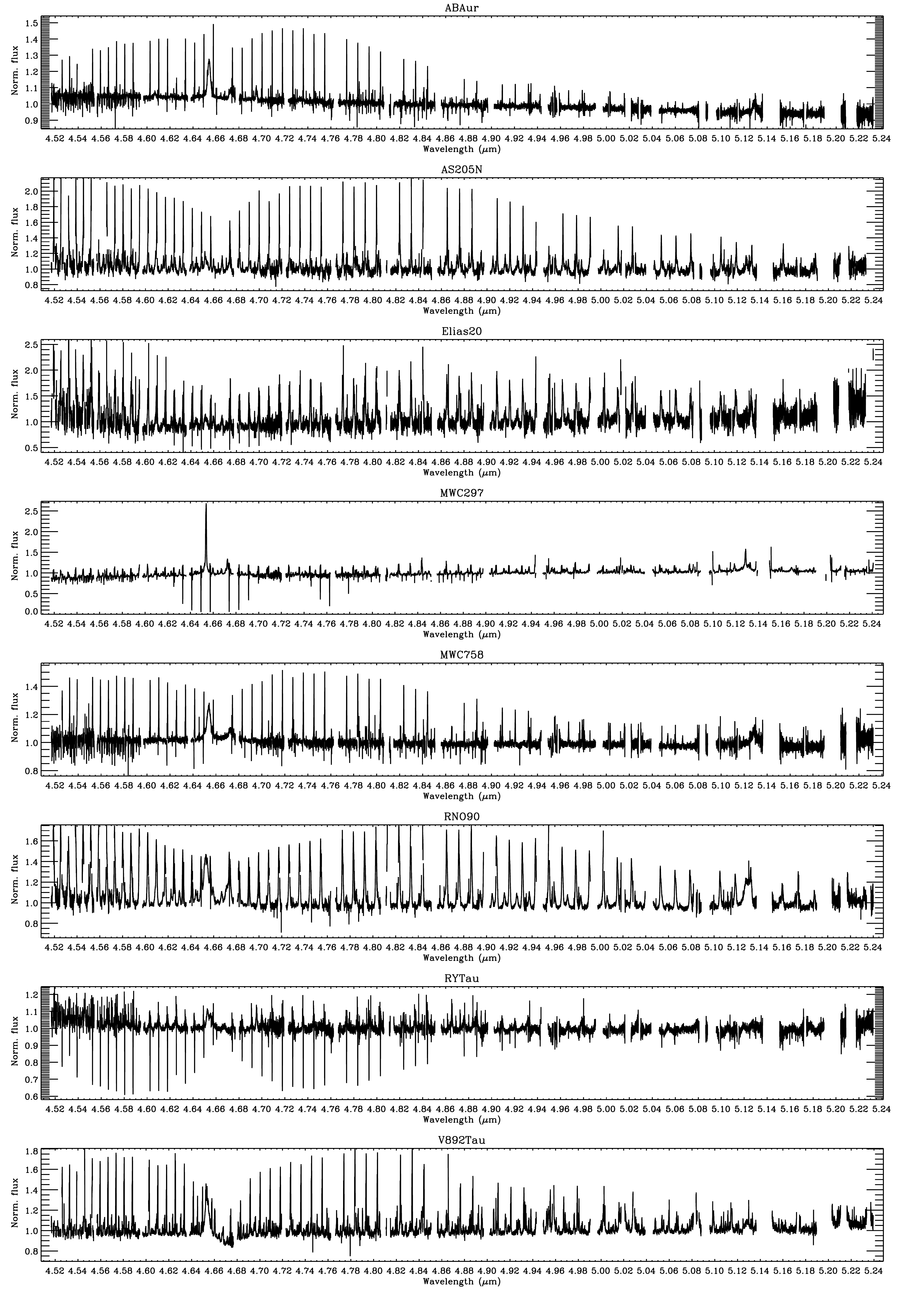} 
\caption{Examples of full iSHELL spectra obtained in this survey. The full sample of spectra is available for interactive visualization at \url{www.spexodisks.com}. }
\label{fig: iSHELL_spec_gallery}
\end{figure*}

Figure \ref{fig: iSHELL_spec_gallery} shows a sample of full iSHELL spectra as obtained in this survey. All the spectra are available for interactive visualization at \url{www.spexodisks.com} \citep[][and Wheeler et al., in prep.]{perez21_spexodisks}.

\section{Peculiarities in individual spectra}
\paragraph{CQ~Tau}
The $M$-band CO spectrum is detected in this source for the first time in this work. The \ce{^{12}CO} low-$J$ $v=1-0$ line profile shows a very narrow central peak and broad line wings, giving the largest shape parameter ($S = 12.3$) measured in the entire combined dataset from the iSHELL and CRIRES surveys. From inspection of the profile and comparison to the high-$J$ and \ce{^{13}CO} profiles, we suspect that this line shape could be the result of a so-far unique situation where two narrow absorption components are carving out the emission profile on the blue and red side of it. This source has also recently been observed with ALMA by \citep{wolfer21}, finding a non-Keplerian velocity structure with spiral arms detected in millimeter CO emission down to 10~au, possibly partly overlapping with the region of $M$-band CO emission. This source should be re-observed to monitor any variability in the line profile that could help support its modeling and interpretation in the future.

\paragraph{MWC~297}
In the iSHELL spectrum of MWC~297, the absorption line FWHM decreases as a function of $J$-level, from a maximum of 10~km/s in the $J=1$ lines down to a minimum of 3.3~km/s in the $v=1-0$ $J=4$ lines (the resolving power of the 0.375" slit). The line shape also changes, becoming less Gaussian and showing saturation effects at the lowest $J$s (see Figure \ref{fig: line_gallery_abs}). 
The \ce{^{13}CO} lines show something similar but with narrower lines and a Gaussian line shape, with FHWM from 7 down to 4~km/s. The mismatch between \ce{^{12}CO} and \ce{^{13}CO} absorption lines widths and shapes is not observed in any other disks in this survey.
It should also be noted that the disk inclination seems to be quite different in different works, as low as 27~deg or 38~deg from VLTI interferometry \citep{lazareff17,kluska20} and as high as 50--65~deg from LBTI observations in \citep{sallum21}, so the geometry of this disk is still somewhat unclear.

\paragraph{MWC~480 / HD~31648}
The iSHELL spectrum taken in 2016 was already published in \citet{banz18}, but by homogeneously reducing with Spextool all the spectra obtained to date we realized that there was an error in the previous reduction. The true line profile does not include the central narrow component shown in Figure B.1 of \citet{banz18}, and it is shown in this work in Figure \ref{fig: line_variability}.

\section{Estimates of inner dust rim radii $R_{\rm{subl}}$} \label{app: dust_radii}
\begin{figure}
\centering
\includegraphics[width=0.45\textwidth]{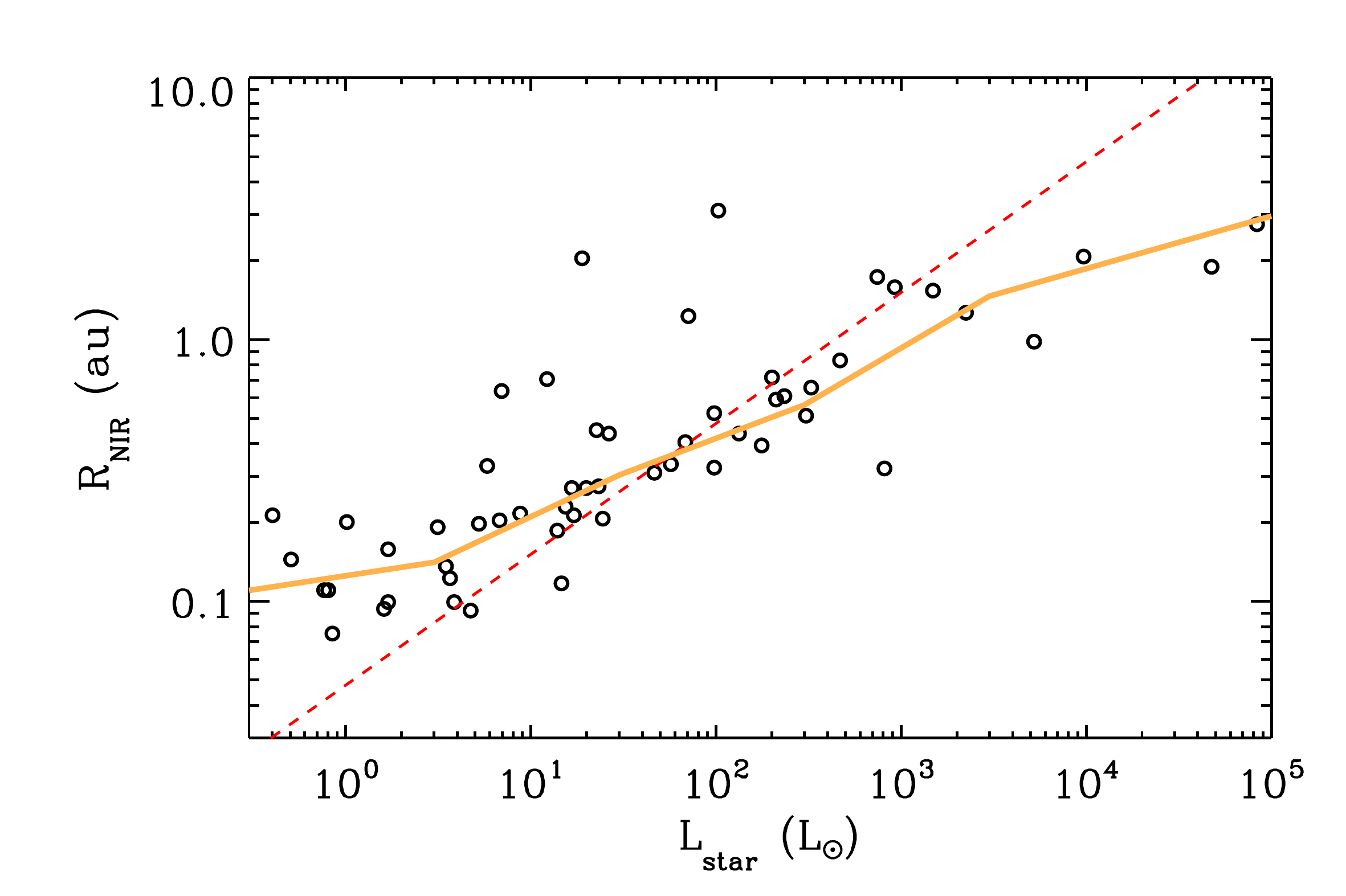} 
\caption{ $R_{\rm{NIR}}$ measurement as collected in \citet{marcosarenal21}, as a function of stellar luminosity. The dashed red line shows the expected dust sublimation radius for a temperature of 1500~K following previous work \citep{monnier02,dullemond10,lazareff17}. The solid orange line shows a moving mean of the measured $R_{\rm{NIR}}$ values, and is used in the paper as an empirical estimate of $R_{\rm{subl}}$ (Table \ref{tab: sample}).
}
\label{fig: Rdust_empirical}
\end{figure}

An estimate of the radial location for the inner dust rim $R_{\rm{subl}}$, typically considered to be set by the disk temperature where dust sublimates, is obtained as a function of luminosity for the whole sample as follows. Dust radii $R_{\rm{NIR}}$ are taken as measured from $K$-band interferometry reported in \citet{marcosarenal21}, which combined the largest sample of NIR interferometry data for Herbig disks (most of them observed with GRAVITY on the VLTI) and an earlier sample of TTauri disks as compiled in \citet{pinte08} and re-derived the physical radii in au from the measured angular sizes using GAIA EDR3 distances. Figure \ref{fig: Rdust_empirical} reports data from Figure~5 in \citet{marcosarenal21} excluding cases where only an upper or lower limit is available. The dashed line shows the expected dust sublimation radius for a sublimation temperature of 1500~K as $R_{\rm{subl}} = \sqrt{L_{\star} / 4 \pi \sigma T_{\rm{subl}}^4}$, following previous work \citep{monnier02,dullemond10,lazareff17}, which can partly provide an explanation but is unable to explain the data across the whole range in luminosities \citep[see e.g. discussion in][]{marcosarenal21}. As an estimate of $R_{\rm{subl}}$, we therefore take and outlier-resistant mean in six logarithmic bins, which is shown with an orange solid line in Figure \ref{fig: Rdust_empirical}, and interpolate that over the measured luminosities in our sample to obtain an estimate of $R_{\rm{subl}}$ in each disk. These estimates are used in Figures \ref{fig: radii_compar} and \ref{fig: discussion} in comparison to $R_{\rm{NIR}}$ and $R_{\rm{CO}}$, and included in Table \ref{tab: sample}.

\section{Luminosity plots for CO excitation tracers} \label{app: co_excit}
\begin{figure}
\centering
\includegraphics[width=0.45\textwidth]{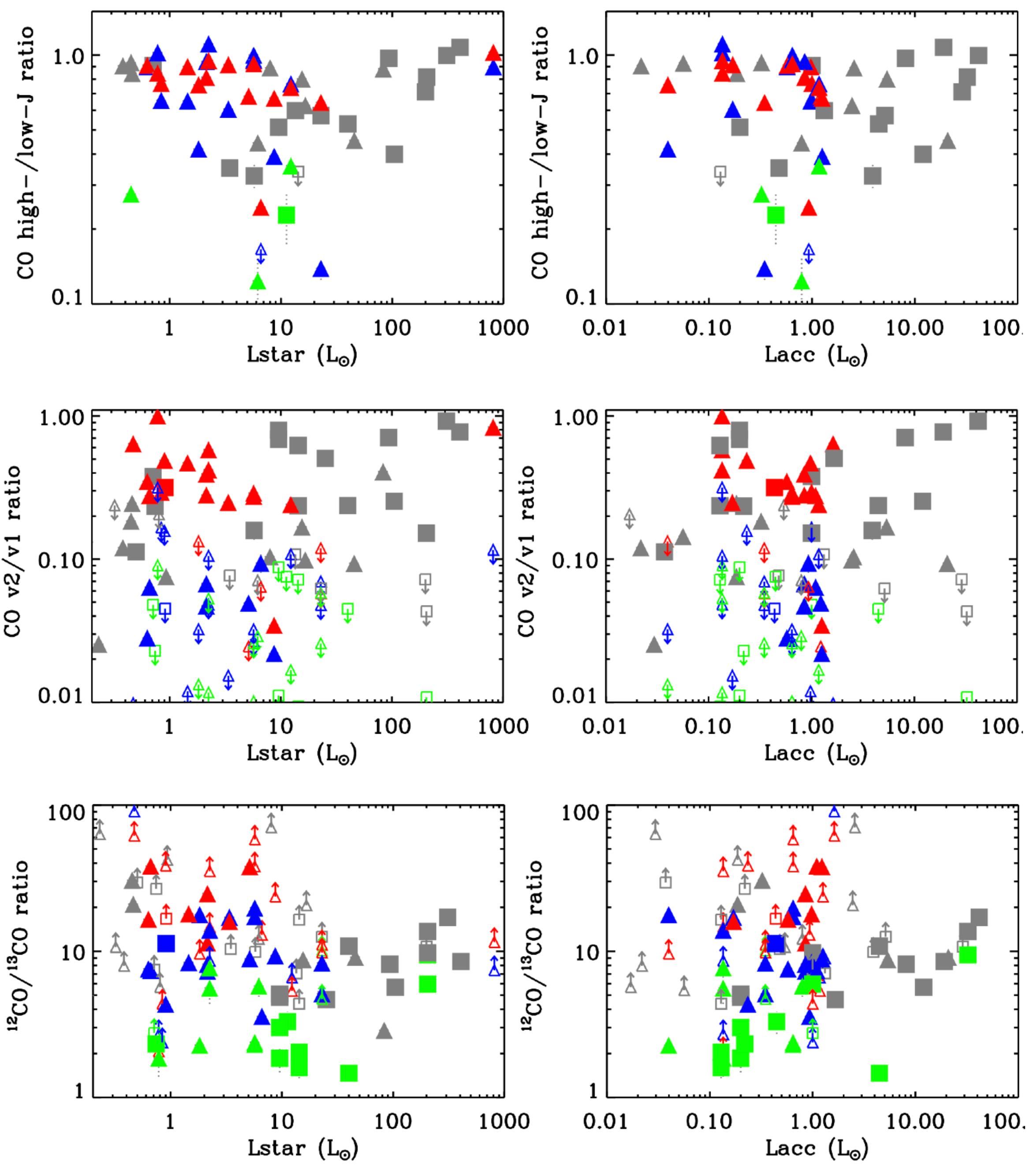} 
\caption{ Plots of the three excitation tracers presented in Section \ref{sec: res_excit}, here shown against the stellar and accretion luminosity. 
}
\label{fig: excit_lum_corr_appendix}
\end{figure}

Figure \ref{fig: excit_lum_corr_appendix} shows scatter plots for the three excitation tracers analyzed in Section \ref{sec: res_excit}, showing no global trends with either the stellar or accretion luminosity. 

\newpage

\section{Observing logs} \label{app: logs}
Table \ref{tab: logs} reports a summary of observing parameters from this iSHELL survey in 2016--2021.

\startlongtable
\begin{deluxetable*}{l c c c c c c c c c}
\tabletypesize{\footnotesize}
\tablewidth{0pt}
\tablecaption{\label{tab: logs} Summary of iSHELL $M$-band observations obtained as part of this survey in 2016-2021.}
\tablehead{ Target & Semester &    Date & Band &  Slit width &  Slit PA &    Time &  Airmass &  Telluric &  Airmass \\ & & $yymmdd$ & & ('') & (deg) & (sec) &  }
\tablecolumns{10}
\startdata
51Oph    & 2017B    & 170802 & M1   & 0.375      & parall  & 474   & 1.5     & HD155125 & 1.31    \\
51Oph    & 2017B    & 170802 & M2   & 0.375      & parall  & 474   & 1.39    & HD155125 & 1.27    \\
ABAur    & 2020B    & 210118 & M1   & 0.75       & 230     & 556   & 1.07    & HR1791   & 1.07    \\
ABAur    & 2020B    & 210118 & M1   & 0.75       & 50      & 556   & 1.12    & HR1203   & 1.06    \\
ABAur    & 2016B    & 161010 & M1   & 0.375      & parall  & 584   & 1.02    & HD39357  & 1.02    \\
ABAur    & 2016B    & 161010 & M2   & 0.375      & parall  & 584   & 1.08    & HD39357  & 1.01    \\
AS205N   & 2021A    & 210417 & M1   & 0.375      & 114     & 3447  & 1.29    & HR6175   & 1.29    \\
CITau    & 2018B    & 190103 & M1   & 0.75       & 11.3    & 2224  & 1.18    & HR1165   & 1.17    \\
CITau    & 2018B    & 190103 & M1   & 0.75       & 11.3    & 3114  & 1.05    & HR1791   & 1.05    \\
CQTau    & 2020B    & 210116 & M1   & 0.375      & 47/227  & 4448  & 1.03    & HR1790   & 1.03    \\
HD141569 & 2017A    & 170419 & M2   & 0.75       & 356/176 & 11388 & 1.11    & HR6556   & 1.08    \\
HD142666 & 2020A    & 200707 & M1   & 0.375      & parall  & 3447  & 1.42    & HR5984   & 1.30    \\
HD142666 & 2021A    & 210415 & M1   & 0.75       & 162     & 2002  & 1.35    & HR5984   & 1.37    \\
HD143006 & 2021A    & 210418 & M1   & 0.75       & 90      & 2836  & 1.38    & HR5984   & 1.37    \\
HD145718 & 2021A    & 210416 & M1   & 0.75       & 90      & 1891  & 1.36    & HR5984   & 1.35    \\
HD150193 & 2020A    & 200707 & M1   & 0.375      & parall  & 556   & 1.39    & HR5984   & 1.32    \\
HD150193 & 2021A    & 210415 & M1   & 0.375      & 176     & 1946  & 1.39    & HR5984   & 1.42    \\
HD163296 & 2021A    & 210415 & M1   & 0.375      & -47/133 & 1557  & 1.35    & HR6378   & 1.36    \\
HD163296 & 2021A    & 210417 & M1   & 0.75       & 43      & 389   & 1.41    & HR6378   & 1.48    \\
HD169142 & 2017A    & 170420 & M2   & 0.75       & 5/185   & 2372  & 1.55    & HR5793   & 1.71    \\
HD169142 & 2017B    & 170826 & M2   & 0.75       & 5/185   & 2135  & 1.55    & HR6879   & 1.71    \\
HD179218 & 2016B    & 161011 & M2   & 0.375      & 23/203  & 1780  & 1.15    & HD177724 & 1.32    \\
HD190073 & 2021A    & 210418 & M1   & 0.375      & -90/90  & 2057  & 1.07    & HR7557   & 1.02    \\
HD259431 & 2020B    & 210117 & M1   & 0.75       & 90      & 556   & 1.01    & HR1791   & 1.08    \\
HD35929  & 2018B    & 190112 & M1   & 0.75       & 90      & 5560  & 1.14    & HR1713   & 1.16    \\
HD35929  & 2018B    & 190113 & M2   & 0.75       & 270     & 3336  & 1.16    & HR1713   & 1.18    \\
HD36917  & 2018B    & 190112 & M1   & 0.75       & 90      & 1112  & 1.55    & HR1713   & 1.67    \\
HD36917  & 2018B    & 190113 & M2   & 0.75       & 270     & 1112  & 1.55    & HR1713   & 1.62    \\
HD37806  & 2020B    & 210116 & M1   & 0.75       & 90      & 834   & 1.19    & HR1790   & 1.22    \\
HD37806  & 2016B    & 161015 & M1   & 0.375      & parall  & 1186  & 1.08    & HD34203  & 1.08    \\
HD58647  & 2018B    & 190113 & M2   & 0.75       & 270     & 1890  & 1.38    & HR2491   & 1.37    \\
IRS48    & 2017B    & 170801 & M2   & 0.375      & 95/275  & 712   & 1.43    & HR5685   & 1.28    \\
LkHa330  & 2017B    & 170826 & M2   & 0.75       & 165/345 & 6405  & 1.11    & HR1220   & 1.07    \\
MWC297   & 2021A    & 210417 & M1   & 0.375      & -75     & 222   & 1.1     & HR7001   & 1.06    \\
MWC297   & 2021A    & 210417 & M1   & 0.375      & 105     & 222   & 1.1     & HR7001   & 1.06    \\
MWC297   & 2021A    & 210417 & M1   & 0.375      & 15      & 222   & 1.1     & HR7001   & 1.06    \\
MWC480   & 2020B    & 210117 & M1   & 0.375      & -32     & 1668  & 1.02    & HR1790   & 1.03    \\
MWC480   & 2016B    & 161011 & M1   & 0.375      & parall  & 2372  & 1.03    & HD39357  & 1.04    \\
MWC480   & 2016B    & 161010 & M2   & 0.375      & parall  & 1168  & 1.12    & HD39357  & 1.01    \\
MWC758   & 2020B    & 210117 & M1   & 0.75       & 90      & 1446  & 1.13    & HR1791   & 1.08    \\
MWC758   & 2020B    & 210118 & M1   & 0.375      & 60      & 3448  & 1.02    & HR1791   & 1.06    \\
MWC758   & 2016B    & 161011 & M1   & 0.375      & parall  & 2372  & 1.01    & HD39357  & 1.04    \\
MWC758   & 2016B    & 161011 & M2   & 0.375      & parall  & 2372  & 1.01    & HD39357  & 1.03    \\
RNO90    & 2021A    & 210417 & M1   & 0.75       & 177     & 2224  & 1.24    & HR6378   & 1.24    \\
RYTau    & 2020B    & 210117 & M1   & 0.75       & 23/-67  & 1668  & 1.1     & HD1203   & 1.11    \\
SR21     & 2017B    & 170801 & M2   & 0.375      & 16/196  & 2372  & 1.4     & HR5685   & 1.28    \\
SUAur    & 2020B    & 210118 & M1   & 0.75       & -30     & 4448  & 1.02    & HR1790   & 1.03    \\
V892Tau  & 2019B    & 200118 & M1   & 0.75       & 90      & 1501  & 1.01    & HR1203   & 1.02    \\
V892Tau  & 2020B    & 210116 & M1   & 0.375      & 235     & 1390  & 1.03    & HR1790   & 1.08    \\
V892Tau  & 2020B    & 210116 & M1   & 0.375      & 55      & 834   & 1.07    & HR1203   & 1.09    \\
VSSG1    & 2021A    & 210418 & M1   & 0.75       & 90      & 2336  & 1.43    & HR5984   & 1.47    \\
VVSer    & 2021A    & 210416 & M1   & 0.75       & 90      & 3058  & 1.07    & HR7001   & 1.10    \\
\enddata
\end{deluxetable*}

\section{Gallery of HI lines in iSHELL spectra}
\begin{figure*}
\centering
\includegraphics[width=1\textwidth]{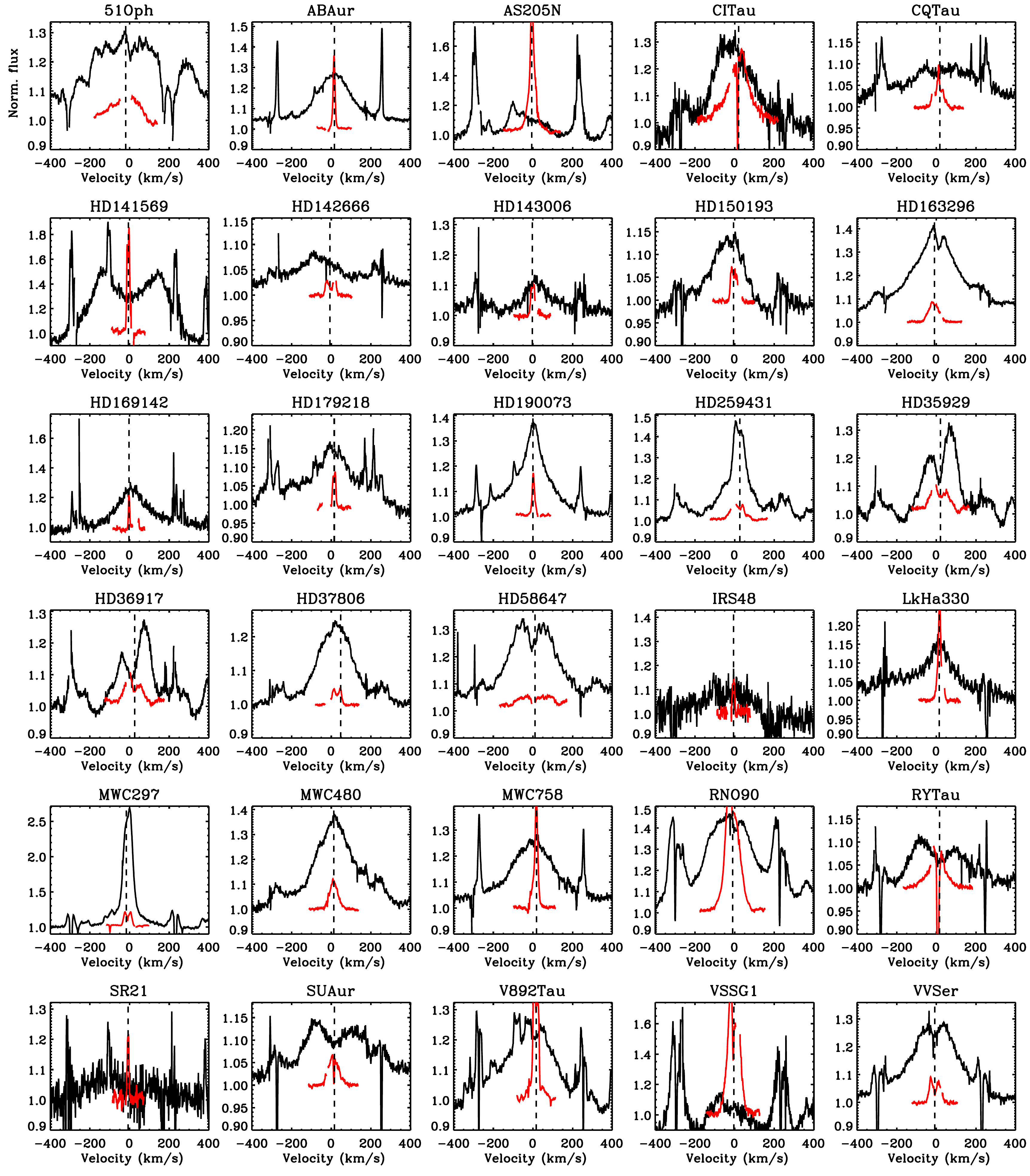} 
\caption{Overview of HI Pfund~$\beta$ line shapes observed in this survey. The stacked line profile of CO $v=1-0$ lines is overplotted in red for reference.}
\label{fig: HI_summary}
\end{figure*}

Figure \ref{fig: HI_summary} shows a gallery of HI Pfund~$\beta$ line shapes observed in this survey. It is worth noting that, while overall HI emission is much broader than CO in each disk as tracing gas closer to the star in the accretion region \citep{salyk13}, in a few cases there is an additional narrower HI component that gets much closer in FWHM to CO (HD~259431 and MWC~297 in this sample); these cases are worth further investigation to clarify the origin of this narrower component. Some double-peak lines are evident in HI (e.g. in HD~141569 and SU~Aur), though with a much lower frequency than the CO lines in this sample. In some cases, narrow absorption seems to be present on top of HI emission (e.g. in HD~163296, HD~35929); this absorption is unrelated to the much broader absorption from stellar photospheres, which in this work we have not corrected for \citep[see more in][]{salyk13}.

\section{Gallery of CO line profiles from CRIRES spectra} \label{app: crires_lines}
\begin{figure*}
\centering
\includegraphics[width=1\textwidth]{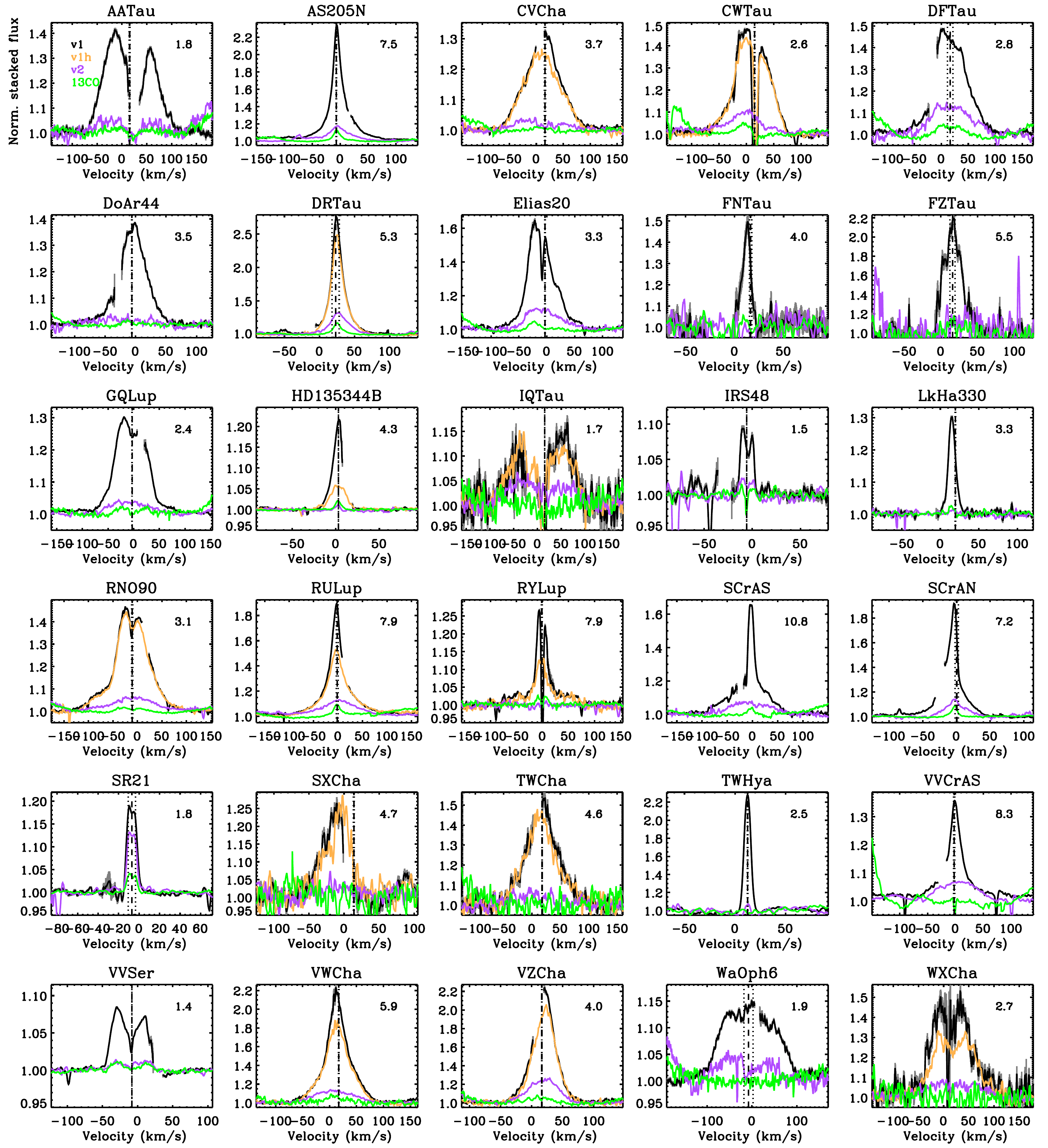} 
\caption{Same as Figure \ref{fig: line_gallery} but for spectra from the CRIRES survey (Section \ref{sec: sample}). Line stacks are shown in different colors for \ce{^{12}CO} $v=1-0$ $J=$~5--18 (black), $v=1-0$ $J=$~25-31 (orange), $v=2-1$ (purple), \ce{^{13}CO} $v=1-0$ (green). The high-$J$ line stack is often missing due to the incomplete spectral coverage of CRIRES. Deviations from a flat continuum at the edges of some lines is due to contamination from nearby lines. The vertical dashed and dotted lines shows the stellar RV and its uncertainty. The measured line shape parameter $S$ is reported at the top right in each plot.}
\label{fig: CRIRES_lines}
\end{figure*}

Figure \ref{fig: CRIRES_lines} shows CO line stacks from the CRIRES survey. These spectra have already been published in previous works \citep[for the entire survey, see][]{brown13}, but here we have re-processed them with the analysis pipeline described in Section \ref{sec: analysis}. 

\end{document}